%% file: mn2eastroph.tex
\documentclass[useAMS,usenatbib]{mn2e}

\usepackage{graphicx}
\usepackage{lscape}
\usepackage{aalongtable}
\usepackage{ulem}

\title[s-Process in Low Metallicity Stars. I.]
{s-Process in Low Metallicity Stars. \\ 
I. Theoretical Predictions}
\author[S. Bisterzo, R. Gallino, O. Straniero, S. Cristallo, F. K\"appeler]
{S. Bisterzo$^{1}$\thanks{E-mail: bisterzo@ph.unito.it (AVR); sarabisterzo@gmail.com (ANO)} 
and 
R. Gallino$^{1}$
and 
O. Straniero$^{2}$
and 
S. Cristallo$^{2,3}$
and 
F. K\"appeler$^{4}$\\
$^{1}$Dipartimento di Fisica Generale, Universit\`{a} 
   di Torino, Via P. Giuria 1, 10125 Torino, Italy\\
$^{2}$INAF Osservatorio Astronomico di Collurania, via M. 
   Maggini, 64100 Teramo, Italy\\
$^{3}$Departamento de Fisica Teorica y del Cosmos, Universidad
de Granada, Campus de Fuentenueva, 18071 Granada, Spain\\
$^{4}$Karlsruhe Institute of Technology, Campus Nord, Forschungszentrum 
Karlsruhe, Institut f$\ddot{\rm u}$r Kernphysik, D-76021 Karlsruhe, Germany\\}

\begin{document}

\date{Accepted 1988 December 15. Received 1988 December 14; in original form 1988 October 11}

\pagerange{\pageref{firstpage}--\pageref{lastpage}} \pubyear{2002}

\maketitle

\label{firstpage}

\begin{abstract}

A large sample of carbon enhanced metal-poor stars enriched in $s$-process 
elements (CEMP-$s$) have been observed in the Galactic halo. 
These stars of low mass ($M$ $\sim$ 0.9 $M_{\odot}$) are located on
the main-sequence or the red giant phase, and do not undergo third 
dredge-up (TDU) episodes.
The $s$-process enhancement is most plausibly due to 
accretion in a binary system from a more massive 
companion when on the asymptotic giant branch (AGB) 
phase (now a white dwarf).
In order to interpret the spectroscopic observations, updated AGB 
models are needed to follow in detail the $s$-process 
nucleosynthesis. We present nucleosynthesis calculations based on 
AGB stellar models obtained with FRANEC (Frascati 
Raphson-Newton Evolutionary Code) for low initial stellar 
masses and low metallicities.
For a given metallicity, a wide spread in the abundances of the 
$s$-process elements is obtained 
by varying the amount of $^{13}$C and its profile in the pocket,
where the $^{13}$C($\alpha$, n)$^{16}$O reaction is 
the major neutron source, releasing neutrons in radiative 
conditions during the interpulse phase. We account also for the second 
neutron source $^{22}$Ne($\alpha$, n)$^{25}$Mg, partially activated 
during convective thermal pulses.
We discuss the surface abundance 
of elements from carbon to bismuth, for AGB 
models of initial masses $M$ = 1.3 -- 2 $M_\odot$, low metallicities 
([Fe/H] from $-$1 down to $-$3.6) and for different $^{13}$C-pockets efficiencies.
In particular we analyse the relative behaviour of the three 
$s$-process peaks: light-$s$ (ls at magic neutron number N = 50), heavy-$s$ (hs at 
N = 82) and lead (N = 126). Two $s$-process indicators, [hs/ls] and [Pb/hs], 
are needed in order to characterise the $s$-process distribution. 
In the online material, we provide a set of data tables 
with surface predictions. 
Our final goal is to provide a full set of theoretical models of low mass
low metallicity $s$-process enhanced stars. In a forthcoming paper, we will
test our results through a comparison with observations of CEMP-$s$ stars.

\end{abstract}

\begin{keywords}
Stars: AGB -- Stars: carbon -- Stars: Population II -- nucleosynthesis
\end{keywords}


\section{Introduction}\label{intro}

In the last years, high-resolution spectroscopic surveys of very metal-poor stars
have acquired particular interest, especially for those
stars with an appreciable carbon enhancement, i.e. [C/Fe] $>$ 1\footnote{The 
standard spectroscopic notation is adopted: for two generic elements A and B, 
[A/B] = log$_{10}$(N$_{\rm A}$/N$_{\rm B}$)$_{\star}$ - 
log$_{10}$(N$_{\rm A}$/N$_{\rm B}$)$_{\odot}$
\citep*{helfer59}.}, the so-called Carbon Enhanced Metal-Poor stars (CEMP, 
\citealt{beers05}). A large number of CEMP stars has been recently 
discovered by several survey projects, as the HK-survey \citep{beers92,beers07}, 
the ESO Large Programme First Stars with the ESO VLT and UVES spectrograph
(e.g. \citealt{cayrel04}), the Hamburg/ESO Survey \citep{christlieb03}, the 
SEGUE survey (Sloan Extension for Galactic Exploration and Understanding), 
the SEGUE Stellar Parameter Pipeline (SSPP; \citealt{lee08a,lee08b}), the Sloan 
Digital Sky Survey (SDSS, \citealt{york00}),
the Chemical Abundances of Stars in the Halo (CASH) Project \citep{roederer08}.
These surveys provide essential means to the understanding of 
the early universe and the Galactic chemical evolution.
 
To interpret the surface abundances of these objects, theoretical 
stellar evolutionary models are needed.
We concentrate here on CEMP stars with enhancement in 
elements produced via $slow$-neutron capture process ($s$-process; 
\citealt{burbidge57,clayton61,clayton68,kaeppeler82}), the so-called
CEMP-$s$ stars.
It is commonly assumed that CEMP-$s$ 
stars belong to binary systems where a more massive asymptotic giant branch 
(AGB) companion, now a white dwarf, had synthesised the $s$-elements and 
polluted the observed star by mass transfer. 
Different studies about AGB nucleosynthesis are available in the literature, 
either with post-process method or with full stellar evolutionary models 
(\citealt{straniero95,straniero97,straniero03}; \citealt{goriely00}; 
\citealt{herwig04}; \citealt{campbell08}; 
\citealt{cristallo09,cristallo09pasa}). 
Our calculations are based on FRANEC (Frascati Raphson-Newton Evolutionary 
Code) \citep{chieffi89}, coupled with a post-process method which includes
a full network up to bismuth \citep{gallino98}.

Late on the AGB, two alternate energy sources are activated, the H-burning 
shell and the He-burning shell. These two shells are separated by a thin 
zone in radiative equilibrium, the so-called He-intershell. The H-burning shell 
progressively erodes the bottom layers of the envelope and produces helium.
Consequently, the He-intershell grows in mass and is progressively 
compressed and heated until the temperature and density become high enough 
to trigger a thermonuclear runaway in the He shell (\textsl{thermal pulse}, 
TP). The sudden energy release causes the above layers to be unstable 
against convection, thus mixing material over the whole He-intershell 
\citep{SH65,SH67}. The consequent expansion of the region above the 
He-burning shell temporarily extinguishes the H-shell. 
In these conditions, the convective envelope can penetrate under the 
H-shell, bringing to the surface newly synthesised material 
(\textsl{third dredge-up}, TDU). During TDU, a chemical discontinuity is 
established at the interface between the H-rich envelope and the 
He-intershell, with a sudden change of the opacity, and then, of the 
temperature gradient. Thus, at the radiative border, the convective velocity 
becomes abruptly zero, whereas it should exist a transition 
region, where the convective velocity smoothly decreases to 0. Therefore, 
a partial mixing may take place. Under this hypothesis,
a small amount of hydrogen from the envelope penetrates into the top 
layers of the He-intershell \citep{iben83}. Then, at H-shell reignition, a 
thin $^{13}$C-pocket forms in the top layers of the He-intershell, 
by proton capture on the abundant $^{12}$C. 
There is an upper limit in the $^{13}$C production in the pocket.
Indeed, a higher proton abundance ingestion would produce $^{14}$N by proton 
capture on $^{13}$C.
This $^{13}$C is \textsl{primary}, 
because the diffused protons are captured by $^{12}$C directly produced by 
helium burning during previous TPs, regardless of the metallicity. 
During the interpulse phase, when the temperature in the 
$^{13}$C-pocket increases up to 1 $\times$ 10$^{8}$ K, neutrons are released 
in radiative conditions via the $^{13}$C($\alpha$, n)$^{16}$O reaction, before 
the occurrence of the next thermal instability. 
This reaction constitutes the major neutron source in low mass AGB 
stars (1.2 $\la$ $M/M_\odot$ $\la$ 3, hereafter LMS), (\citealt{straniero95};
\citealt{gallino98}).
\\ 
During the convective episode originated by a TP, the $s$-process material 
synthesised in the pocket is diluted over the whole He-intershell, while 
the abundant $^{14}$N nuclei in the H ashes are converted 
to $^{22}$Ne via the $^{14}$N($\alpha$, $\gamma$)$^{18}$F($\beta$$^{+}$$\nu$)$^{18}$O($\alpha$, 
$\gamma$)$^{22}$Ne chain. 
In LMS, the maximum temperature at the bottom of the TP slightly increases 
with pulse number, depending on the initial mass and metallicity, reaching 
$T$ $\sim$ 3 $\times$ 10$^{8}$ K toward the end of the AGB phase.
At this temperature the $^{22}$Ne($\alpha$, n)$^{25}$Mg 
reaction is partially activated, producing a neutron burst of small neutron 
exposure\footnote{The neutron exposure is the time integrated neutron flux 
$\tau$ = $\int n_n v_{th} dt$, where $n_n$ is the neutron density and $v_{th}$ 
is the thermal velocity.} but with a high neutron density peak $n_n^{peak}$ 
$<$ 10$^{11}$ cm$^{-3}$, followed by a rapid decline (neutron freezout). 
The time dependence of this neutron burst is particularly 
important for defining the freezeout conditions for most of the branchings,
that are sensitive to temperature and neutron density.
In intermediate mass AGB stars (4 $\la$ $M/M_\odot$ $\la$ 8, hereafter 
IMS), the maximum temperature during a TP reaches about 3.5 $\times$ 10$^{8}$ K, 
thus leading to a substantial activation of the $^{22}$Ne($\alpha$, n)$^{25}$Mg. 

The aim of this paper is to present in detail theoretical predictions of LMS 
AGB stars ($M$ = 1.3, 1.4, 1.5 and 2.0 $M_\odot$) as a function 
of metallicity, starting from solar down to [Fe/H] = $-$3.6\footnote{The metallicity
is defined as 
\begin{equation}
[{\rm Fe/H}] = {\rm log}_{10}({\rm Fe/H}) - {\rm log}_{10}({\rm Fe/H})_{\odot}.
\end{equation}},
focusing on stars with [Fe/H] $\leq$ $-$1.
Additional models of IMS in the range of $-$1.6 $\leq$ [Fe/H] $\leq$ 0
are discussed to highlight the effects that the mass has on the stellar 
structure and evolution.
Comparison with spectroscopic observations of CEMP-$s$ stars will be presented 
in a forthcoming paper (Bisterzo et al., in preparation, hereafter Paper II).
The models on which we base our calculations 
are described in Section~\ref{models}. We study the behaviour of all the 
elements from carbon to bismuth, the three $s$-process peaks (light $s$-process 
elements \textit{ls}, heavy $s$-process elements \textit{hs}, and 
[Pb/Fe])\footnote{We define \textit{ls} and \textit{hs} as 
\begin{equation}
\centering
\label{eqls}
[{\rm ls/Fe}] = 1/2([{\rm Y/Fe}] + [{\rm Zr/Fe}]),
\end{equation}
\begin{equation}
\centering
\label{eqhs}
[{\rm hs/Fe}] = 1/3([{\rm La/Fe}] + [{\rm Nd/Fe}] + [{\rm Sm/Fe}]).
\end{equation}
We exclude Sr from the ls elements and Ba from the 
hs elements because they are affected by spectroscopic 
uncertainties, as discussed by \citet{busso95} (see also Paper II). 
Moreover, non-local
thermodinamical equilibrium (NLTE) corrections, can be significant for 
Sr and Ba in particular at low metallicities (\citealt{andr09};
\citealt{mashonkina08}; \citealt{sh06}).},
and the $s$-process indicators [hs/ls] and [Pb/hs] by varying the  
metallicity and the $^{13}$C-pocket efficiency, which is a 
parameterisation of the amount of $^{13}$C and its profile in the pocket
 (Section~\ref{results}).
We also discuss the important effect of $^{22}$Ne at low metallicities
(Section~\ref{ne22}), and the relevance of Na (and Mg) abundances 
as indicators of the initial AGB mass (Section~\ref{Na}).
In Section~\ref{conclusions} we give our conclusions.


\section{Description of the models} \label{models}

\begin{figure}
  \centering
\includegraphics[angle=-90,width=8cm]{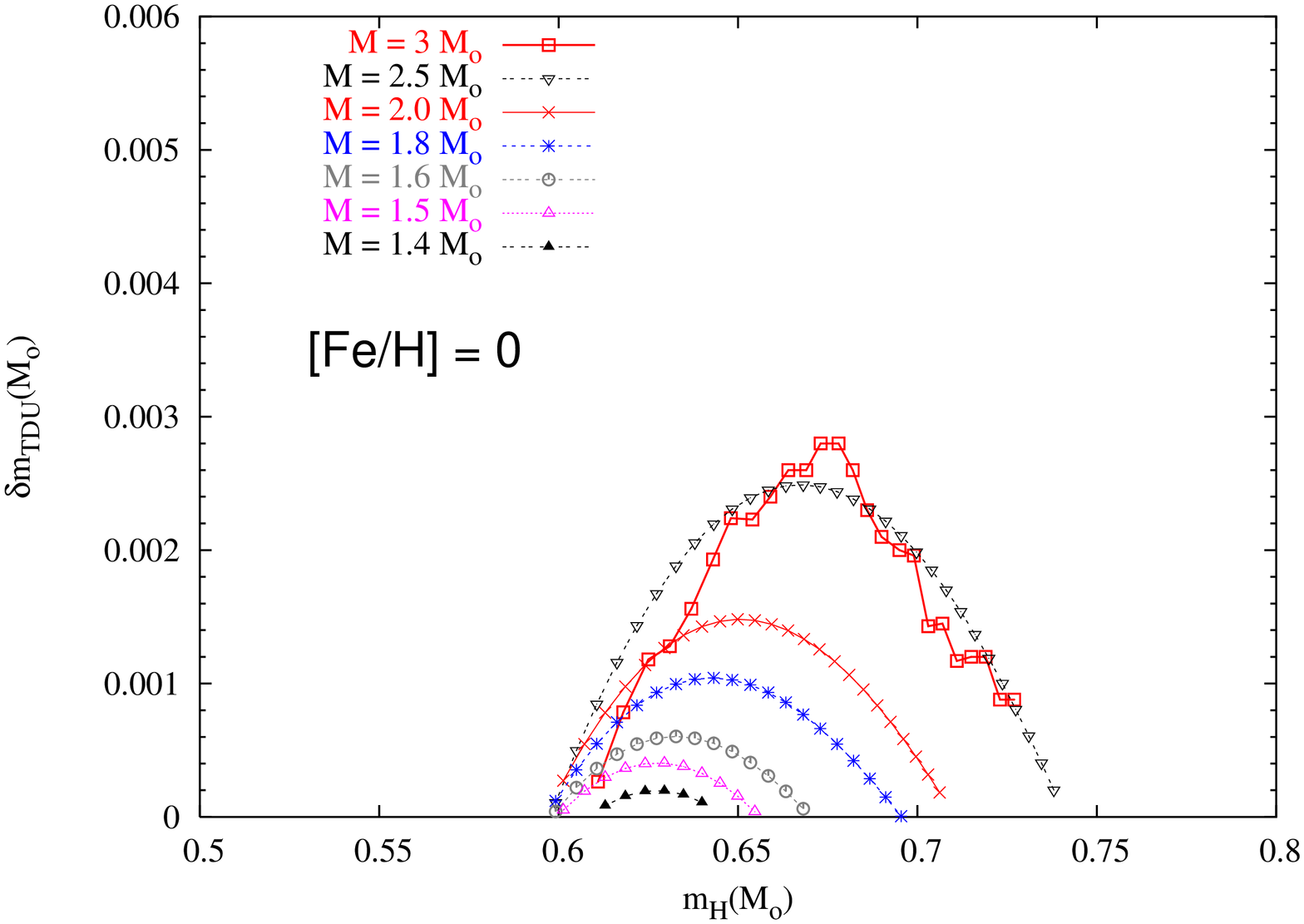}
\includegraphics[angle=-90,width=8cm]{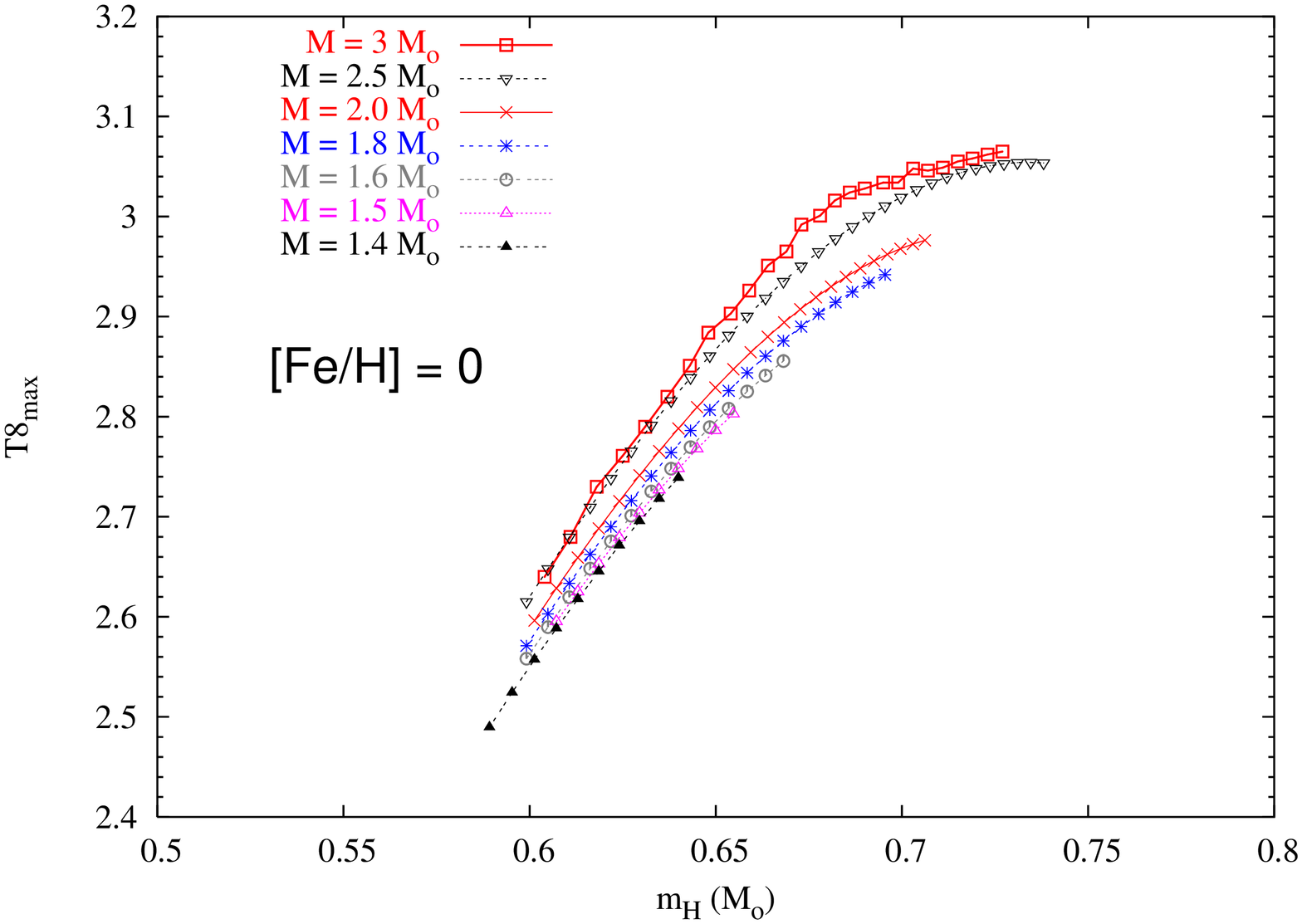}
\caption{\textit{Top panel:} the mass of the $^{12}$C-rich and 
$s$-process-rich layer, 
$\delta m_{\rm TDU}$, that is mixed into the envelope after each TDU 
episode, as a function of the position in mass of the H shell,
$m_{\rm H}$, for AGB models of [Fe/H]~=~0 and 
different initial masses. 
Each symbol on the curves indicates
 the appearance of the various TDU episodes. 
The location of $m_{\rm H}$ grows in time due to the activation 
of the H-burning shell in the interpulse phase. 
\textit{Bottom panel:} evolution of the maximum temperature at the 
base of the convective shell generated by the TPs as a
function of $m_{\rm H}$ for AGB models at [Fe/H]~=~0 and
different masses. $T_{\rm 8}$ stand for 1 $\times$ 10$^{8}$ K.
(\textit{See the electronic paper for a colour version 
of this and the following figures.})}
\label{MTDU}
\end{figure} 
\begin{figure}
   \centering
\includegraphics[angle=-90,width=8cm]{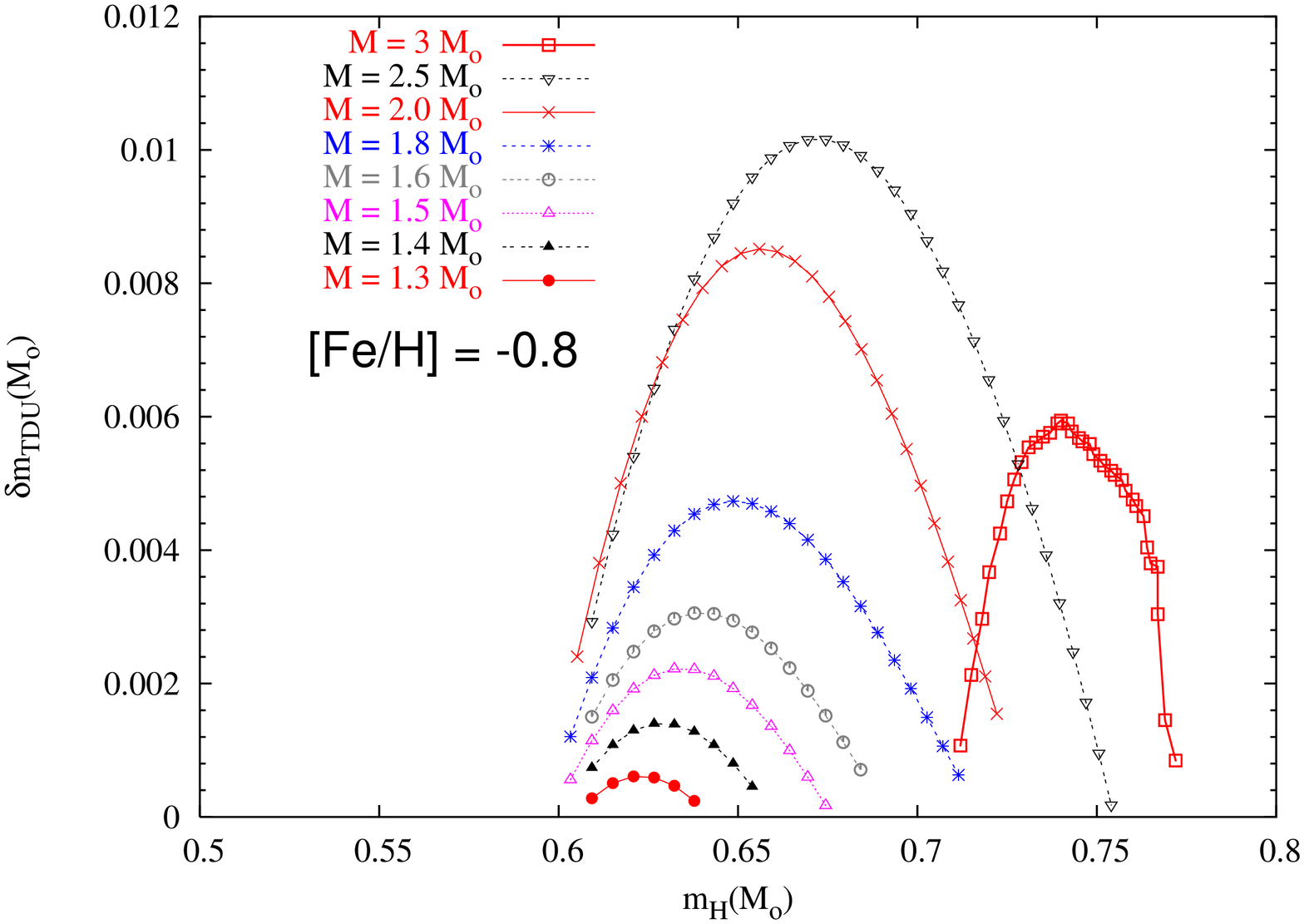}
\includegraphics[angle=-90,width=8cm]{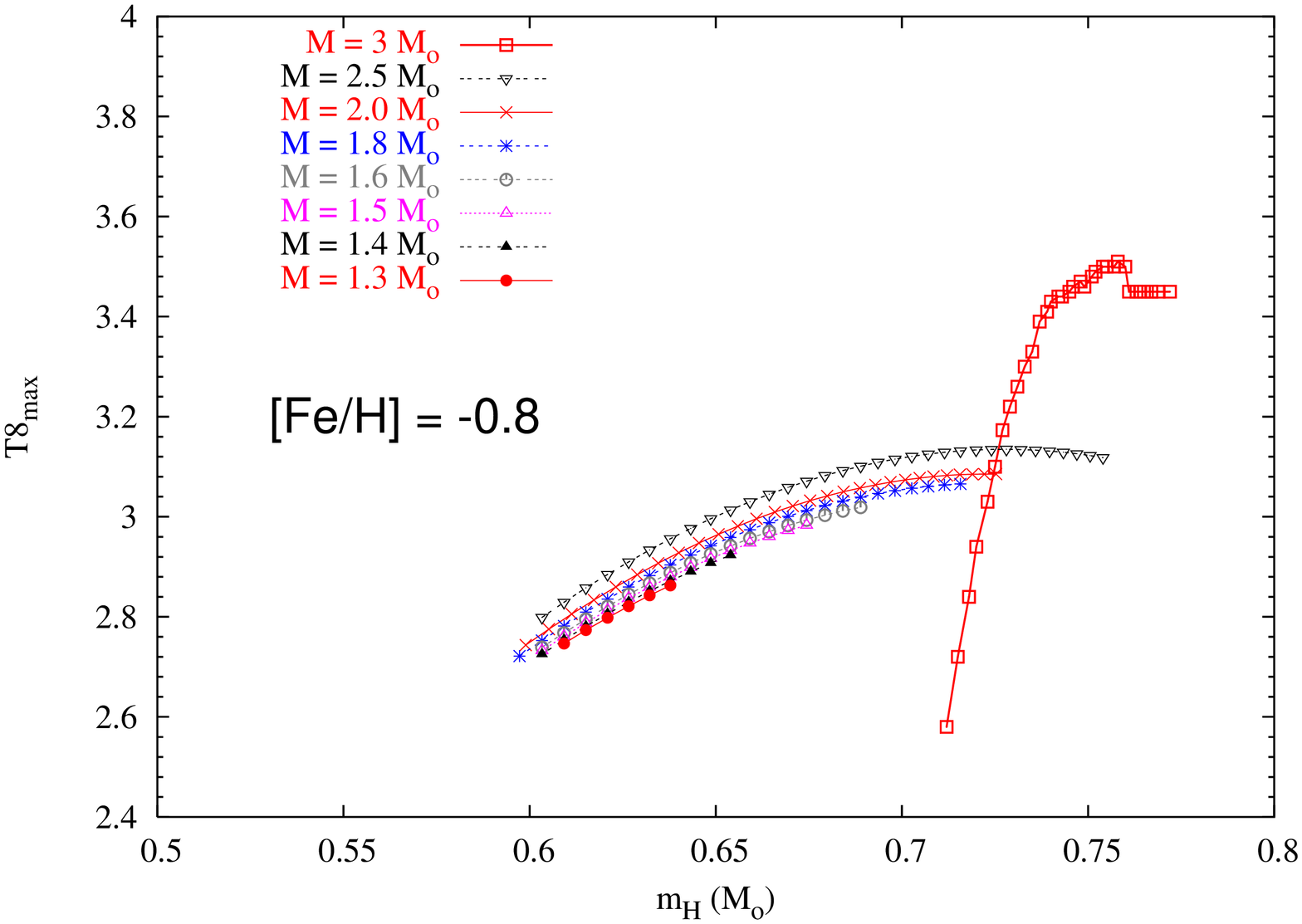}
\caption{The same as Fig.~\ref{MTDU}, but at [Fe/H] = $-$0.8.
Note the different scale adopted with respect
to Fig.~\ref{MTDU}, top panel. By decreasing the 
metallicity, the mass involved in the TDU significantly
increases.}
\label{MTDU1}
\end{figure} 
\begin{figure}
   \centering
\includegraphics[angle=-90,width=8cm]{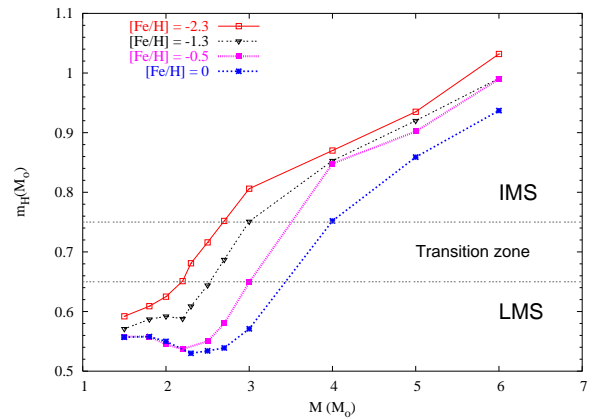}
\caption{The mass of the core at the first TP versus the initial
mass of the star, at different metallicities. The two horizontal lines
correspond to a transition limit between LMS and IMS.
Values are provided by Tables~1 --~4 of \citet{dominguez99}.}
\label{Mcore}
\end{figure} 

We perform the $s$-process nucleosynthesis with a post-process 
method \citep{gallino98} based on full evolutionary models obtained with FRANEC code
(\citealt{straniero97,straniero00,straniero03}).
Our post-processing calculations follow
the convective instabilities generated by TPs, the 
$^{13}$C-pocket burning and the TDU episodes.
The nuclear reaction rates are updated to 2009
(see KADoNiS\footnote{Karlsruhe Astrophysical Database of Nucleosynthesis in 
Stars, web address http://nuclear-astrophysics.fzk.de/kadonis/.} and 
further references given in Appendix~\ref{arlandiniupdated}).
The main differences with respect to previous publications is the 
discovery and the correction of an old bug in the models, which affected
the treatment of $^{16}$O produced in the pocket via $^{13}$C($\alpha$, 
n)$^{16}$O.
In addition, protons captures are now included to improve the prediction 
of the light elements. In particular $^{19}$F: reaction rates of isotopes 
involving charged particles are taken from the NACRE compilation \citep{angulo99} 
(web address http://pntpm3.ulb.ac.be/Nacre/barre$\_$database.htm).
We follow the prescriptions given by \citet{straniero03}
for input data such as:
\begin{itemize}
	\item  the mass of the material dredged-up at each TDU episode,
 	\item  the temporal history of the temperature and density during the 
 	TP and their distribution in the various mass zones of the convective TP, 
	\item  the mass of the H-shell,
	\item  the mass of the He-intershell, 
	\item  the \textsl{overlapping factor} $r$ between successive TDUs, 
	\item  the mass of the envelope, which decreases with the TP number because of
mass loss and by the activation of the H-burning shell, which
progressively erodes the bottom of the envelope.
\end{itemize}
The interpolation formulae by \citet{straniero03} were available
in a grid of masses up to $M$ = 2 $M_\odot$ and in the metallicity range of
$-$1 $\la$ [Fe/H] $\la$ 0.
We have further extrapolated the stellar parameters down to [Fe/H] = $-$3.6.

\begin{figure*}
   \centering
\includegraphics[angle=-90,width=10cm]{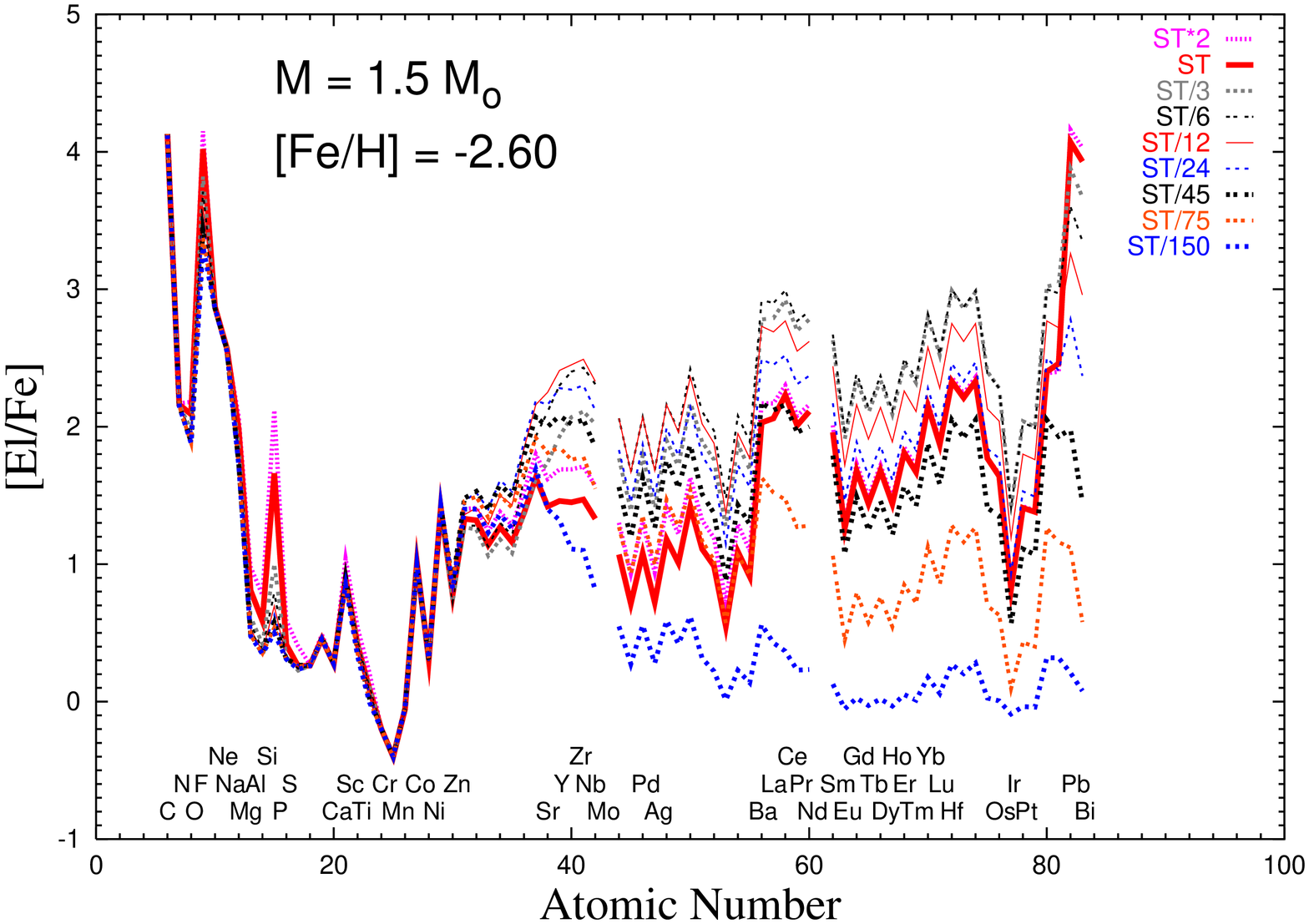}
\includegraphics[angle=-90,width=10cm]{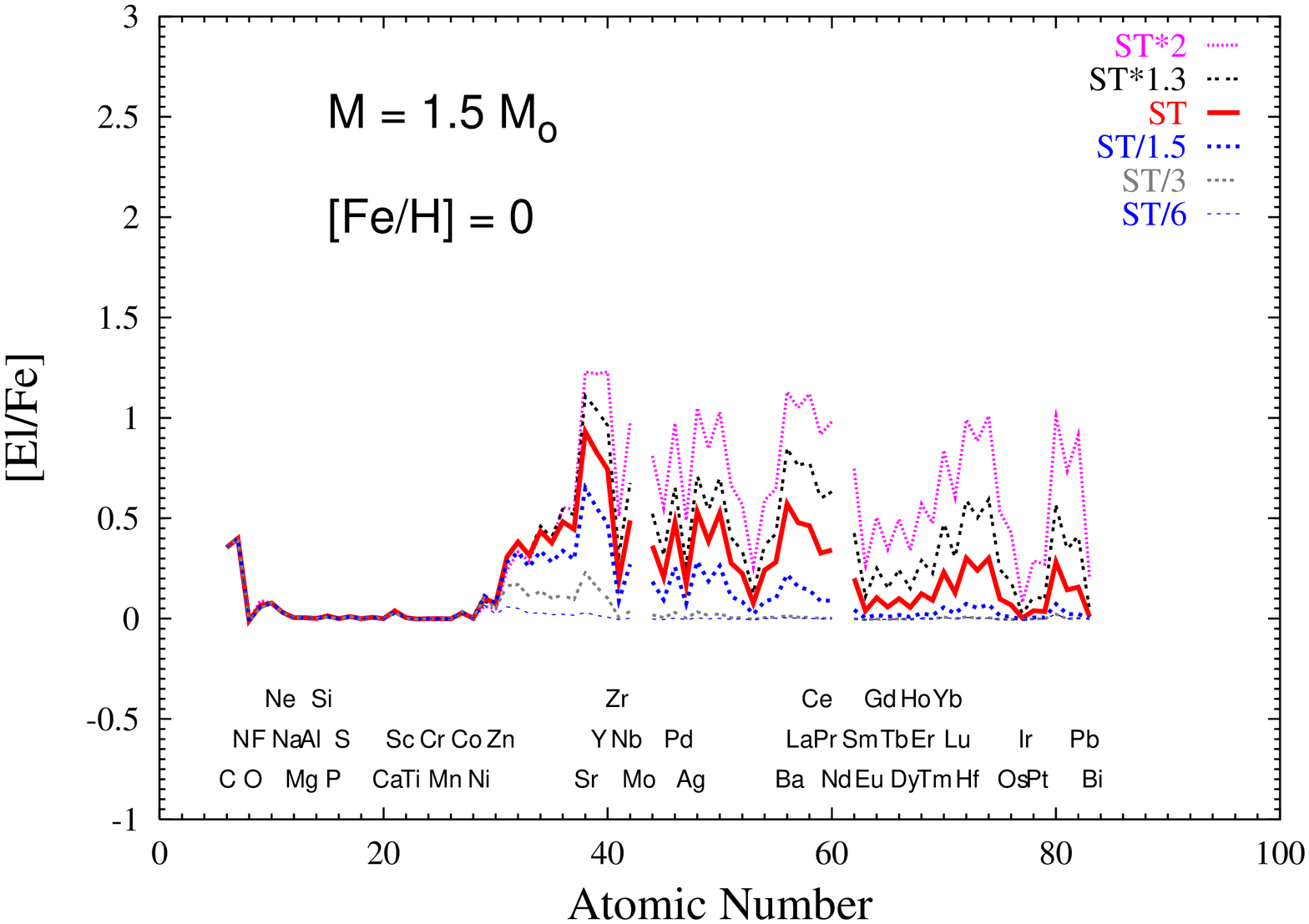}
\caption{\textit{Top panel}: elemental composition in the envelope 
at the last TDU, for AGB models of initial mass $M$ = 1.5 $M_{\odot}$, 
initial metallicity [Fe/H] = $-$2.6, and different choices of the 
$^{13}$C-pocket efficiency.
Similar results are predicted for $M$ = 2.0 $M_{\odot}$.
In the on line material, Fig.~\ref{m1p5z5m5nro16eq_alcuniST_n5}, 
the same plot is shown for an AGB model of initial mass $M$ = 
1.3 $M_{\odot}$.
\textit{Bottom panel}: the same as top panel, but at [Fe/H] = 0.
These models are for intrinsic AGBs
([Zr/Nb] $\sim$ 1, see Sect.~\ref{binary}).}
\label{m1p5z5m5nro16eq_alcuniST}
\end{figure*}

For a given metallicity, a spread in the $s$-process 
elements is observed in MS, S, C(N) and Ba stars of the 
Galactic disk (see \citealt{busso95,busso01}; \citealt{abia01,abia02}; 
\citealt{gallino05}; \citealt{husti08}). 
At low metallicities, recent high-resolution spectroscopic 
measurements of CEMP-$s$ stars showed an even larger spread 
(\citealt{ivans05}; \citealt{aoki06}; \citealt{thompson08}; 
\citealt{roederer08}; \citealt{bisterzo08a,bisterzo08b}; 
 \citealt*{SCG08}). 
A range of $s$-process efficiencies is therefore needed in order 
to interpret the observations.  In our calculations, we artificially 
introduce the $^{13}$C-pocket, which is treated as a free parameter 
and kept constant pulse by pulse. 
Starting from the ST case adopted by \citet{gallino98} and 
\citet{arlandini99}, which was shown to reproduce the solar $s$-process
main component as the average of the 1.5 and 3.0 $M_{\odot}$ models at 
half solar metallicity (as described in Appendix~\ref{arlandiniupdated}),
we multiply or divide the $^{13}$C (and $^{14}$N) abundances in the 
pocket by different factors.
The pocket extends in mass for 9.4 $\times$ 10$^{-4}$ $M_{\odot}$
(for LMS about 1/20 of the typical mass involved in a TP),
and contains 4.7 $\times$ 10$^{-6}$ $M_{\odot}$ of $^{13}$C
and 1.6 $\times$ 10$^{-7}$ $M_{\odot}$ of $^{14}$N.
These masses are the integrated masses in the pocket accounting 
for the $^{13}$C and $^{14}$N profiles, respectively.
 $^{14}$N has a resonant neutron
capture reaction rate ($\sigma$[$^{14}$N(n, p)$^{14}$C]$_{({\rm 30 keV})}$ = 
1.85 mbarn, \citealt{koehler89}): $^{14}$N(n, p)$^{14}$C preventing 
the captured neutrons to be available for the $s$-process nucleosynthesis 
(while $^{14}$C has a very low neutron capture cross section). 
We exclude the possible upper zone 
close to the border of the TDU where higher proton injection would 
produce $^{14}$N $>$ $^{13}$C.
The case ST$\times$2 corresponds to an upper limit, because 
further proton ingestion leads to the formation of $^{14}$N at expenses
of $^{13}$C. 
The $s$-process efficiency of the $^{13}$C-pocket increases by 
decreasing the metallicity because of the lower number of iron seeds. 
We may define a minimum $^{13}$C-pocket as the one which significantly 
affect the final $s$ distribution, below which the $s$-process path 
mostly depends on the $^{22}$Ne neutron exposure only.
As first approximation, for solar metallicity the lowest $^{13}$C-pocket
is ST/6, and ST/150 for [Fe/H] = $-$2.6 (see Section~\ref{results}, 
Fig.~\ref{m1p5z5m5nro16eq_alcuniST} for further details).
In IMS the He-intershell mass involved is smaller than LMS by one 
order of magnitude and the TDU efficiency is reduced.
For this reason, a different choice of $^{13}$C-pocket is 
used for these stars: we defined the case ST-IMS, with
 $M$($^{13}$C$_{\rm ST-IMS}$) = 10$^{-7}$ $M_{\odot}$.
\\
The crude approximation in the treatment of the 
$^{13}$C-pocket reflects a significant uncertainty
which affects AGB models: in particular, the mixing processes at 
radiative/convective interfaces during TDU episode.
In fact, the penetration of protons in the He-intershell during 
TDUs is currently matter of study and one of the most debated issues. 
Models including rotation \citep{langer99}, or gravity waves 
\citep{denissenkov03} have obtained a partial mixing zone at the 
base of the convective envelope during TDU episode, which leads 
to the formation of a $^{13}$C-rich layer of limited mass extension. 
Similar results have been obtained by \citet{herwig97} and \citet{herwig00},
who introduced an exponential diffusive overshoot at the 
borders of all convective zones.
A formally similar algorithm, but based on a different mixing scheme
(i.e. not diffusive) has been proposed by \citet{straniero06}, 
(see also \citealt{cristallo09}).
Finally, we remind that the efficiency of the TDU itself is 
significantly affected by the different treatment of the mixing algorithm, 
and it is currently a main problem with AGB computations. 
Different evolutionary codes do not reproduce the same results, 
since they are based on different treatment of the mixing algorithm
(see, for instance \citealt{straniero03,herwig04,karakas07,stancliffe07,cristallo09},
and references therein).
\\
In AGB stars, photospheric $s$-process enrichment is observed 
only when the TDU takes place. 
According to FRANEC model, for LMS it occurs after a limited 
number of pulses, when the mass of the H-exhausted core reaches $\sim$ 0.6 
$M_\odot$ \citep{straniero03}, bringing freshly 
synthesised material from the He-intershell into the envelope.
The TDU efficiency 
depends on the envelope mass and on the H-burning rate
(see \citealt{straniero00}).
The amount of the mass dredged-up by each TDU episode, $\delta$$m_{\rm TDU}$,
first increases as the core mass $m_{\rm H}$ increases, 
and then decreases until vanishing when the envelope mass is reduced 
by mass loss to a fraction of it (about 0.5 $M_\odot$ at solar metallicity).
\\
The minimum envelope mass for the occurrence of the TDU depends upon 
the metallicity, the core mass and the mass loss adopted. 
The minimum initial mass for a star that undergoes recursive 
TDU episodes depends on the metallicity. For a solar composition 
model, $M^{\rm min}_{\rm TDU}$ $\sim$ 1.4 $M_\odot$, and it
decreases with metallicity \citep{straniero03}. 
In Fig.~\ref{MTDU}, we show the mass involved in TDUs (top 
panel) and the maximum temperature at the base of the convective 
shell generated by the TPs (bottom panel) as a function 
of the position in mass of the H-shell, $m_{\rm H}$, 
for AGB models of solar metallicity and different initial masses.
Calculations of $M$ $\leq$ 2.5 $M_\odot$ have been computed with interpolation 
formulae published by \citet{straniero03}.
Note that the mass involved in the TDU increases with the initial AGB mass 
at [Fe/H] = 0. 
At solar metallicity $M$ = 3 $M_\odot$ shows a peculiar
behaviour of the TDU mass. Indeed, its maximum TDU mass is of the same order
of $M$ = 2.5 $M_\odot$, and does not further increase.
The same of Fig.~\ref{MTDU} is shown in Fig.~\ref{MTDU1}, but for 
[Fe/H] = $-$0.8, where the mass involved in the TDU increases. 
Significant differences are shown for $M$ = 3 $M_\odot$ at [Fe/H] = 
$-$0.8, with respect to LMS models: the first TP with TDU occurs when 
the mass of the H shell is $m_{\rm H}$ $\sim$ 0.7 $M_\odot$ (instead 
of 0.6 $M_\odot$ as for $M$ $\leq$ 2.5 $M_\odot$) and the TDU is less
efficient.
It was shown by \citet{dominguez99} (see also \citealt{boot88}) 
that for a given initial mass, 
the core mass increases by decreasing [Fe/H] (Fig.~\ref{Mcore}). 
Then, for [Fe/H] $<$ $-$1.3 an initial mass 
$M$ = 3 $M_\odot$ approaches the behaviour of IMS stars, 
characterised by a thinner He-intershell and a weaker TDU. 
Therefore, $M$ = 3 $M_\odot$ behaves as a LMS for disk 
metallicities and as a IMS for halo metallicities.
For $-$1.3 $\leq$ [Fe/H] $\leq$ $-$0.8, we adopt the TDU mass 
$\delta$$m_{\rm TDU}$ shown in Fig.~\ref{MTDU1} for each initial mass.
Then, we assume an increase by a factor of two 
in the metallicity range $-$2 $<$ [Fe/H] $<$ $-$1.3, 
and a further factor of two for [Fe/H] $\leq$ $-$2 for AGBs with
$M$ = 1.3, 1.4 and 1.5 $M_\odot$.
For $M$ = 2 $M_\odot$, we adopt an increase
of a factor of two in all the metallicity range $-$3.6 $\leq$ [Fe/H] 
$<$ $-$1.3, following detailed results of AGB models \citep{cristallo09}.
Indeed, as shown in Fig.~\ref{Mcore}, $M$ = 2 $M_\odot$ at [Fe/H] = $-$2.3
is not far from the transition zone between LMS and IMS stars.
For initial masses in the range 1.3 $\leq$ $M$ 
$\leq$ 2.0 $M_\odot$ and $-$3.6 $\leq$ [Fe/H] $\leq$ $-$1.3, we assumed 
the values of models with [Fe/H] = $-$0.8 for the other input 
data (like the temperature and density
during the TP, the mass of the H-shell and the He-intershell,
the overlapping factor and the residual mass of the envelope).
\\
Another debated problem is the evaluation of the mass loss, which 
plays a key role during the AGB phase.
Many stellar properties of the AGB phase depend on the mass loss: for
instance, the efficiency of the TDU, the number of pulses 
and, therefore, the duration of this evolutionary phase.
Different methods were developed to estimate the mass loss rate
 based on:
\begin{itemize}
	\item  observations of circumstellar envelope in giants and supergiants \citep{Rei77}, 
 	\item  dynamical calculations for atmospheric Mira-like stars \citep{bloecker95}, 
 	\item  the observed mass loss-period and the period-luminosity relations of 
 	AGB stars (\citealt{vassiliadis93}; recently revised by \citealt{straniero06} 
 	who account for most recent infrared data).
\end{itemize}
In the models adopted here, the mass loss is estimated with the Reimers formula:
\begin{equation}
\centering
\label{eq1}
\frac{dM}{dt} = - \eta \times (1.34\cdot10^{5})\frac{L^{\frac{3}{2}}}{MT^{2}_{\rm eff}} (M_\odot \cdot yr^{-1}),
\end{equation}
where $\eta$ is the Reimers parameter; $M$, $R$ and $L$ are in 
solar units.
We set $\eta$ = 0.3 for $M$ = 1.3 -- 1.5 $M_\odot$,
 $\eta$ = 0.5 for $M$ = 2.0 $M_\odot$, and $\eta$ = 1 for $M$ = 3.0 $M_\odot$.
\\
With this mass loss, at [Fe/H] = $-$2.6, $M$ = 1.3 $M_{\odot}$ 
shows 5 TPs followed by TDU, $M$ = 1.5 $M_{\odot}$ experiences 20 TPs 
with TDU, while $M$ = 2.0 $M_{\odot}$ shows 26 TDUs.
The poor knowledge of the 
actual mass loss rate remains one of the major uncertainties of 
AGB stellar models. In particular, observational constraints for the 
calibration of the mass loss rate at very low metallicities are 
lacking (see \citealt{straniero06}).
Using the prescription of Reimers as compared with the one by 
\citet{vassiliadis93}, few differences in the model
results are obtained during the AGB phase, with the exception of the lasts TPs
where superwinds take place \citep{iben83}. 
On the other hand, in our models the TDU 
episodes cease when the envelope mass is still substantial, 
and all the superwind phase will take place afterward without 
further variations in the chemical composition.
\\
IMS AGB models are strongly affected by the  
mass loss evaluation (see \citealt{straniero00}; \citealt{vanloon06}). 
For IMS stars, we based our calculations on FRANEC AGB models of 
$M$ = 5.0 $M_\odot$ at [Fe/H] = 0 and $-$1.3 \citep{straniero00}.
These two models show very similar characteristics during TP phase, as
the maximum temperature at the bottom of the TP (3.6 and 3.7 $\times$ 
10$^{8}$ K, respectively), the mass of the TDU almost constant 
pulse by pulse (1.2 -- 1.3 $\times$ 10$^{-3}$ $M_{\odot}$), and the 
mass of the He-intershell ($\sim$ 3 $\times$ 10$^{-3}$ $M_{\odot}$).
Extrapolating, this might imply that for halo metallicities, 
the structure of $M$ = 5.0 $M_\odot$ is barely distinguishable from solar.
Similar characteristics are obtained for a $M$ = 7 $M_{\odot}$ model
\citep{straniero00},
with the exception of the TDU mass which is reduced by a factor
of six with respect to $M$ = 5.0 $M_\odot$ at [Fe/H] = 0.
With our mass loss choice, the $M$ = 5 $M_{\odot}$ 
model experience 24 TPs followed by TDU, leaving a residual total mass 
of 1.67 $M_{\odot}$ and a core mass of 0.9 $M_{\odot}$.
Since in IMS the temperature at the bottom of the convective 
pulse and the peak neutron density are higher than in LMS ($T_8$ $\sim$ 
3.5; $n_n$ $\sim$ 10$^{11}$ cm$^{-3}$), the neutron source 
$^{22}$Ne($\alpha$, n)$^{25}$Mg is more efficiently activated.
Here, neutron-rich isotopes involved in important branchings 
along the $s$-process, like $^{86}$Kr, $^{87}$Rb and $^{96}$Zr,
are strongly enhanced.
As already demonstrated by \citet{travaglio04}, a 
remarkable $s$-process contribution to the isotopes up to
zirconium is provided by IMSs during thermally pulsing phases
via $^{22}$Ne($\alpha$, n)$^{25}$Mg, the main neutron
source in these stars. 
\\
Other mixing and nuclear burning phenomena not treated here are find to occur 
in IMS, as hot bottom burning (\citealt{sugimoto71,iben73,karakas03,ventura05}), 
or, for [Fe/H] $\la$ $-$2.3, hot third dredge-up (\citealt{herwig04,goriely04,campbell08,lau09}).
When hot bottom burning occurs (during the interpulse phase) the bottom of the convective 
envelope reaches the top of the H$-$burning and the temperature at the 
base is large enough ($T$ $>$ 5 $\times$ 10$^{7}$ K) 
to activate the CNO, NeNa, and MgAl cycles of H burning.
The hot third dredge-up, instead, is an extremely efficient 
dredge-up occurring at low metallicities, in which the envelope 
reaches almost the bottom of the intershell (or perhaps beyond it, 
if overshooting is included at low metallicities \citealt{herwig04,lau09}). 
This last phenomenon modifies the efficiency of the TDU and, then, 
may influence the structure of the star and its evolution. Nucleosynthesis
models which include hot third dredge-up are still matter of study.
For this reason only IMS AGB models of disk metallicities 
are discussed here ($M$ = 5.0 and 7.0 $M_\odot$; $-$1.6 $\leq$ [Fe/H] $\leq$ 0).
\\
At [Fe/H] $\la$ $-$2.3, also LMS may experience a single and 
anomalous TDU episode. It occurs when the convective zone generated 
by the first fully-developed TP extends till the H-rich zone, bringing 
protons at high temperature and causing the development of a violent 
thermonuclear runaway \citep{hollowell90,fujimoto00,iwamoto04,campbell08,lau09,cristallo09pasa}.
In the present calculations we do not consider such an extreme case.

\begin{figure*}
   \centering
\includegraphics[angle=-90,width=10cm]{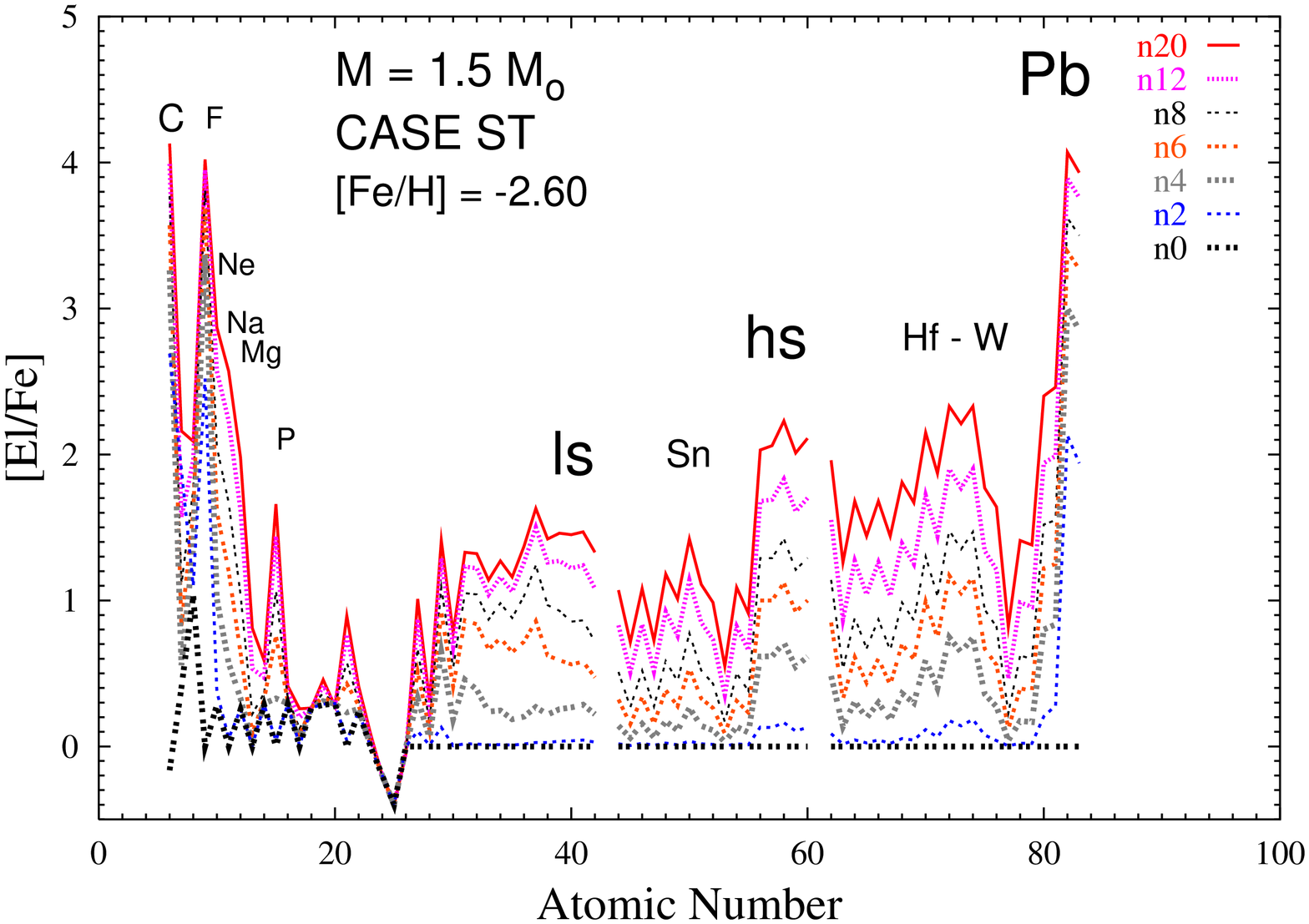}

\includegraphics[angle=-90,width=10cm]{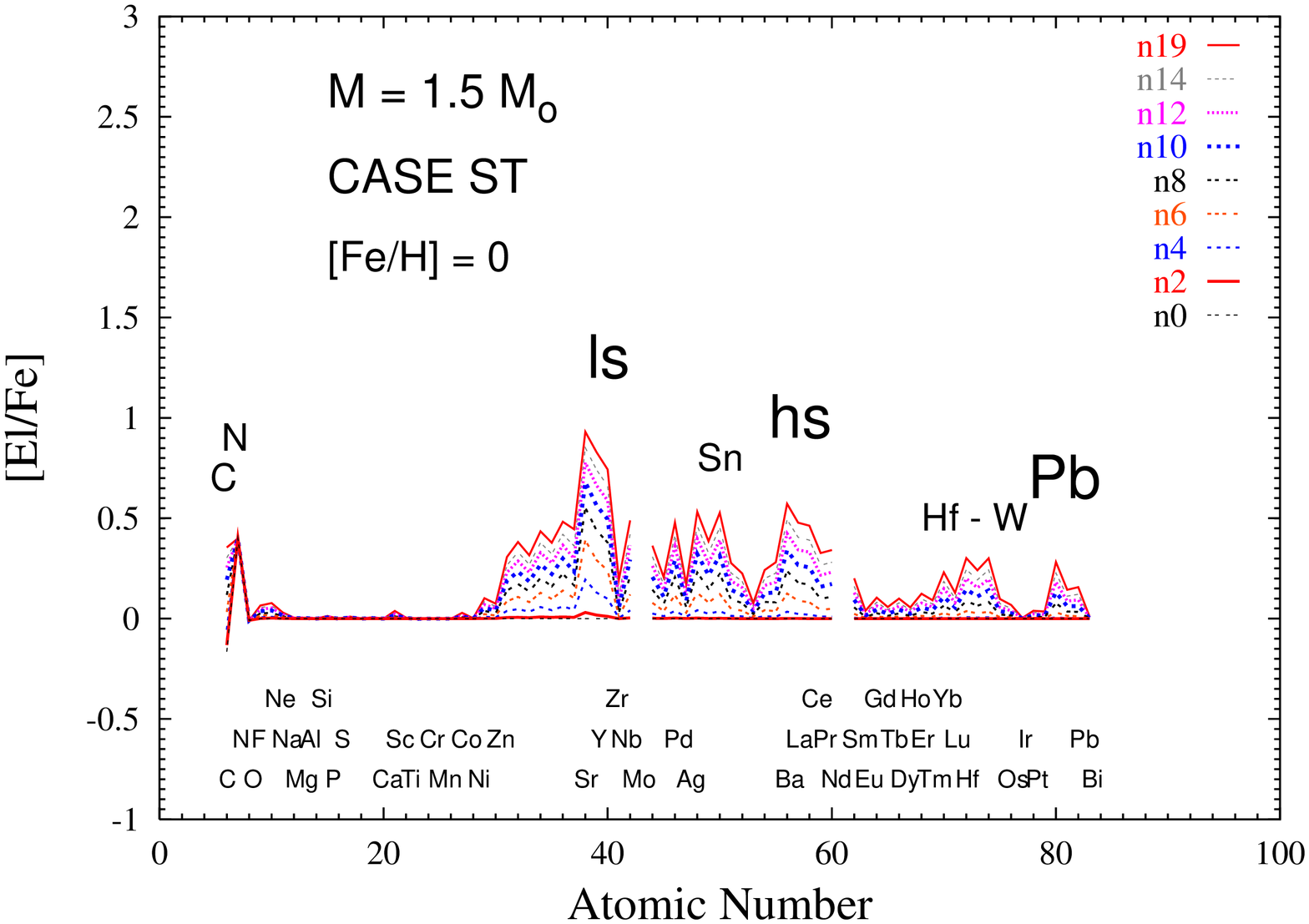}
\caption{\textit{Top panel}: elemental composition in the envelope
after different TDU episodes (labeled as $n(i)$), for an AGB model 
of initial mass $M$ = 1.5 $M_{\odot}$, initial metallicity 
[Fe/H] = $-$2.6, and standard choice of the $^{13}$C-pocket 
(case ST, \citealt{gallino98}). The labels `n0' stands for the 
initial composition. Note that after 15 TPs, the abundances 
[El/Fe] reach an asymptotic value (the 
label `El' stands for a generic element).
The initial Cr and Mn abundances are chosen in agreement with
the observations of unevolved halo stars (\citealt{cayrel04,francois04}).
The assumed value [Mn/Fe] = $-$0.4 may increase up to [Mn/Fe] $\sim$ 
$-$0.1 due to NLTE corrections \citep{bergemann08}.
\textit{Bottom panel}: for comparison, we report the similar case
for a model with the same initial mass, but at [Fe/H] = 0.
These models are for intrinsic AGBs
([Zr/Nb] $\sim$ 1, see Sect.~\ref{binary}).
Note the difference of scale between the two plots.}
\label{m1p5z1m2z5m5_tuttipulsiST}
\end{figure*} 
  
\begin{figure*}
   \centering
\includegraphics[angle=-90,width=10cm]{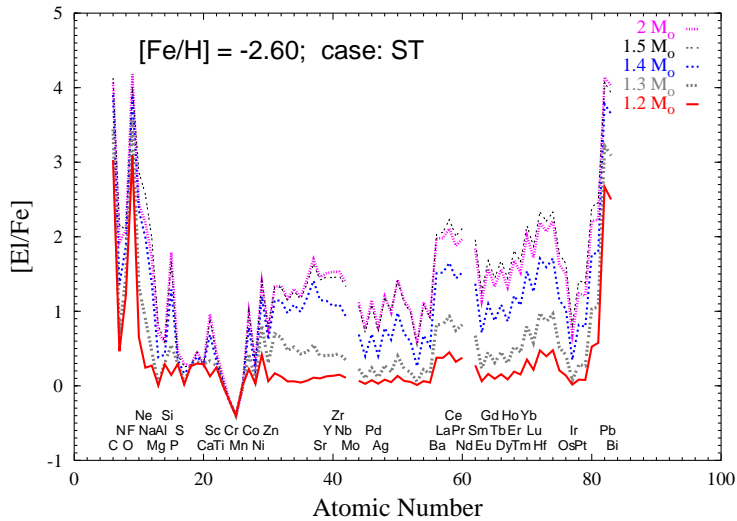}
\caption{Elemental composition in the envelope at the last TDU, 
for AGB models 
of initial metallicity [Fe/H] = $-$2.6, but different initial 
masses: 1.2 $\leq$ $M/M_{\odot}$ $\leq$ 2.0 (cases ST).}
\label{ba5_p1p5diffMz5m5_nro16eq}
\end{figure*}

\begin{figure}
   \centering
\includegraphics[angle=-90,width=8cm]{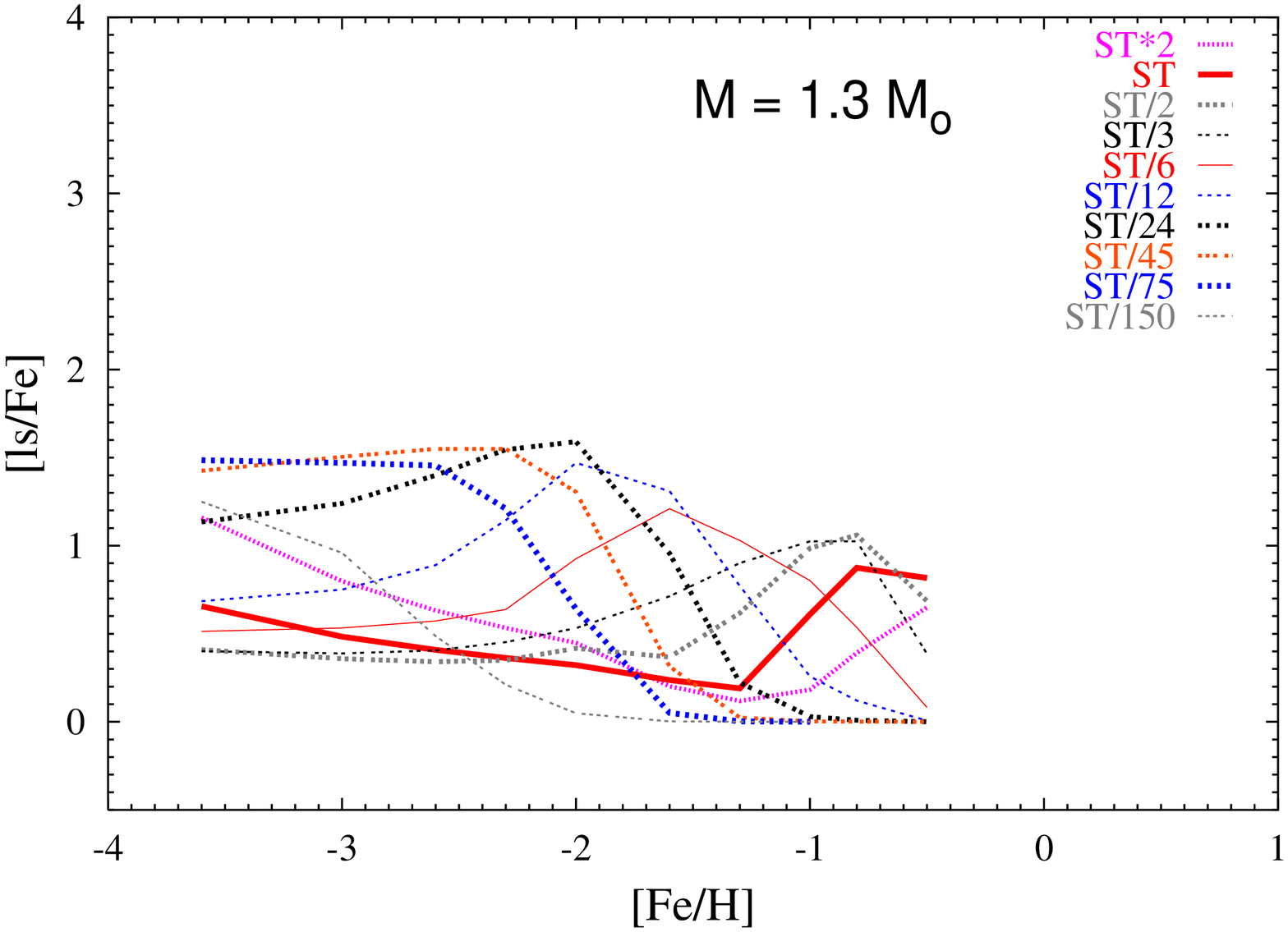}
\includegraphics[angle=-90,width=8cm]{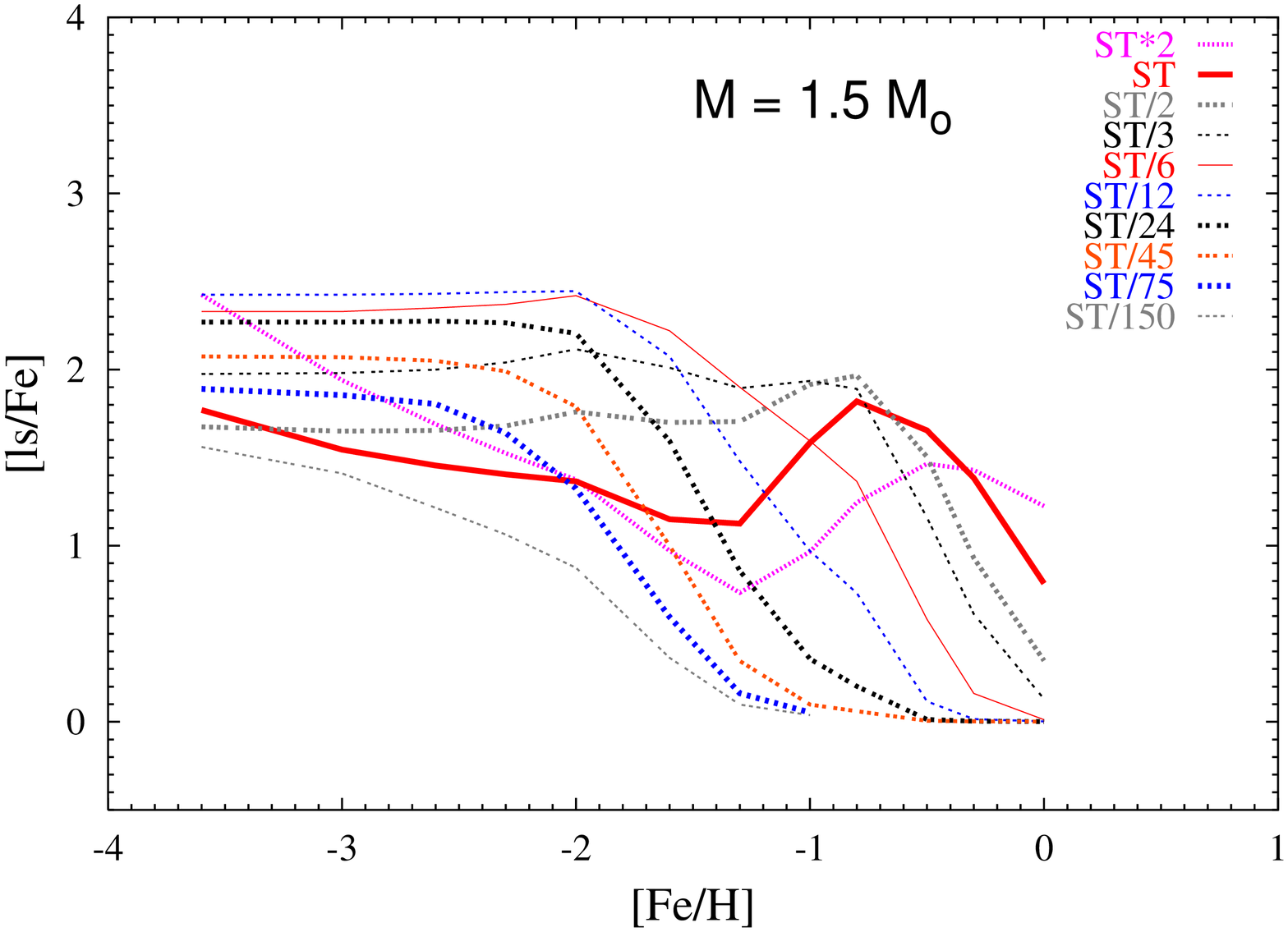}
\includegraphics[angle=-90,width=8cm]{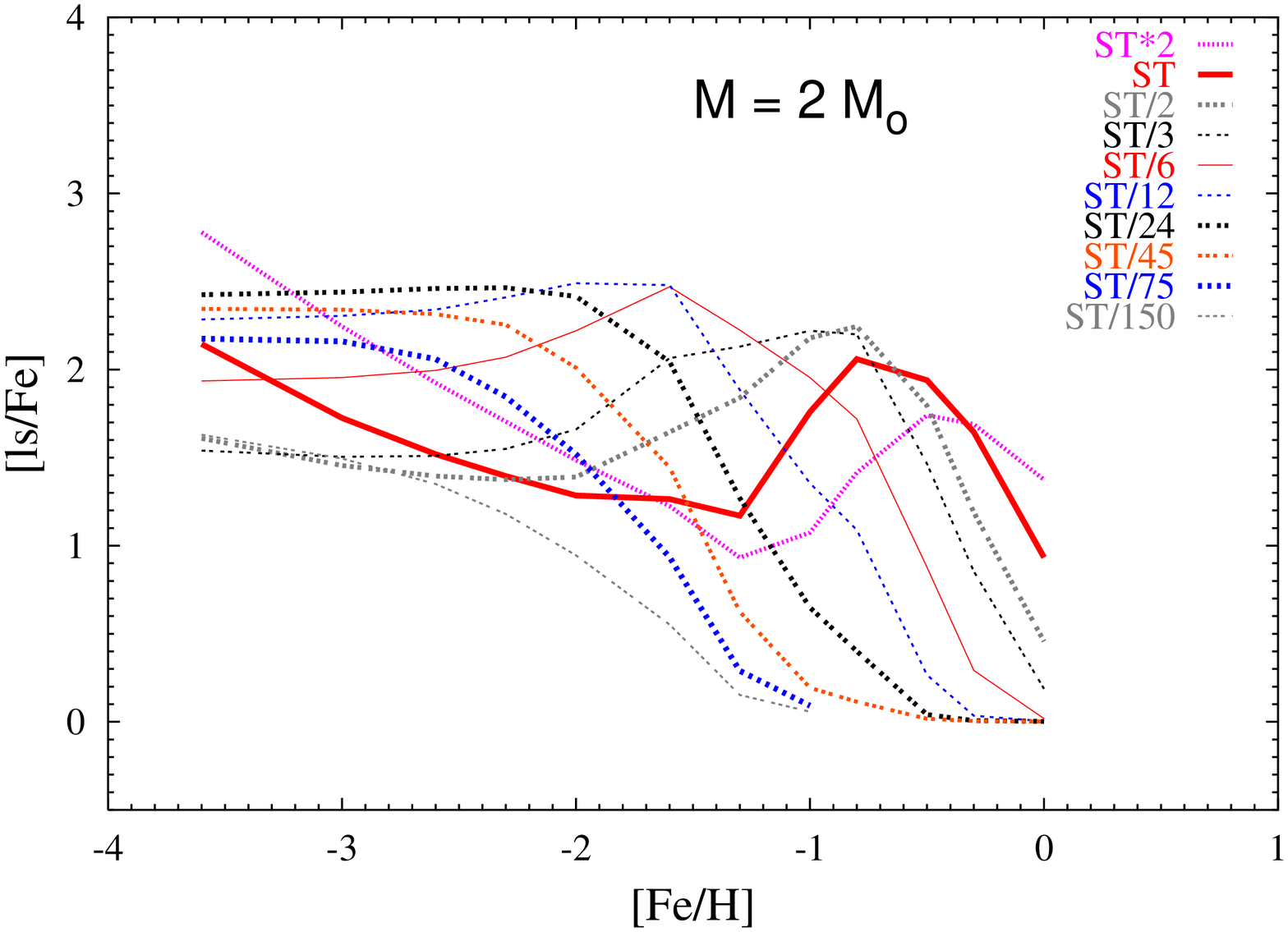}
\caption{Theoretical results of [ls/Fe] versus metallicity, 
for AGB models of initial mass $M$ = 1.3 $M_{\odot}$ (\textit{top panel}),
$M$ = 1.5 $M_{\odot}$ (\textit{middle panel}) and $M$ = 2.0 $M_{\odot}$
(\textit{bottom panel}). A wide range of $^{13}$C-pocket 
efficiencies is presented. 
A 1.3 $M_{\odot}$ model undergoes TDU episodes starting from [Fe/H] 
$\leq$ $-$0.6 
(see Figs.~2 and~3 of \citealt{straniero03}).}
\label{AA_lssufe_m1p3m1p5m2_noobs}
\end{figure}

\begin{figure}
   \centering
\includegraphics[angle=-90,width=8cm]{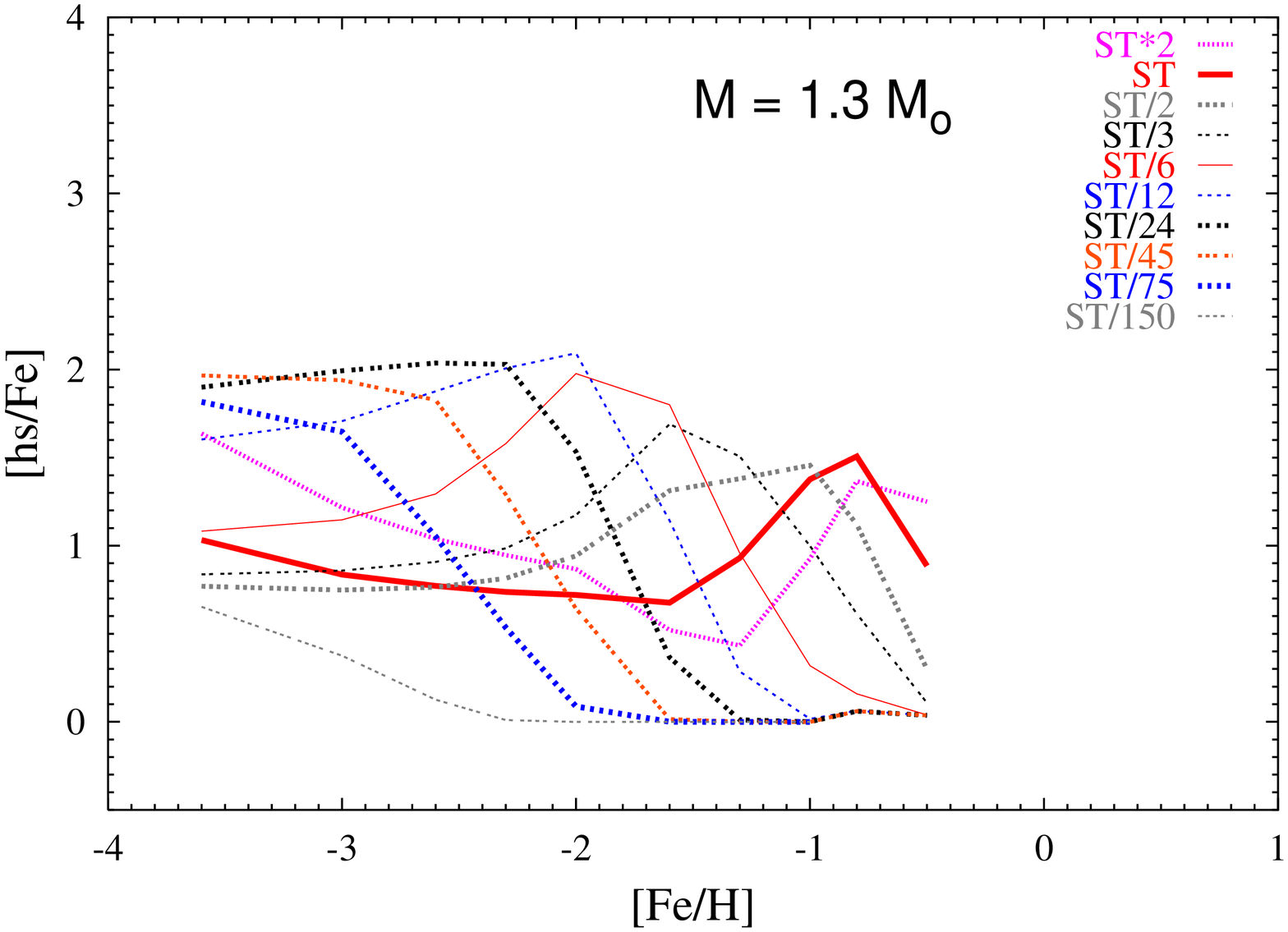}
\includegraphics[angle=-90,width=8cm]{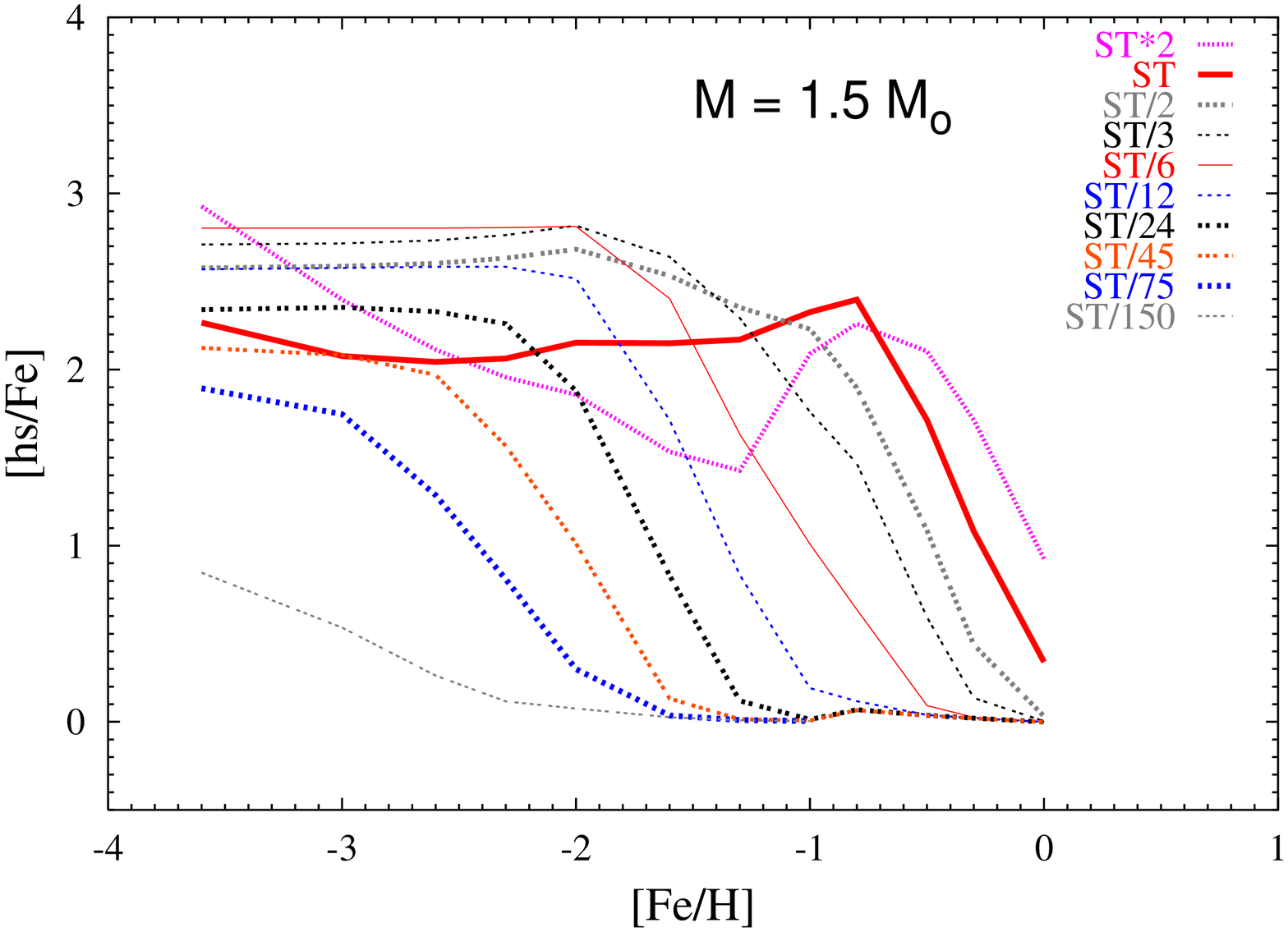}
\includegraphics[angle=-90,width=8cm]{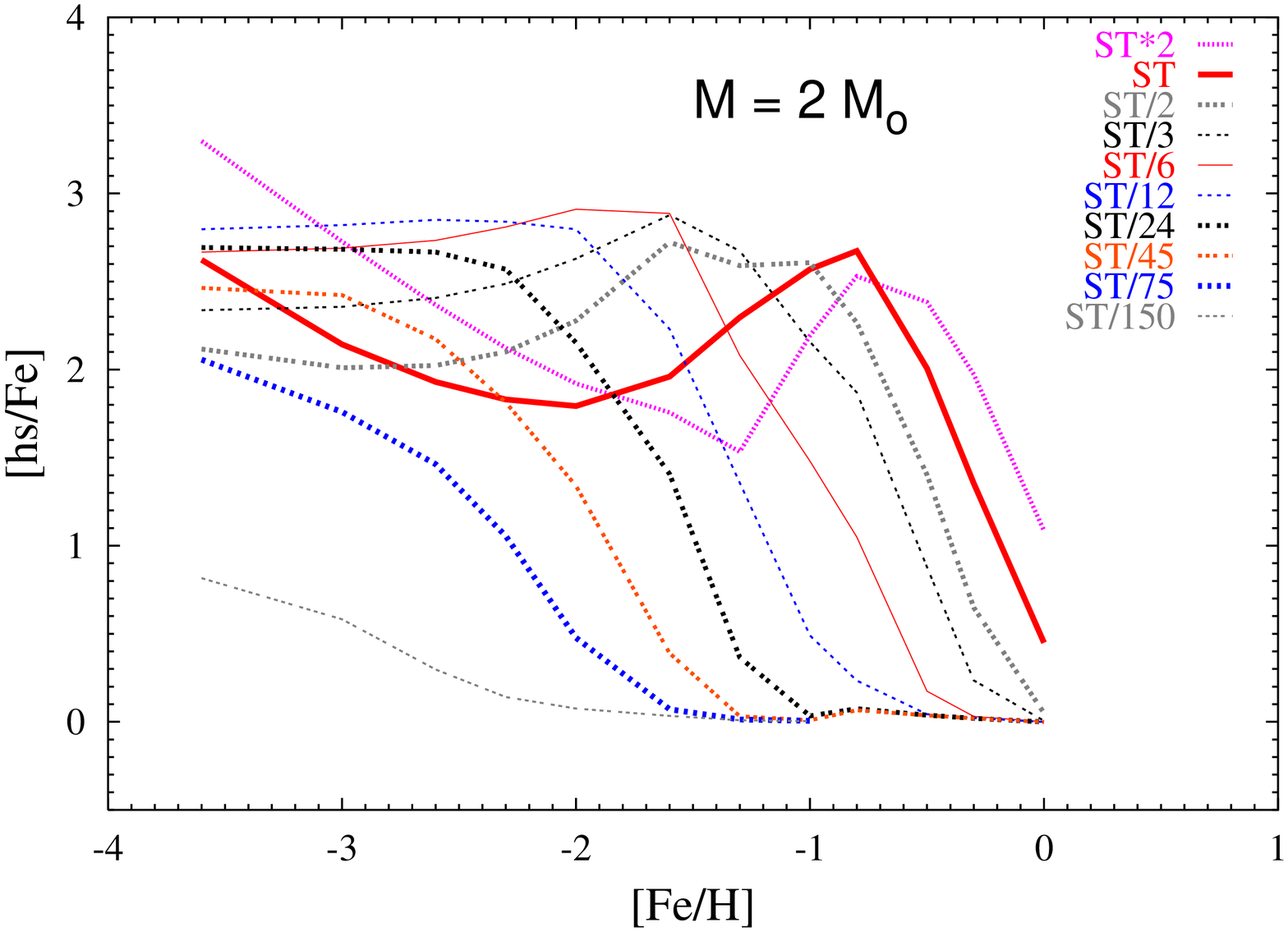}
\caption{The same as Fig.~\ref{AA_lssufe_m1p3m1p5m2_noobs}, but for
[hs/Fe].}
\label{AA_hssufe_m1p3m1p5m2_noobs}
\end{figure}

\begin{figure}
   \centering
\includegraphics[angle=-90,width=8cm]{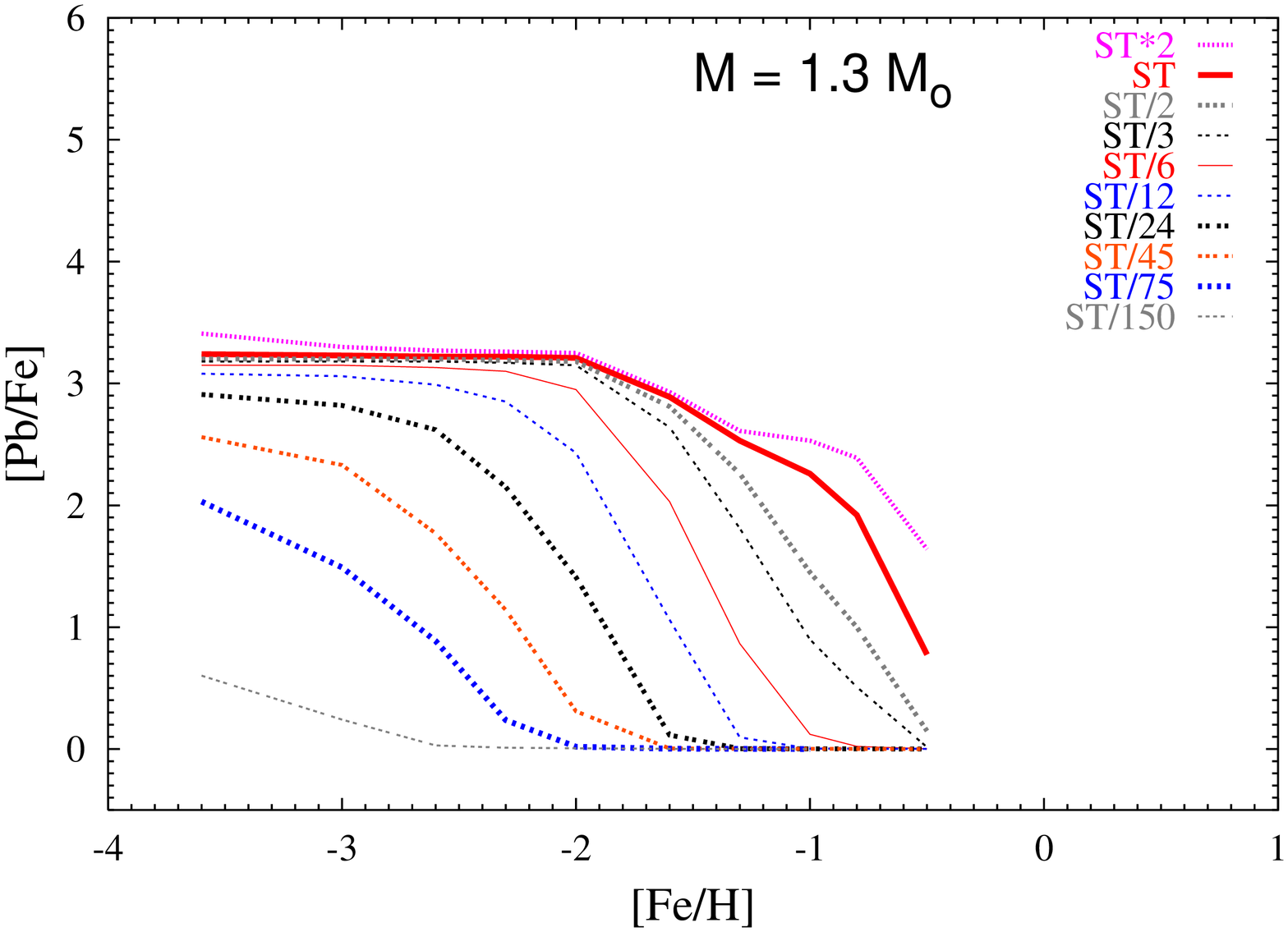}
\includegraphics[angle=-90,width=8cm]{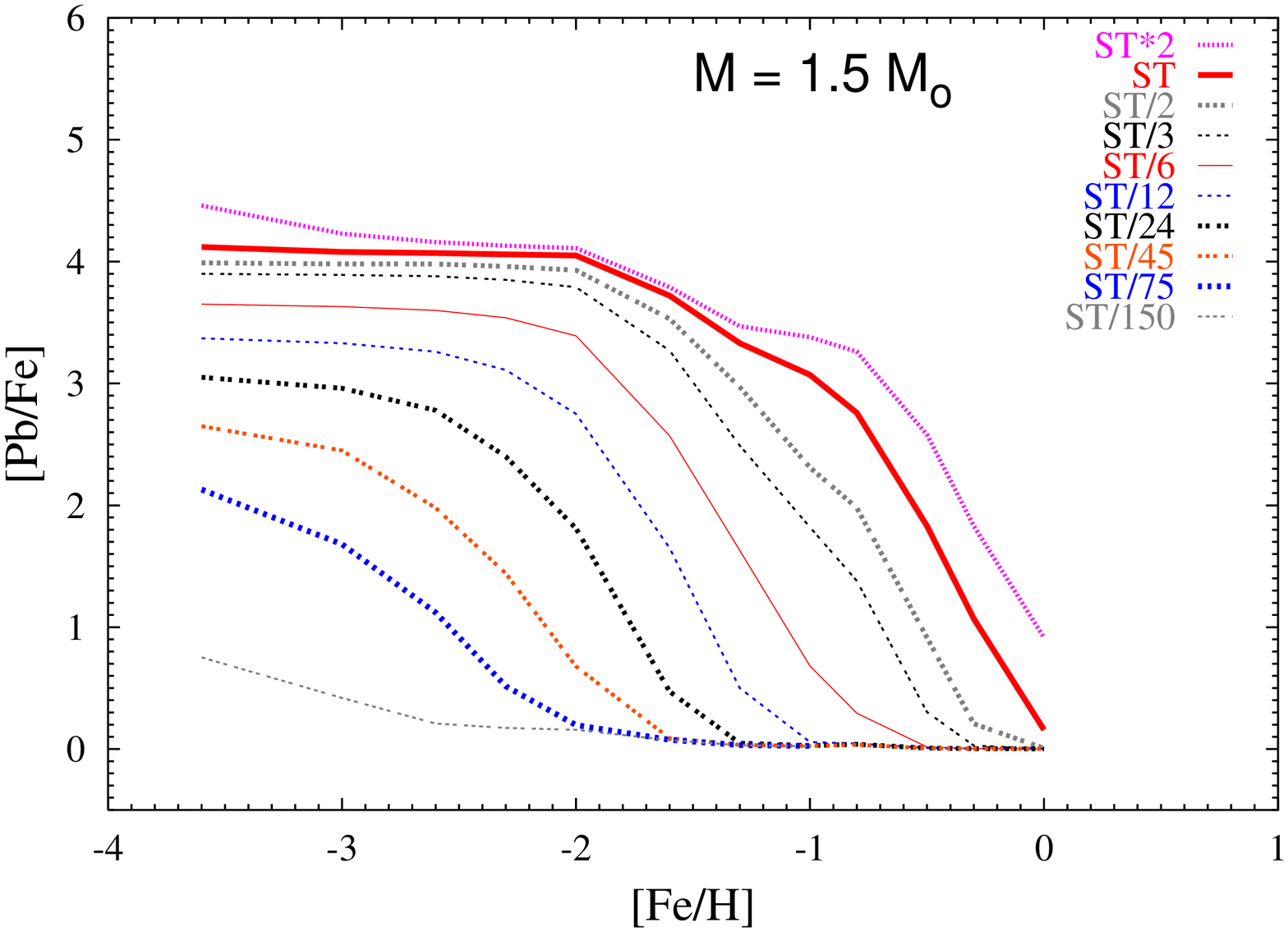}
\includegraphics[angle=-90,width=8cm]{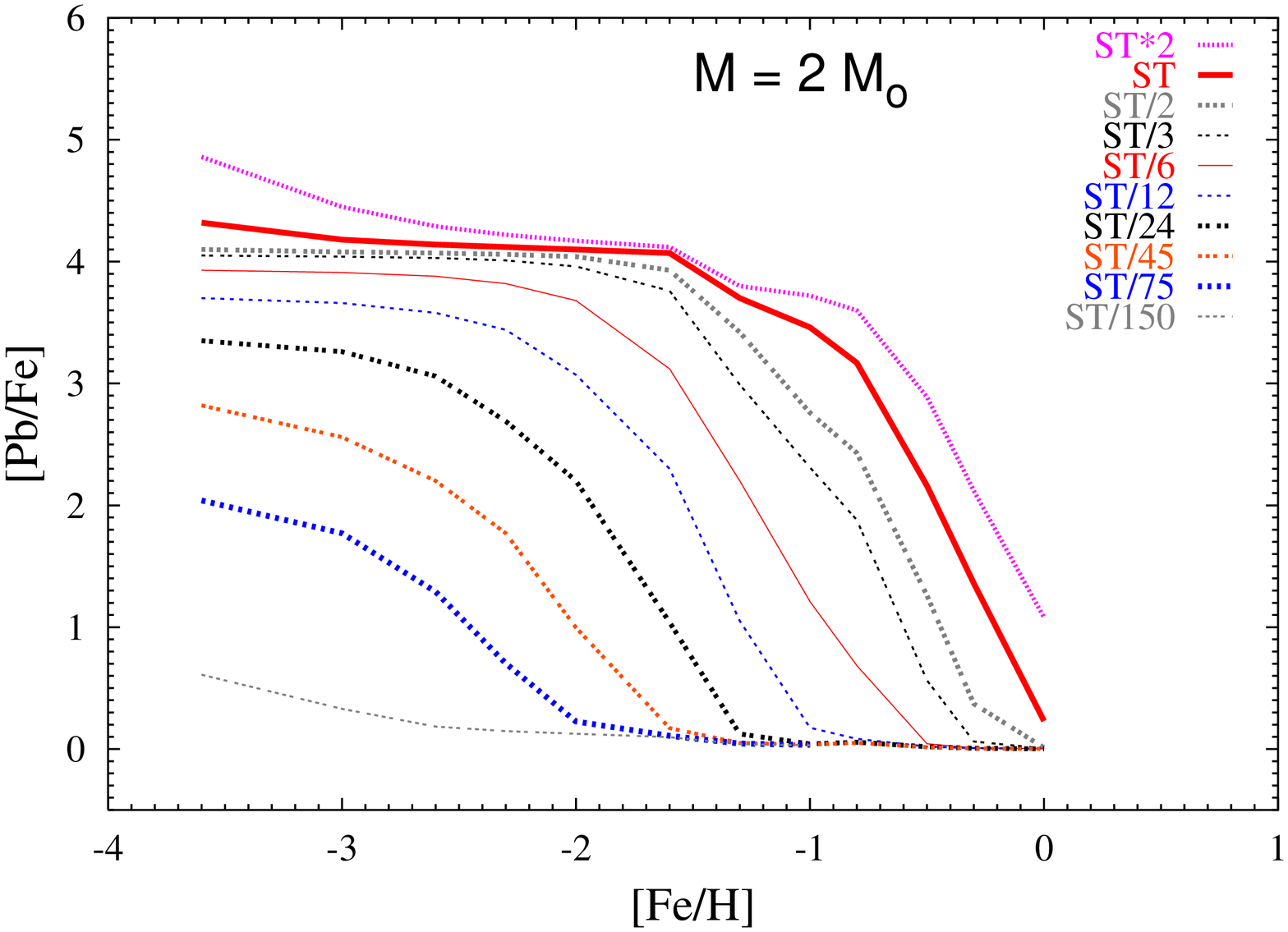}
\caption{The same as Fig.~\ref{AA_lssufe_m1p3m1p5m2_noobs}, but for
[Pb/Fe].}
\label{AA_pbsufe_m1p3m1p5m2_noobs}
\end{figure}

\subsection{Initial chemical composition} \label{ini}

The initial compositions (in mass fractions) used for He are:
$Y$ = 0.28 for $-$0.3 $\leq$ [Fe/H] $\leq$ 0.0;
$Y$ = 0.27 for $-$0.8 $\leq$ [Fe/H] $<$ $-$0.3;
$Y$ = 0.26 for $-$1.3 $\leq$ [Fe/H] $<$ $-$0.8;
$Y$ = 0.25 for [Fe/H] $<$ $-$1.3.
The initial isotopic composition of the other isotopes is based on
 the solar meteoritic values of \citet{AG89},
except for C, N, and O isotopes, which are 
upgraded to \citet{lodders03}.
For metallicity lower than solar we use solar scaled abundances,
with the exception of the $\alpha$ elements.
For oxygen, we assumed a linear increase [O/Fe] with decreasing [Fe/H] 
\citep{abia01}, ([O/Fe] = $-$0.4 [Fe/H]).
For the other $\alpha$ elements ($^{20}$Ne, $^{24}$Mg, 
$^{28}$Si, $^{32}$S, $^{36}$Ar, 
$^{40}$Ca, $^{48}$Ti) we adopt a linear relation [$\alpha$/Fe] = $-$0.3 
[Fe/H] at disk metallicities ($-$1 $\leq$ [Fe/H] $\leq$ 0), 
and an average constant [$\alpha$/Fe] = 0.3 dex for [Fe/H] $\leq$ $-$1,
following the trend observed in unevolved stars (\citealt{cayrel04};
\citealt{francois04}).
\\
We took into account the effect of the first dredge-up (\citealt*{busso99}) 
and second dredge-up (SDU occurring in IMS, 
\citealt{lattanzio96}; \citealt{busso99}) in the initial C, N, O 
composition of our AGB models.
Concerning nitrogen, we assumed an initial value after the 
first dredge up [N/Fe] = 0.3 -- 0.4, depending on the mass of the AGB, 
as predicted by full evolutionary models by \citet{cristallo09,cristallo09pasa}. 
The higher surface predicted value for [N/Fe] in 
the AGB phase derives from dredging to the surface also the ashes of the 
temporary inactivated H-burning shell, where all CNO isotopes are almost 
converted to $^{14}$N, including the primary $^{12}$C that is present in 
the envelope due to previous TDU episodes. As a consequence, [N/Fe] 
increases with the number of TDU episodes.


\section{Results and discussion} \label{results}

\subsection{The three $s$-process peaks} \label{3peaks}

\input{table1.tex}

Nuclei with a magic neutron number ($N$ = 50, 82, and 126) 
are essentially produced by the $s$-process, since their 
low neutron capture cross sections act as a bottleneck for the 
$s$-path. 
\\
In Fig.~\ref{m1p5z1m2z5m5_tuttipulsiST},
we shown the envelope abundances after each TDU 
for an AGB model of initial mass $M$ = 1.5 $M_{\odot}$,
a case ST and [Fe/H] = $-$2.6 (top panel).
As a comparison, we plot the envelope abundances for the same case, but at [Fe/H] = 0
(bottom panel). 
We underline the completely different shapes of the two distributions.
At low metallicities a distribution weighted
toward heavier elements emerges: in particular, already after the first TDU, 
lead receives an extraordinary contribution
([Pb/Fe] $\sim$ 4.1), producing also a huge amount of 
 $^{209}$Bi by neutron captures on $^{208}$Pb ([Bi/Fe] $\sim$ 3.9).
This occurs because $^{56}$Fe (the seed of the $s$-process) 
decreases with metallicity, thus the number of neutrons 
available per iron seed increases, 
 due to the fact that the $^{13}$C neutron source is primary 
\citep{clayton88}. Hence, the neutron fluence overcomes the first two 
peaks, directly feeding $^{208}$Pb 
(\citealt{gallino98}; \citealt{goriely00}; \citealt{travaglio01}).
Note that the increase of heavy elements
is more pronounced during the first TPs, achieving
an asymptotic trend beyond the 10$^{\rm th}$ TDU.
\\
During the TDU, also the material of the H-shell
 is carried to the surface: this zone is rich of 
primary $^{14}$N produced from primary $^{12}$C during the CNO cycle.
The nitrogen surface prediction takes into account for this 
primary $^{14}$N, and, therefore, [N/Fe] increases with the number of
TDU episodes and by decreasing the metallicity.
Carbon is efficiently enhanced in the envelope
at [Fe/H] = $-$2.6 already at the first TDUs (for the case 
ST we predict [C/Fe] $\sim$ 2.7, Fig.~\ref{m1p5z1m2z5m5_tuttipulsiST}, 
top panel). The overproduction of carbon, in fact, increases by 
decreasing the metallicity due to the  primary amount of $^{12}$C 
and to the deeper TDUs as well (Section~\ref{models}).
Two further peaks are predicted: at Sn (before hs) and
at Hf/W (before Pb/Bi). 
These elements 
receive an $s$-process contribution to the solar abundances of 
Sn$_{\rm s}$ = 66\%, 
Hf$_{\rm s}$ = 59\%, Ta$_{\rm s}$ = 45\%, W$_{\rm s}$ = 64\% 
(see Appendix~\ref{arlandiniupdated}).

Fig.~\ref{m1p5z5m5nro16eq_alcuniST}, top panel, shows AGB models 
of $M$ = 1.5 $M_{\odot}$ at [Fe/H] = $-$2.6, by varying the 
efficiency of the $^{13}$C-pocket. 
Similarly, we show in Fig.~\ref{m1p5z5m5nro16eq_alcuniST}, bottom panel, 
the same result at solar metallicity. 
The solid heavy line represents the case ST. 
Decreasing the $^{13}$C-pocket efficiency to ST/24, 
lead is reduced of more than 1 dex with respect to the ST case. 
At [Fe/H] = $-$2.6, for $^{13}$C-pockets $\leq$ ST/75, the 
$^{13}$C is negligible and the $s$-process production
 is principally due to the $^{22}$Ne($\alpha$, n)$^{25}$Mg reaction
activated during the TPs.
Some production of the ls elements is still observed with 
a case ST/150.
\\
In Table~\ref{bab9ltHTZlow}, we list the theoretical predictions
for elements from Sr to Bi at the last TDU for an initial mass $M$ = 1.5 $M_{\odot}$ 
at [Fe/H] = $-$2.6 and different $^{13}$C-pocket efficiencies (ST, ST/12, ST/45, ST/75).
In Col.s~3 to~6 we shown the [El/Fe] predictions, 
which, in Col.s~7 to~10, are normalised to
europium in logarithmic scale [El/Eu] 
(the label `El' stands for a generic element). 
The case ST predicts [hs/ls] = 0.6 (Col.~3), while [hs/ls] reaches $-$0.25 
by decreasing the $s$-process efficiency down to $\sim$ ST/70 (similarly to
the case ST at [Fe/H] = $-$0.3, see Appendix~\ref{arlandiniupdated}).
Since europium is an element with dominating $r$-process abundances, 
the lanthanum/europium ratio provides a measure of the $s$- and $r$-process components 
in halo stars.
A pure $s$-process contribution predicts 0.8 $\leq$ [La/Eu]$_{\rm s}$ $\leq$ 
1.1 at [Fe/H] = $-$2.6 (Cols.~7 to~10).
Values [La/Eu]$_{\rm s}$ $\la$ 0.4 indicate stars that have experienced 
an important $r$-process contribution in addition to the $s$-process 
enhancements (see Paper II).
\\
Another parameter affecting the overabundances is the initial AGB mass,
which determines not only the $s$-process element abundances, but also some of 
the light element abundances (like Na and Mg),
which are not affected by the $^{13}$C-pocket efficiency (see Section~\ref{Na}
for a detailed description).
In Fig.~\ref{ba5_p1p5diffMz5m5_nro16eq}, we show a comparison 
of the theoretical results obtained for 1.2 $\leq$ $M/M_{\odot}$ $\leq$ 2.0. 
The limited number of TDUs for the lowest mass models explains the smaller abundances.
Changing the initial mass, we predict variations up to 2.5 dex for 
the light elements (as Na and Mg, as discussed in Section~\ref{Na}), 1.5 dex 
for the ls peak, about 2 dex for the hs peak and 1.5 dex for Pb.
In Appendix~B (Section~\ref{dataappendixa}), we list surface elemental predictions
 at [Fe/H] = $-$2.6 for different initial AGB masses: 
$M^{\rm AGB}_{\rm ini}$ = 1.3, 1.4, 1.5 and 2.0
$M_{\odot}$.

\input{table2.tex}

In order to extend our theoretical analysis to different stellar populations, we 
discuss the AGB results by varying the metallicity. 
In Figs.~\ref{AA_lssufe_m1p3m1p5m2_noobs} to~\ref{AA_pbsufe_m1p3m1p5m2_noobs},
we present theoretical predictions of [ls/Fe], [hs/Fe] and [Pb/Fe] versus [Fe/H]
for AGB models with initial masses $M$ = 1.3, 1.5 and 2.0 $M_{\odot}$ and 
different $^{13}$C-pocket efficiencies (from ST$\times$2 down to ST/150).
In Appendix~B (Section~\ref{dataappendixb}), we give tabulated surface 
predictions, for AGB initial masses $M$ = 1.3 and 1.5
$M_{\odot}$, $-$3.6 $\leq$ [Fe/H] $\leq$ $-$1.0, 
and two choices of the $^{13}$C-pocket, (ST and ST/12).
Obviously, models with more TDU (Fig.~\ref{MTDU} and~\ref{MTDU1}) 
show larger surface s-process enhancements. In Table~\ref{pulsenumber} we list 
the number of pulses with TDU experienced by each model.
Results from 1.5 and 2 $M_{\odot}$ models are in general quite similar
(Fig.~\ref{AA_lssufe_m1p3m1p5m2_noobs}, middle and bottom panels). 
Depending on the metal content and on the efficiency of the
$^{13}$C-pocket, the behaviour of the [ls/Fe] peak is highly non
linear as a function of the metallicity and on the $^{13}$C 
amount. 
For example in the 
ST case of the $M$ = 1.5 $M_{\odot}$ model, the [ls/Fe] ratio 
increases starting from solar down to [Fe/H] $\sim$ $-$0.8,
where there is a local maximum. 
Then, [ls/Fe] decreases down [Fe/H] $\sim$ $-$1.5, because the
 $s$-path overcomes the ls (and hs) elements feeding lead.
Further decreasing the metallicity ([Fe/H] $\la$ $-$1.5), [ls/Fe]
increases again.
This is the consequence of the primary $^{22}$Ne contribution, which 
provides an additional neutron exposure. 
In fact, by decreasing the metallicity, a progressively high amount of primary 
$^{22}$Ne is produced in the advanced pulses, by the conversion of primary 
$^{12}$C to primary $^{14}$N in the H-burning ashes followed by double 
$\alpha$ capture on the $^{14}$N in the early phases of the next TP
(see Section~\ref{ne22} for further explanations).
For less efficient $^{13}$C-pockets, e.g. the case ST/12, the highest [ls/Fe]
value is shifted toward lower metallicities ([Fe/H] $\sim$ $-$2), because of the 
lower number of neutrons produced.
For [Fe/H] $\leq$ $-$2, a flat behaviour is achieved.
\\
The [hs/Fe] trend is similar to the behaviour previously discussed
for [ls/Fe], but with slightly different slopes (Fig.~\ref{AA_hssufe_m1p3m1p5m2_noobs}).
As for [ls/Fe], a primary $^{22}$Ne contribution to [hs/Fe] is 
observed for [Fe/H] $\la$ $-$2.
When the hs-peak reaches a local maximum, the hs nuclei are used 
for the production of lead. 
\\
{[Pb/Fe]} increases by decreasing the metallicity, covering a range of 
4 orders of magnitude (Fig.~\ref{AA_pbsufe_m1p3m1p5m2_noobs}).
The results for [Pb/Fe] varying the initial stellar mass is very similar, 
with higher overabundances according to the rising number of TPs.

The differences between models presented here and previous publications 
\citep{bisterzo09pasa,SCG08} are principally due to the correction of a
 past bug in the code, where we did not account for the poisoning effect 
 of $^{16}$O in the pocket produced by the  $^{13}$C($\alpha$, n)$^{16}$O 
 reaction. While no variations in the $s$-process abundances are found from 
 solar metallicity down to [Fe/H] = $-$1.6, at halo metallicities, the 
 $s$-process abundances may be strongly reduced for high $^{13}$C-pocket 
 efficiencies.
For cases ST$\times$2 down to ST variations up to 0.5 dex are obtained for 
 $M$ = 1.5 and 2 $M_\odot$. Even higher variations (up to 0.8 dex) are obtained
 for AGB models of $M$ = 1.3 and 1.4 $M_\odot$. 
Changes lower than 0.1 dex are observed for cases $<$ ST/1.5. 

The addition of proton captures in the network does not involve variations 
in the $s$-process elements production, however have a key role  in improving the
prediction of fluorine (Sect.~\ref{Na}).

\subsection{Nb and Zr predictions and their significance.} \label{binary}

\begin{figure}
   \centering
\includegraphics[angle=-90,width=8cm]{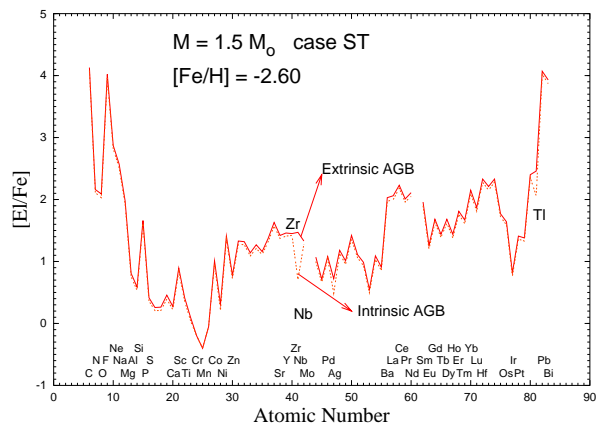}
\caption{Theoretical predictions for [Zr/Fe] and [Nb/Fe]
in an extrinsic (solid line) and an intrinsic (dashed line) AGB, with 
initial mass $M$ = 1.5 $M_{\odot}$, and [Fe/H] = $-$2.6.
The efficiency of the $^{13}$C-pocket has been set to the ST case.}
\label{p1p5m1p5z5m5nro16eq_ZrNb_n14Hshell}
\end{figure} 

An intrinsic AGB is a star with high luminosity and low log $g$
that lies on the TP-AGB or Post-AGB phase.
An extrinsic AGB is located on the main-sequence or on the 
red giant branch.
In this case, which is the case for CEMP-s stars, the 
$s$-process abundances are due to mass transfer from a more 
massive AGB companion, which now is cooling along the white dwarf 
sequence \citep{WD88}.
These two classes can be discerned by studying the abundances
of two elements: Tc and Nb.
Long-lived isotopes fed by the $s$-process, like $^{99}$Tc 
($t_{\rm 1/2}$ = 2$\times$ 10$^{5}$ yr),
can be observed in intrinsic AGB stars, as first discovered by
\citet{merrill52}.
As the interpulse phase in LMS has a typical period of the order of 
10$^{4}$ yr, $^{99}$Tc, brought into the surface after the TDU, 
does not have time to completely decay.
Spectroscopic observations of disk metallicity stars were
done in the past for MS, S, C(N) stars, in order to search for 
intrinsic AGBs by detecting Tc lines (\citealt{SL88,SL90};
\citealt{brown90}; \citealt{busso01}; \citealt{abia02}).
\\
Another key element to discriminate between intrinsic and extrinsic
AGB stars is niobium.
According to \citet{arlandini99}, solar 
Nb ($^{93}$Nb), which is bypassed by the $s$-process fluence,
receives a contribution of 85\% from the $s$-process (see also Appendix~A) 
by radiogenic decay of $^{93}$Zr ($t_{\rm 1/2}$ = 1.5 Myr).
This latter isotope is strongly fed by the $s$-process because of its 
low neutron caption cross section \citep{bao00}.
The long lived isotope $^{93}$Zr partially decays
into $^{93}$Nb during the interpulse phase and, at each 
TDU, newly synthesised $^{93}$Zr is mixed within the envelope. 
In intrinsic AGB stars, a fraction of Nb is produced in the
envelope by the decay of $^{93}$Zr.
Instead, for extrinsic AGBs, all the $^{93}$Zr produced
by the $s$-process already decayed into $^{93}$Nb. 
Then, for an intrinsic AGB, such as CEMP-s stars, we expect [Zr/Nb] $\sim$ 1 (dashed line of 
Fig.~\ref{p1p5m1p5z5m5nro16eq_ZrNb_n14Hshell}), while for an extrinsic AGB 
[Zr/Nb] $\sim$ 0 (solid line of Fig.~\ref{p1p5m1p5z5m5nro16eq_ZrNb_n14Hshell}).

\subsection{Intrinsic $s$-process indexes} \label{binaryindicators}

\begin{figure}
   \centering
\includegraphics[angle=-90,width=8cm]{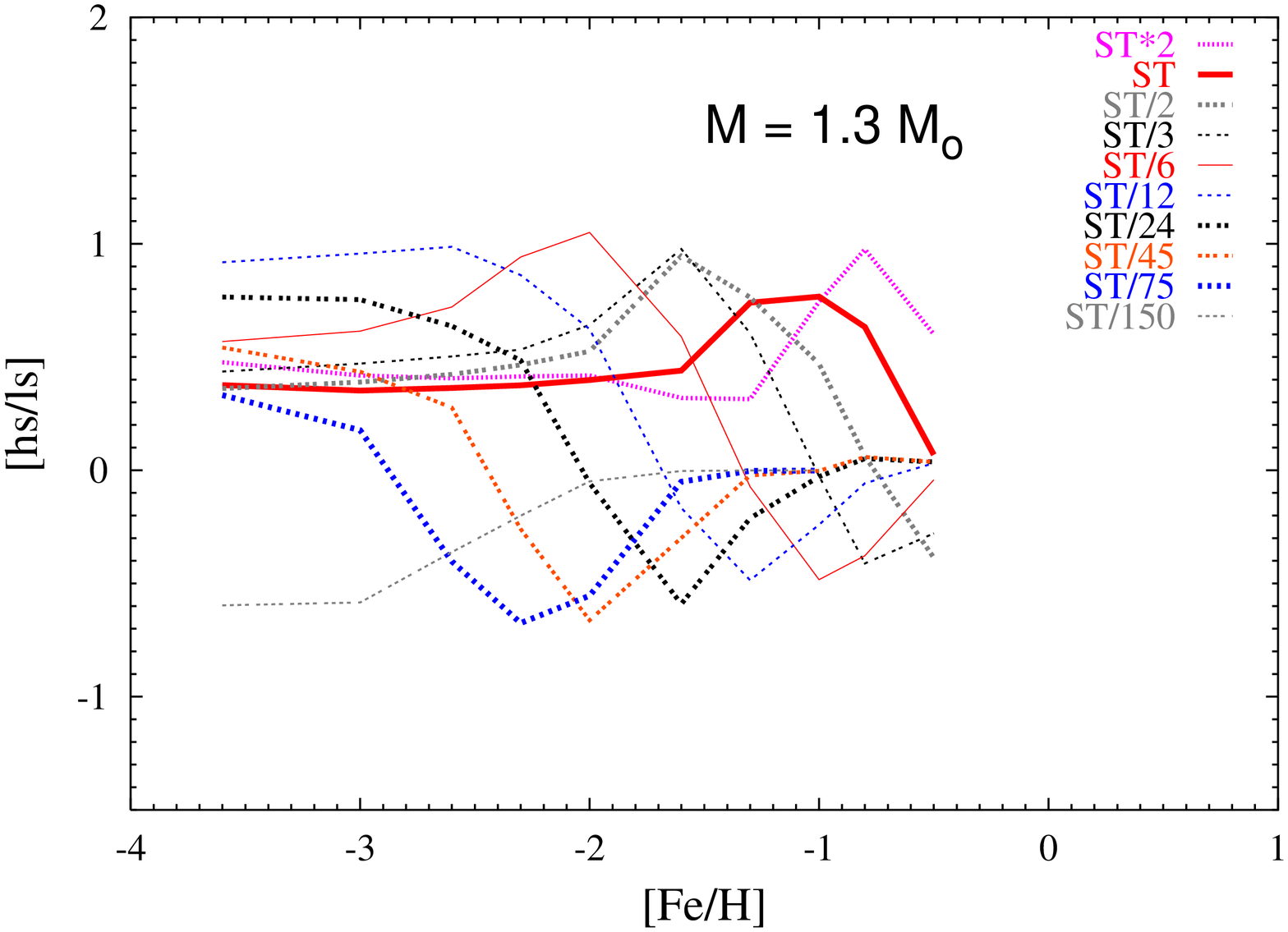}
\includegraphics[angle=-90,width=8cm]{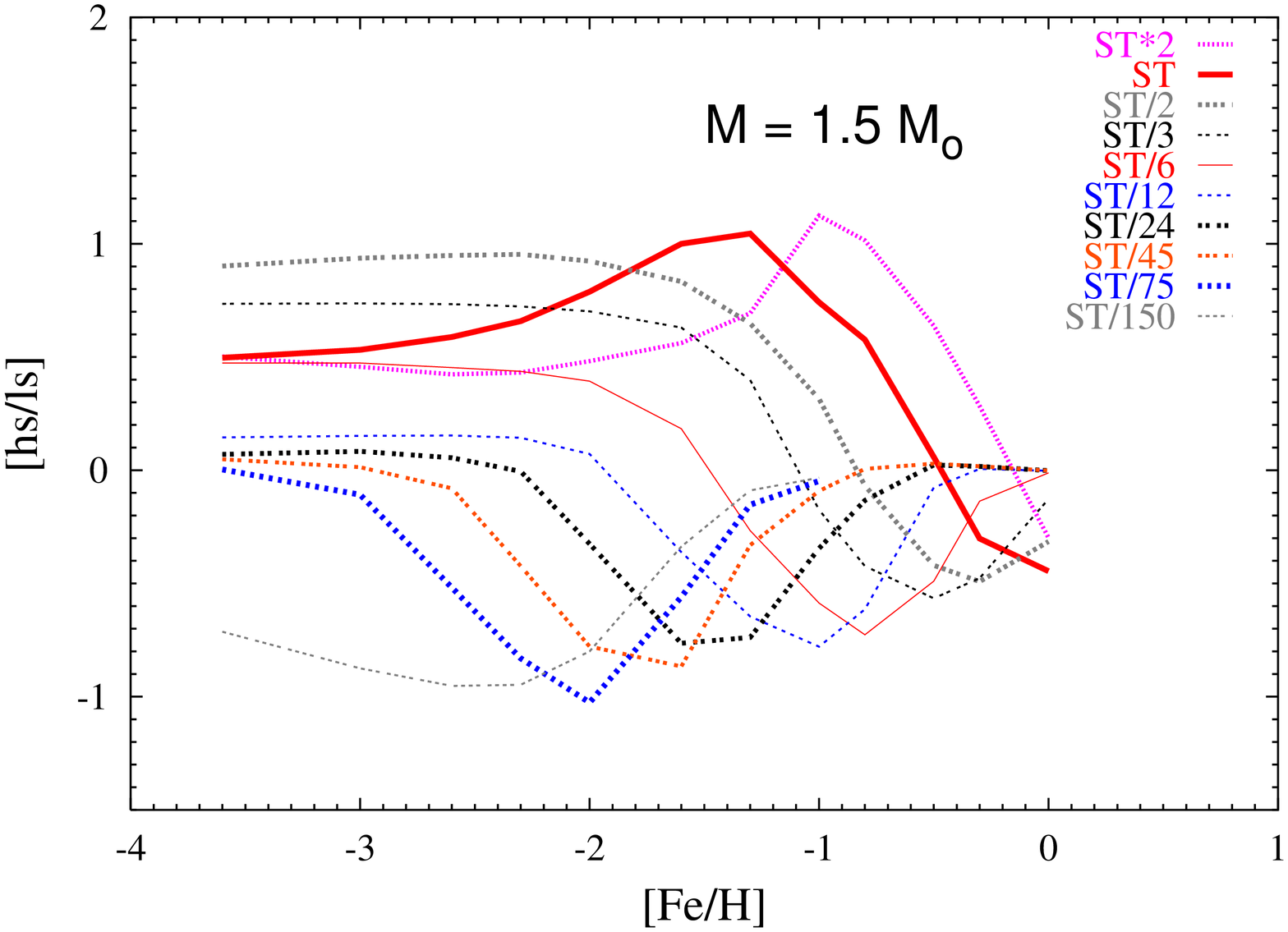}
\includegraphics[angle=-90,width=8cm]{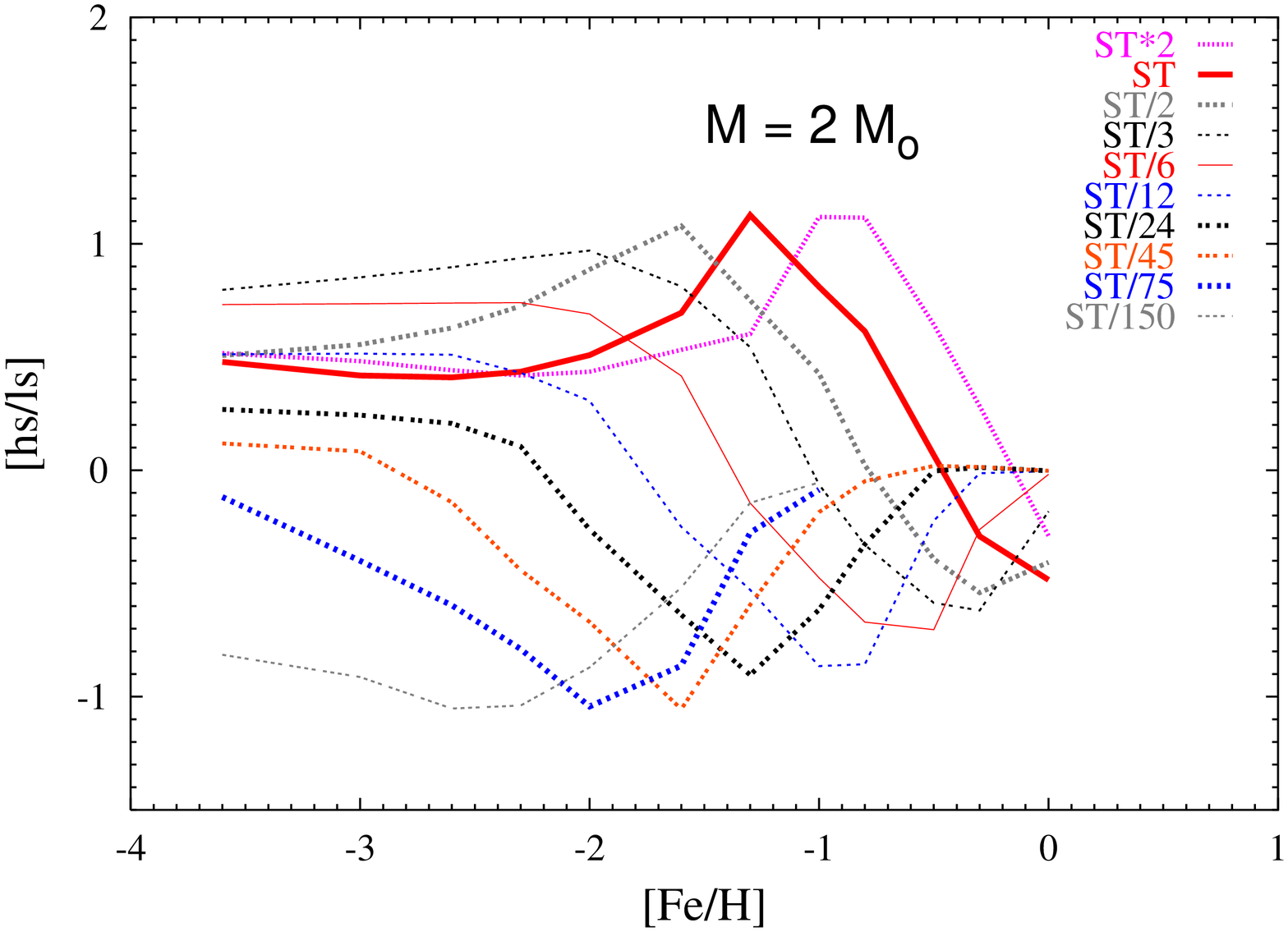}
\caption{The same as Fig.~\ref{AA_lssufe_m1p3m1p5m2_noobs}, but for
[hs/ls].}
\label{AA_hssuls_m1p3m1p5m2_noobs}
\end{figure}

\begin{figure}
   \centering
\includegraphics[angle=-90,width=8cm]{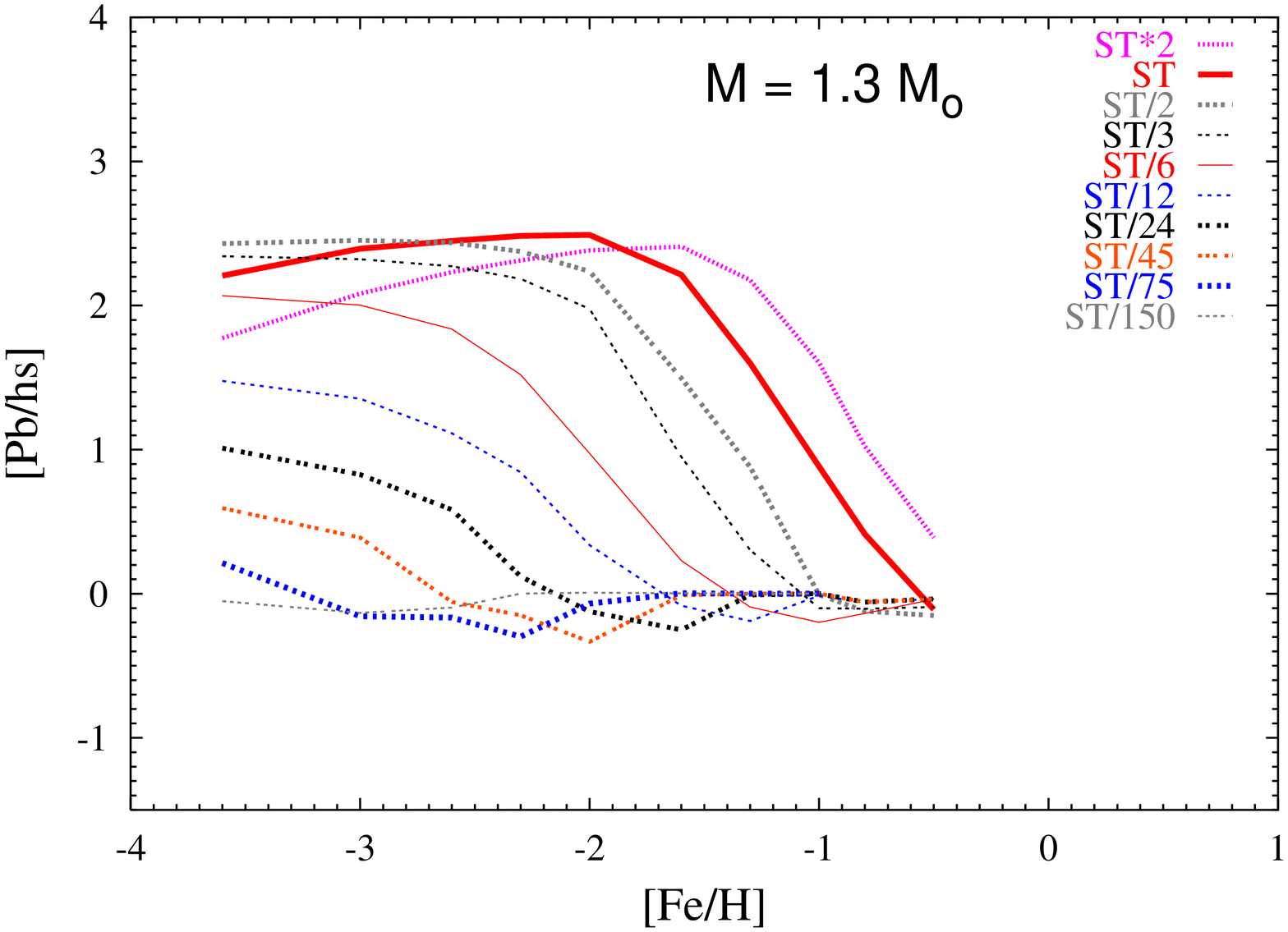}
\includegraphics[angle=-90,width=8cm]{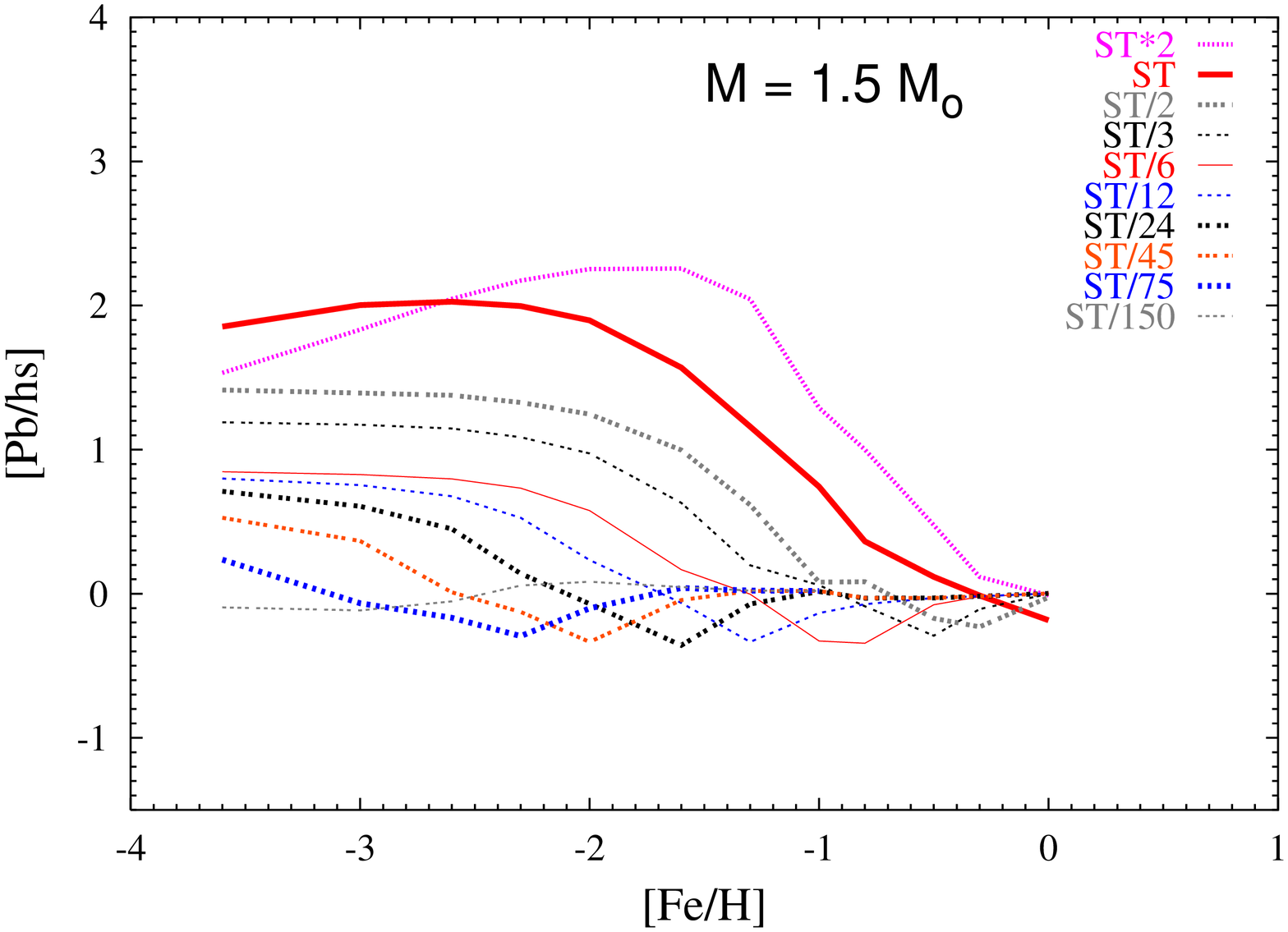}
\includegraphics[angle=-90,width=8cm]{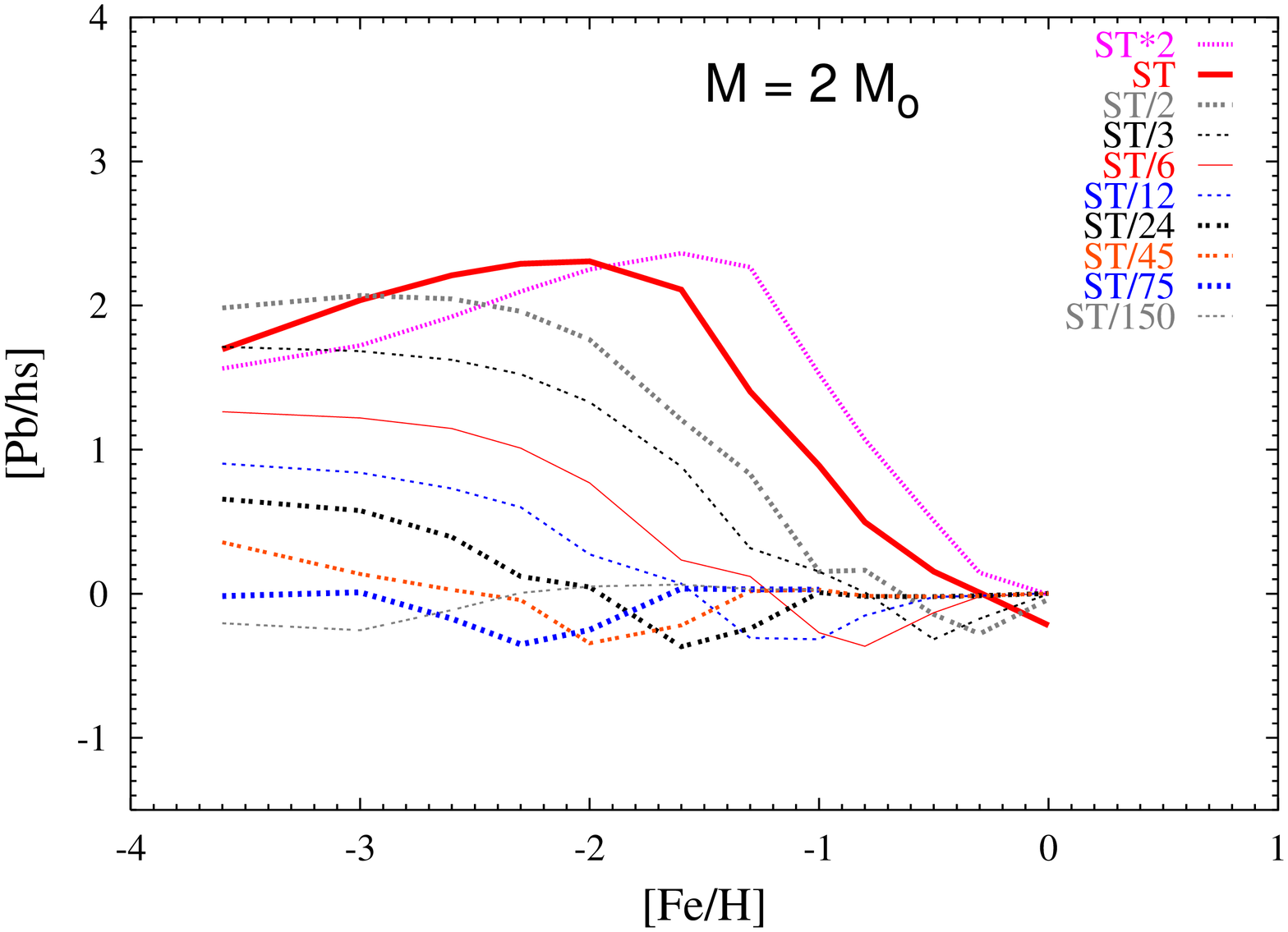}
\caption{The same as Fig.~\ref{AA_lssufe_m1p3m1p5m2_noobs},
but for [Pb/hs].}
\label{AA_pbsuhs_m1p3m1p5m2_noobs}
\end{figure}

\input{table3.tex}

As demonstrated in Sect.~\ref{3peaks}, in AGB stars s-rich material
 coming from the 
He-intershell is mixed with the AGB envelope during each TDU and thus
[ls/Fe], [hs/Fe] and [Pb/Fe] increase with the number
of TDUs. The elemental distribution after each TDU represents therefore
the surface composition of the star.
\\
Instead, in extrinsic AGBs, we do not observe the pristine amount of
material synthesised since this material may undergo further mixing 
with the envelope of the extrinsic AGB star. Low mass main-sequence stars 
have a very thin convective envelope. In this case thermohaline mixing or 
gravitational settling could dilute the original AGB material \citep{denissenkov08,denissenkov09,stancliffe08,charbonnel07,eggleton06,vauclair04}. 
Red giants instead have extended convective envelopes in which the original AGB material is diluted.
The dilution factor of the AGB material can be calculated using the formula\footnote{This equation 
is obtained starting from the simple hypothesis that 
$$\rm{El^{obs}_{\star} \cdot \textit{M}^{obs}_{\star} = El^{env}_{AGB} \cdot
 \Delta \textit{M}^{trans}_{AGB} + El^{env}_{\star} \cdot \textit{M}^{env}_{\star}},$$
\noindent where 
$\rm{El^{obs}_{\star}}$ is the mass fraction of the element `El' 
observed in the extrinsic star,  
$\rm{El^{env}_{AGB}}$ is the mass fraction of the element `El' in
 the material transferred from the AGB to the companion, 
$\rm{El^{env}_{\star}}$ is the mass fraction of the element `El' 
in the envelope of the observed star before the mass transfer,
$M^{\rm obs}_{\star}$ is the envelope mass of the observed star, 
$\Delta M^{\rm trans}_{AGB}$ is the material transferred from the
 AGB to the companion, and
$M^{\rm env}_{\star}$ is the envelope mass of the observed star 
before the mass transfer.}:
\begin{equation}
\centering
\label{eq2}
[{\rm El}^{\rm obs}/{\rm Fe}] = \log (10^{[{\rm El}_{\rm AGB}/{\rm Fe}] - dil} + 
10^{[{\rm El}_{\star}/{\rm Fe}]} (1 - {10}^{- dil})),
\end{equation}
where [El/Fe]$^{\rm obs}$ is the overabundance measured in the 
extrinsic star, [El/Fe]$^{\rm AGB}$ is the amount of the element 
`El' in the AGB envelope.
The dilution \textit{dil}, chosen 
for each star in order to obtain a best fit 
of the experimental data, is defined as the mass of the convective
envelope of the observed star ($M^{\rm obs}_{\star}$) over 
the material transferred from the AGB to the companion ($M^{\rm trans}_{\rm AGB}$):
\begin{equation}
\centering
\label{eq3}
dil = \log\left(\frac{M^{\rm obs}_{\star}}{\Delta M^{\rm trans}_{\rm AGB}}\right).
\end{equation}

In this context, the [hs/ls] and [Pb/hs] ratios are extremely valuable 
indexes for the $s$-process as they are independent both on the TDU 
efficiency in the AGB star as well as the dilution of the AGB material 
onto the companion. Moreover, the [hs/ls] ratio may give a constrain on 
the initial AGB mass because, with increasing the mass
(and decreasing the metallicity), the elements of the first $s$-peak receive
a significant contribution from the $^{22}$Ne($\alpha$, n)$^{25}$Mg neutron 
source during the convective TP.
In Table~\ref{bab9ltHThssuls_z5m5_nro16eq_diffM}, we 
list the values of the intrinsic index [hs/ls] at [Fe/H] = $-$2.6
by changing the $^{13}$C-pocket, for various AGB initial 
masses.
The [hs/ls] ratio may give a constraint on the initial mass,
and it can be an indicator of the $s$-process efficiency 
as well. 
For example, by comparing $M$ = 1.3 and 1.5 $M_{\odot}$, the case 
ST/12 predicts [hs/ls] $\sim$ 1.0 dex and $\sim$ 0.2 dex, respectively.
In Fig.~\ref{AA_hssuls_m1p3m1p5m2_noobs}, we show the [hs/ls] 
theoretical predictions versus metallicity for AGB models 
of $M$ = 1.3, 1.5 and 2.0 $M_{\odot}$.
\\
While in disk metallicity stars ([Fe/H] $>$ $-$1.5), the ratio 
[hs/ls] provides strong constraints on the mass and on the 
$^{13}$C-pocket efficiency, at low metallicities ([Fe/H] $\leq$ $-$1.5)
models characterised by the same [hs/ls] value could show different
[Pb/Fe] values.
For example, for a $M$ = 1.3 $M_{\odot}$ model at [Fe/H] = $-$2.5, 
the ST/3 and ST/30 cases predict
the same [hs/ls] $\sim$ 0.5, while they have [Pb/Fe] $\sim$ 3.2 and 2.4,
respectively (see Figs.~\ref{AA_pbsufe_m1p3m1p5m2_noobs} 
and~\ref{AA_hssuls_m1p3m1p5m2_noobs}, top panels).
Another $s$-process indicator is therefore required to discern between
different theoretical models: [Pb/hs] (or 
[Pb/ls]), \citep{delaude04,vaneck03}. In Fig.~\ref{AA_pbsuhs_m1p3m1p5m2_noobs}, 
theoretical predictions of [Pb/hs] versus metallicity 
are shown for different $^{13}$C-pockets and different masses
($M$ = 1.3, 1.5 and 2.0 $M_{\odot}$). 
These values are also listed in
 Table~\ref{bab9ltHThssuls_z5m5_nro16eq_diffM} at [Fe/H] = $-$2.6.
[Pb/hs] ratio covers a large spread  (0 $\la$ [Pb/hs] $\la$ 2, 
Fig.~\ref{AA_pbsuhs_m1p3m1p5m2_noobs}). 
The maximum amount of $^{13}$C and $^{14}$N in the pocket and different 
hydrogen profiles (then the amount of $^{13}$C and $^{14}$N in the tails 
of the pocket) modify the final $s$-process distribution. This 
explains possible differences between the range of [hs/ls]
and [Pb/hs] predicted by our models and other results presented 
in the literature (see Bisterzo et al., in preparation).
\\
In Figs.~\ref{AA_lssufe_m5_noobs} and~\ref{AA_pbsuhs_m5_noobs}
(Appendix~C, online material), we show [ls/Fe], [hs/Fe], [Pb/Fe] and 
[hs/ls] and [Pb/hs] for $M$ = 5 $M_{\odot}$ in the metallicity range 
$-$1.6 $\leq$ [Fe/H] $\leq$ 0.
Note as the range of $^{13}$C-pockets adopted has a negligible influence
on the [ls/Fe] ratio ($\sim$ 0.3 dex).
As expected, lead is more sensible to the $^{13}$C-pocket efficiency,
showing however a reduced spread with respect to lower stellar masses 
(Fig.~\ref{AA_lssufe_m5_noobs}, bottom panel, Appendix~C, online material).
For $M$ = 3 $M_{\odot}$ a behaviour similar to IMS stars is expected
for halo metallicities (see Section~\ref{models}, Fig.\ref{Mcore}). 
We show in Figs.~\ref{AA_m3novlz_noobs_disk} and~\ref{AA_m3novlz_noobs_disk1} 
(Appendix~C, online material), the three $s$-peak predictions in the metallicity 
range $-$1.6 $\leq$ [Fe/H] $\leq$ 0.



\section{The importance of the primary $^{22}$Ne at low 
metallicities} \label{ne22}

\begin{figure}
   \centering
\includegraphics[angle=-90,width=8cm]{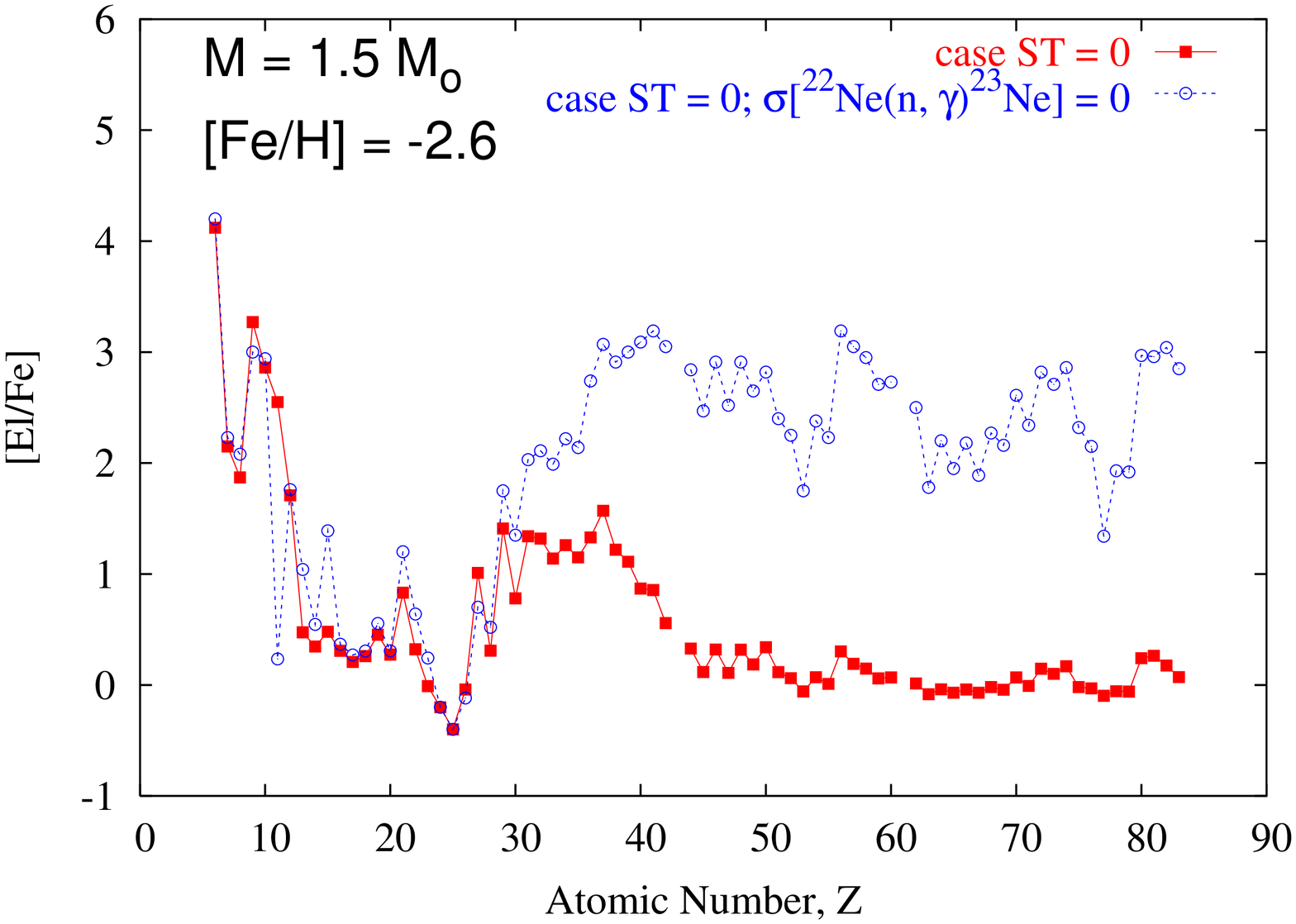}
\includegraphics[angle=-90,width=8cm]{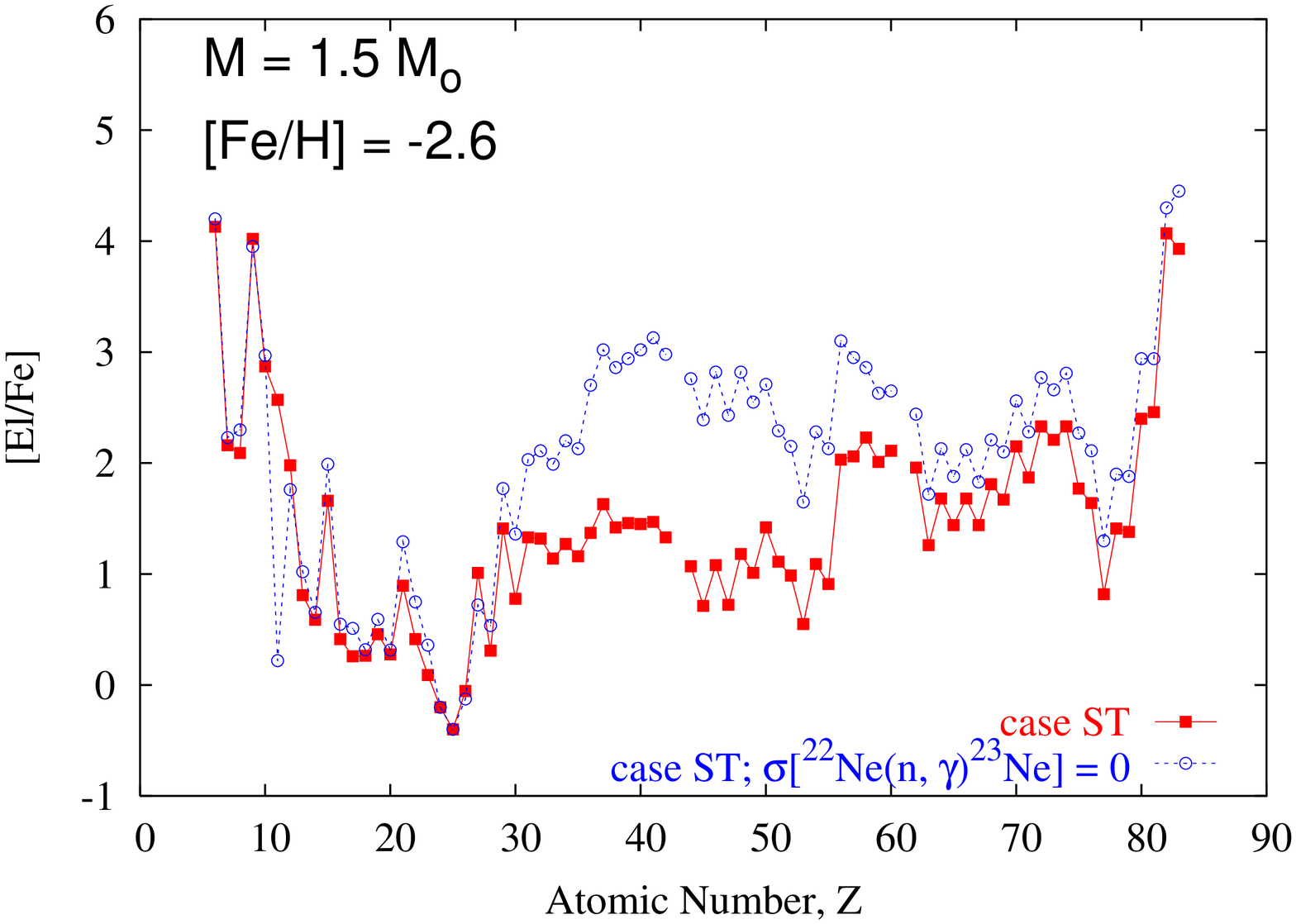}
\caption{{\it Top panel:} theoretical predictions of AGB models 
of $M$ = 1.5 $M_{\odot}$ for a test case with no $^{13}$C-pocket
(case ST = 0).
The line with squares stands for the ordinary case, while the  
line with circles corresponds to a case with the cross section of 
$^{22}$Ne(n, $\gamma$)$^{23}$Ne put to zero. {\it Bottom panel:}
The same as top panel but with a ST $^{13}$C-pocket.}
\label{Ne22m1p5}
\end{figure}

\begin{figure}
   \centering
\includegraphics[angle=-90,width=8cm]{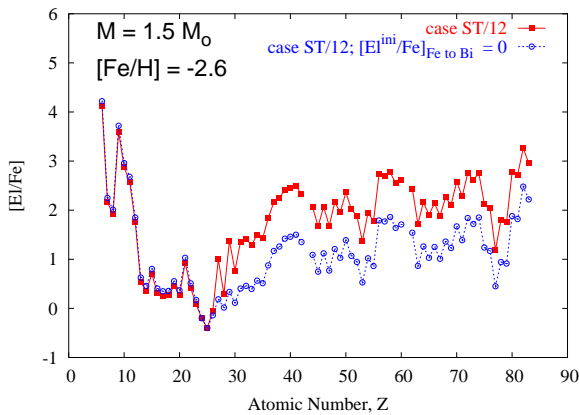}
\caption{Theoretical predictions of AGB models of $M$ = 1.5
 $M_{\odot}$ and case ST/12. The line with full squares stands for the
  ordinary case, while the line with empty circles corresponds to a case
   with the initial abundances of elements from $^{56}$Fe to
    $^{209}$Bi put to zero.}
\label{Ne22fetobi}
\end{figure}

At very low metallicities, a primary production of 
$^{22}$Ne, which increases with the initial mass, results in the 
advanced pulses by the conversion of primary $^{12}$C to 
primary $^{14}$N in the H-burning ashes, via 
$^{14}$N($\alpha$, $\gamma$)$^{18}$F($\beta^{+}$$\nu$)$^{18}$O 
and $^{18}$O($\alpha$, $\gamma$)$^{22}$Ne reaction during TPs
 (\citealt{mowlavi99}; \citealt{gallino06}; \citealt{husti07}).
The production of primary $^{22}$Ne is more efficient at lower
metallicities because a larger abundance of primary $^{12}$C 
is mixed with the envelope at each TDU episode. 
The subsequent activation of the H-shell converts almost 
all CNO nuclei into $^{14}$N.
This 
amount of primary $^{14}$N is left in the H-burning ashes and 
during the subsequent TP is converted into primary $^{22}$Ne.
  
To test the effect of neutron captures on $^{22}$Ne, 
we show in Fig.~\ref{Ne22m1p5}, the theoretical 
prediction for AGB models of $M$ = 1.5 $M_{\odot}$ and [Fe/H] = 
$-$2.6, without $^{13}$C-pocket (case ST = 0, top panel) and
with a case ST (bottom panel). Solid lines represent the ordinary 
cases, while the dashed lines correspond to cases
with the cross section of $^{22}$Ne(n, $\gamma$)$^{23}$Ne set to zero.
Concerning light elements, this test mainly affect $^{23}$Na, 
which decreases more than 2 dex, thus highlighting the primary
 production of $^{23}$Na via $^{22}$Ne(n, $\gamma$).
For Z $\ga$ 30, the larger amount of available neutrons 
feeds the production of the elements up to lead (blue dotted lines). 
This makes $^{22}$Ne, as well as $^{23}$Na, the main neutron
 poison at low metallicity during the $^{22}$Ne neutron burst. 
As illustrated in Fig.~\ref{Ne22m1p5}, bottom panel, for a 
ST case of $s$-process efficiency, $^{22}$Ne remains the most
 efficient neutron poison even inside the $^{13}$C-pocket 
 during the interpulse phase. 
Decreasing the initial AGB mass, the poison effect is visible
 only for efficient $^{13}$C-pocket.
For higher metallicities this effect decreases, becoming
 negligible at [Fe/H] = $-$1 \citep{gallino06}.
\\
Primary $^{22}$Ne plays a second role at low metallicities, 
becoming an iron seed producer for a primary $s$-process chain, 
depending on the adopted $^{13}$C-pocket.
Even if the neutron capture cross section of $^{22}$Ne is very 
small, not only the light elements as $^{23}$Na and the Mg 
isotopes are produced, but the neutron capture chain extends 
up to $^{56}$Fe, replenishing the starting seed for the production of 
$s$-elements.
This is shown in Fig.~\ref{Ne22fetobi}, for a low $^{13}$C-pocket 
(ST/12) and [Fe/H] = $-$2.6. Even if we set the initial abundances from 
$^{56}$Fe to $^{209}$Bi to zero, some primary iron 
and a consistent amount of primary $s$-elements are produced.
When neutrons are largely produced by the $^{13}$C($\alpha$, n)$^{16}$O 
reaction as tests with higher $^{13}$C-pockets demonstrate (e.g. a 
case ST $\times$ 2), a high
 percentage of $s$ elements is primary (produced directly by 
 neutron capture on $^{22}$Ne), and the $s$-element abundances are about
  0.2 dex lower than the standard case.
 The primary $^{16}$O and $^{12}$C produced at low metallicities act also
as neutron poison during the pocket (see online material,
Fig.~\ref{o16c12m1p3}).
We tested also the poison effect of other light elements that show minor 
or negligible differences.


\section{Na and Mg as initial mass indicators} \label{Na}

\begin{figure}
   \centering
\includegraphics[angle=-90,width=8cm]{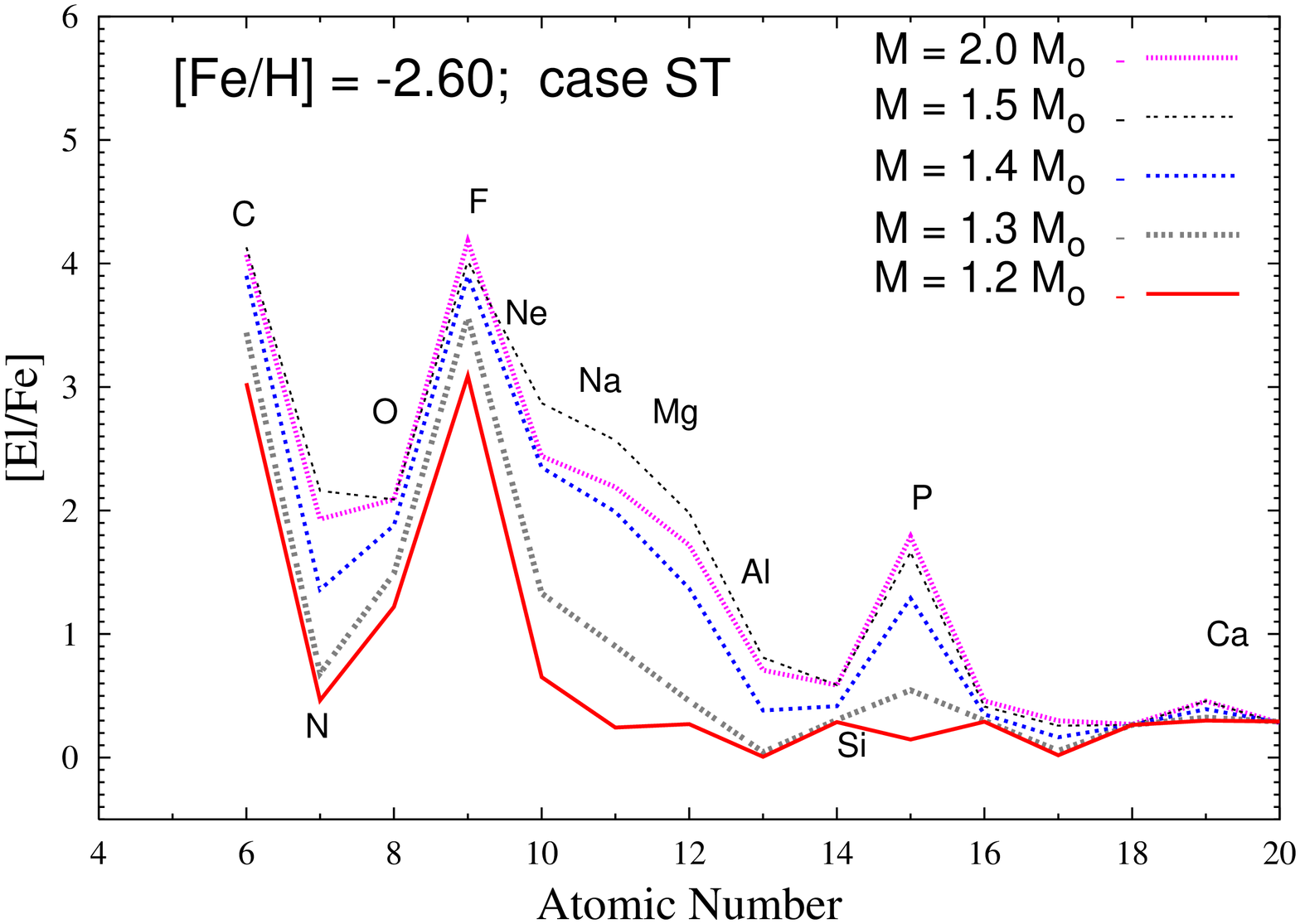}
\includegraphics[angle=-90,width=8cm]{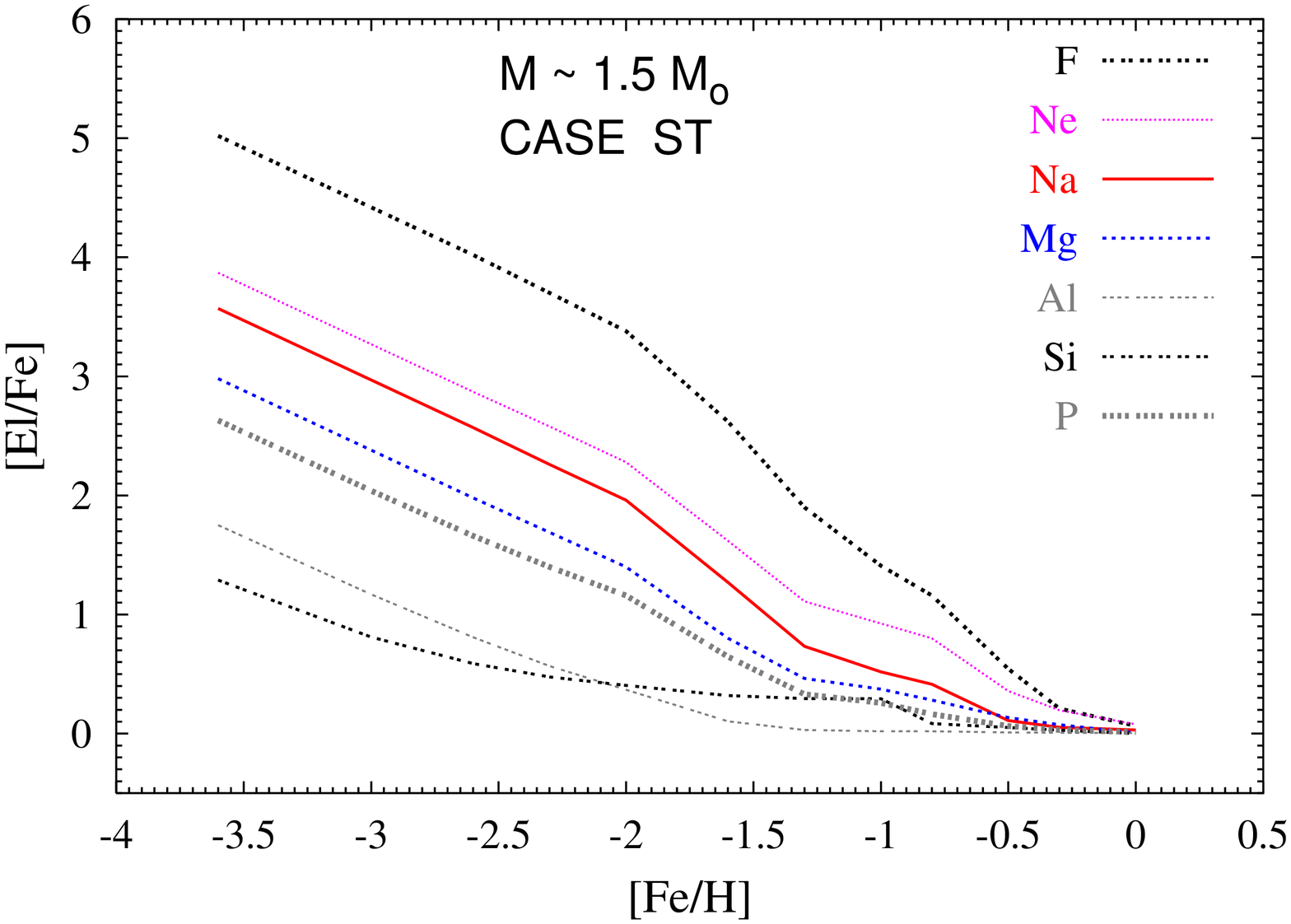}
\caption{{\it Top panel:} theoretical predictions of light 
elements for different initial masses, [Fe/H] = $-$2.6 
and a ST case. 
 {\it Bottom panel:} theoretical predictions of the light 
 elements in the region between Ne up to Si as a function
 of metallicity, for an AGB model of $M$ = 1.5 $M_{\odot}$ and 
ST case.}
\label{NaMg}
\end{figure}

As discussed in Section~\ref{ne22}, a progressive amount of primary 
$^{22}$Ne is produced in the He-intershell from the CNO cycle during 
the H-burning shell. 
The effect is a primary production of $^{23}$Na via
$^{22}$Ne(n,$\gamma$)$^{23}$Na, followed by the nucleosynthesis of Mg 
via $^{23}$Na(n,$\gamma$)$^{24}$Mg, $^{22}$Ne($\alpha$,n)$^{25}$Mg and
$^{22}$Ne($\alpha$,$\gamma$)$^{26}$Mg (\citealt{mowlavi99}; \citealt{gallino06}).
This chain involves light elements up to P, including Al and Si.
In Fig.~\ref{NaMg}, top panel, we show the [El/Fe] predictions in the 
region of the light elements, in a stellar mass range 1.2 $\leq$ $M/M_{\odot}$ $\leq$ 
2, for a case ST and [Fe/H] = $-$2.6. 
The production of the light elements as Ne, Na, Mg, and P 
 is strongly dependent on the initial AGB mass. In particular, 
differences higher than 2.5 dex are observed between $M$ =  1.2 and 1.5 $M_{\odot}$. 
The production of primary $^{22}$Ne increases at lower metallicity 
(Fig.~\ref{NaMg}, bottom panel, for a $M$ = 1.5 $M_{\odot}$ and a case ST), 
due to the higher production of $^{22}$Ne in the H-burning ashes, thus
explaining the slope predicted for [Na/Fe], [Mg/Fe], and
 other light elements.
In particular, Na (and Mg) are indicator of the 
mass of the AGB companion which polluted the observed CEMP-s stars
(\citealt{bisterzo06}; Paper II).
\\
We also remind that Na can be produced in IMS models by hot bottom 
burning (\citealt{sugimoto71,iben73,karakas03,ventura05}), which is not 
included in our models.

Fig.~\ref{NaMg}, top panel, shows a huge amount of fluorine produced in AGB stars via the chain $^{18}$O(p, $\alpha$)$^{15}$N($\alpha$, $\gamma$)$^{19}$F,
where protons are produced via $^{14}$N(n, p)$^{14}$C, neutrons via $^{13}$C($\alpha$, n)$^{16}$O reaction and $^{18}$O via the chain
$^{14}$N($\alpha$, $\gamma$)$^{18}$F($\beta^+$)$^{18}$O \citep{lugaro04,lugaro08,abia09}.
With the addition of proton captures in the network, the prediction of [F/Fe] increased by about 2 dex. We estimate a further increase by about 0.3 dex due to the primary $^{13}$C from H shell \citep{cristallo09}. Further discussions and 
improvement about fluorine, accounting for recent experimental measurements of reaction rates involved \citep{lacognata10}, will be given in a forthcoming paper.

\section{Conclusions}\label{conclusions}

We presented theoretical results of updated AGB stellar 
nucleosynthesis models in a mass range 1.3 $\leq$ $M/M_{\odot}$ $\la$ 3.0 
and metallicities $-$3.6 $\leq$ [Fe/H] $\leq$ $-$1.
We analysed the behaviour of the three $s$-process peaks and their ratios 
[hs/ls] and [Pb/hs] as the initial AGB mass and the $^{13}$C-pocket
 change. 
Then, we presented in detail the abundance composition in the 
envelope of all the elements, from carbon to bismuth.
\\
We find a high [C/Fe] already 
after the first TDU episode of models at low metallicities, due to the 
increase with the decrease in metallicity of the primary $^{12}$C 
and to an efficient TDU.
We discuss the deep impact of $^{22}$Ne as neutron seed and 
neutron poison by decreasing the metallicity. 
  This primary $^{22}$Ne increases with the initial mass and as a 
  consequence Na (and Mg) are strongly produced by decreasing the metallicity.
This makes of Na and Mg good indicators for the initial mass
of the AGB star.
\\ 
The three $s$-peaks are strongly dependent on the choice of the
 $^{13}$C-pocket as well as the initial mass and the metallicity.
[ls/Fe], [hs/Fe] and [Pb/Fe] do not have a linear behaviour with decreasing metallicity, and, depending on the $^{13}$C-pocket efficiency, they can cover a large range of values. Especially [Pb/Fe], largely produced by decreasing the metallicity, can reach 4.5 dex. 
By increasing the AGB mass and then the temperature at the bottom of the thermal pulse, [ls/Fe] receives an increasing contribution by the $^{22}$Ne($\alpha$, n)$^{25}$Mg reaction.
Two intrinsic $s$-process indexes, [hs/ls] and [Pb/hs], are needed in order to characterise the $s$ distribution, independent on the dilution of the AGB material onto the companion and on the TDU efficiency. 
The [hs/ls] ratio may give a constraint on the initial mass,
and it can be an indicator of the $s$-process efficiency as well, while [Pb/hs] gives information about the efficiency of the $^{13}$C-pocket only.
\\
As well as ls, hs, and Pb, we predict two peaks at Sn 
and Hf -- W, which receive a significant contribution from the $s$-process.
Also bismuth is synthesised in these objects via neutron capture 
on $^{208}$Pb.
We remark the importance of Nb measurement in CEMP-s stars, because an 
extrinsic AGB has [Zr/Fe] = [Nb/Fe] (e.g. CS 29497-030 by 
\citealt{ivans05}, see Paper II).
We refer to Paper II for a comparison between our AGB models and 
spectroscopic observations in CEMP-$s$ stars. 

This study together with Paper II are part of a series of works, 
planned to provide a complete picture
of the $s$-process nucleosynthesis in AGB stars.
We are planning to extend the analysis from CEMP-$s$ stars
up to solar metallicities.
We will compare AGB predictions presented here 
with spectroscopic observations of $s$-enhanced disk stars of different
spectral classification (e.g~CH and Barium binary stars, intrinsic post-AGBs,
extrinsic or intrinsic MS, S, C stars; for previous analysis see
\citealt{busso95,busso01} and \citealt{abia01}).
As last analysis, we will present update Galactic chemical evolution results,
with AGB yields of different masses and metallicities.
From the whole spectrum of these studies, we aim at better understanding
the AGB nucleosynthesis.

\section*{Acknowledgments}

We thank Dr. Maria Lugaro for helpful comments and discussions
which have helped improving the clarity and focus of this paper,
and, in particular, which motivated us to improve the network 
by introducing protons. 
This work was supported by the Italian MIUR-PRIN 2006 Project 
`Late Phases of Stellar Evolution: Nucleosynthesis in Supernovae,
AGB Stars, Planetary Nebulae'.


\onecolumn

\appendix

\newpage

\input{AppendixA.tex}\label{arlandiniupdated}

\clearpage
\newpage

\input{AppendixB.tex}

\clearpage
\newpage

\twocolumn

\input{AppendixC.tex}

\bsp

\label{lastpage}

\end{document}

%% file: table1.tex
\begin{table*}
\caption{Theoretical predictions of [El/Fe] and [El/Eu] for elements from Sr to Bi
(the label `El' stands for a generic element), at [Fe/H] = $-$2.6,
for $M$ = 1.5 $M_{\odot}$ models and various choices of the $^{13}$C-pocket
(ST, ST/12, ST/45, ST/75).}
\label{bab9ltHTZlow}
\centering
\resizebox{10cm}{!}{\begin{tabular}{|lcc|ccccc|cccc|} 
\hline
   &  & &  \multicolumn{7}{c}{$M$ = 1.5 $M_{\odot}$} & & \\[0.5ex]
   &  &    \multicolumn{7}{c}{[Fe/H] = $-$2.6} & \\[0.5ex]
\hline
El& Z & & & \multicolumn{3}{c}{[El/Fe]}   & &     \multicolumn{3}{c}{[El/Eu]} & \\                                  
 &     &    &  	ST	& ST/12	& ST/45	& ST/75     & &   ST	 &  ST/12&	ST/45&	ST/75     \\  
 (1)&(2)& &(3)&(4)&(5)&(6)& &(7)&(8)&(9)&(10)\\        
\hline                                                                                      
Sr & 38 & & 1.42	&  2.25 	& 2.00	& 1.81   &   & 0.16  &0.53  &0.92 &	1.36      \\
Y  & 39 & & 1.46	&  2.41 	& 2.08	& 1.85   &   & 0.20  &0.69  &1.00 &	1.40      \\
Zr & 40 & & 1.45	&  2.45 	& 2.02	& 1.76   &   & 0.19  &0.73  &0.94 &	1.31      \\
Nb & 41 & & 1.47	&  2.49 	& 2.05	& 1.78   &   & 0.21  &0.77  &0.97 &	1.33      \\
Mo & 42 & & 1.33	&  2.33 	& 1.85	& 1.55   &   & 0.07  &0.61  &0.77 &	1.10      \\
Ru & 44 & & 1.07	&  2.06 	& 1.56	& 1.27   &   & -0.19 &0.34  &0.48 &	0.82      \\
Rh & 45 & & 0.71	&  1.67 	& 1.19	& 0.92   &   & -0.55 &-0.05 &0.11 &	0.48      \\
Pd & 46 & & 1.08	&  2.07 	& 1.63	& 1.35   &   & -0.18 &0.35  &0.55 &	0.90      \\
Ag & 47 & & 0.72	&  1.68 	& 1.26	& 1.00   &   & -0.54 &-0.04 &0.18 &	0.55      \\
Cd & 48 & & 1.18	&  2.16 	& 1.76	& 1.47   &   & -0.08 &0.44  &0.68 &	1.02      \\
In & 49 & & 1.01	&  1.96 	& 1.56	& 1.27   &   & -0.25 &0.24  &0.48 &	0.82      \\
Sn & 50 & & 1.42	&  2.36 	& 1.87	& 1.54   &   & 0.16  &0.64  &0.79 &	1.09      \\
Sb & 51 & & 1.11	&  2.02 	& 1.51	& 1.17   &   & -0.15 &0.30  &0.43 &	0.72      \\
Te & 52 & & 0.99	&  1.88 	& 1.37	& 1.03   &   & -0.27 &0.16  &0.29 &	0.58      \\
I  & 53 & & 0.55	&  1.37 	& 0.88	& 0.57   &   & -0.71 &-0.35 &-0.20&	0.13      \\
Xe & 54 & & 1.09	&  1.95 	& 1.43	& 1.06   &   & -0.17 &0.23  &0.35 &	0.61      \\
Cs & 55 & & 0.91	&  1.77 	& 1.29	& 0.92   &   & -0.35 &0.05  &0.21 &	0.48      \\
Ba & 56 & & 2.03	&  2.73 	& 2.17	& 1.63   &   & 0.77  &1.01  &1.09 &	1.18      \\
La & 57 & & 2.06	&  2.69 	& 2.12	& 1.52   &   & 0.80  &0.97  &1.04 &	1.07      \\
Ce & 58 & & 2.23	&  2.77 	& 2.16	& 1.47   &   & 0.97  &1.05  &1.08 &	1.02      \\
Pr & 59 & & 2.01	&  2.55 	& 1.96	& 1.27   &   & 0.75  &0.83  &0.88 &	0.82      \\
Nd & 60 & & 2.11	&  2.62 	& 2.00	& 1.28   &   & 0.85  &0.90  &0.92 &	0.83      \\
Sm & 62 & & 1.96	&  2.44 	& 1.79	& 1.06   &   & 0.70  &0.72  &0.71 &	0.61      \\
Eu & 63 & & 1.26	&  1.72 	& 1.08	& 0.45   &   & 0.00  &0.00  &0.00 &	0.00      \\
Gd & 64 & & 1.68	&  2.16 	& 1.49	& 0.79   &   & 0.42  &0.44  &0.41 &	0.34      \\
Tb & 65 & & 1.44	&  1.91 	& 1.25	& 0.57   &   & 0.18  &0.19  &0.17 &	0.13      \\
Dy & 66 & & 1.68	&  2.14 	& 1.46	& 0.76   &   & 0.42  &0.42  &0.38 &	0.31      \\
Ho & 67 & & 1.44	&  1.89 	& 1.21	& 0.55   &   & 0.18  &0.17  &0.13 &	0.10     \\ 
Er & 68 & & 1.81	&  2.26 	& 1.57	& 0.85   &   & 0.55  &0.54  &0.49 &	0.41      \\
Tm & 69 & & 1.67	&  2.11 	& 1.42	& 0.72   &   & 0.41  &0.39  &0.34 &	0.27      \\
Yb & 70 & & 2.15	&  2.58 	& 1.87	& 1.13   &   & 0.89  &0.86  &0.79 &	0.68      \\
Lu & 71 & & 1.87	&  2.28 	& 1.58	& 0.85   &   & 0.61  &0.56  &0.50 &	0.41      \\
Hf & 72 & & 2.33	&  2.75 	& 2.05	& 1.28   &   & 1.07  &1.03  &0.97 &	0.83      \\
Ta & 73 & & 2.21	&  2.62 	& 1.92	& 1.16   &   & 0.95  &0.90  &0.84 &	0.71      \\
W  & 74 & & 2.33	&  2.75 	& 2.05	& 1.27   &   & 1.07  &1.03  &0.97 &	0.82      \\
Re & 75 & & 1.77	&  2.13 	& 1.41	& 0.70   &   & 0.51  &0.41  &0.33 &	0.25      \\
Os & 76 & & 1.64	&  2.04 	& 1.35	& 0.63   &   & 0.38  &0.32  &0.27 &	0.18      \\
Ir & 77 & & 0.82	&  1.19 	& 0.56	& 0.09   &   & -0.44 &-0.53 &-0.52&	-0.35     \\
Pt & 78 & & 1.41	&  1.80 	& 1.11	& 0.43   &   & 0.15  &0.08  &0.03 &	-0.02     \\
Au & 79 & & 1.38	&  1.76 	& 1.07	& 0.40   &   & 0.12  &0.04  &-0.01&	-0.05     \\
Hg & 80 & & 2.40	&  2.77 	& 2.05	& 1.26   &   & 1.14  &1.05  &0.97 &	0.81      \\
Tl & 81 & & 2.46	&  2.72 	& 1.93	& 1.16   &   & 1.20  &1.00  &0.85 &	0.71      \\
Pb & 82 & & 4.07	&  3.26 	& 1.98	& 1.12   &   & 2.81  &1.54  &0.90 &	0.67      \\
Bi & 83 & & 3.93	&  2.96 	& 1.45	& 0.58   &   & 2.67  &1.24  &0.37 &	0.13      \\ 
\hline                                                                                   
{[hs/ls]}&  & &     0.59 & 0.15	& -0.08	& -0.52  &    &		&		 &        &        \\
{[Pb/hs]}&  & &     2.03 & 0.68	& 0.01	& -0.17  &    &		&		 &        &        \\
\hline                                         
\end{tabular}}
\end{table*}

%% file: table2.tex
\begin{table*}
\caption{We listed the number of pulses with TDU experienced by each model.}
\label{pulsenumber}
\centering
\resizebox{15cm}{!}{\begin{tabular}{|lcc|ccccc|cccc|} 
\hline
Stellar Mass ($M_{\odot}$) & [Fe/H] & Number of TDUs & [Fe/H] &  Number of TDUs \\
1.3 & 0; $-$0.3 & 0  & from $-$0.5 down to $-$3.6 & 5 \\
1.4 & 0; $-$0.3 & 10 & from $-$0.5 down to $-$3.6 & 10 \\
1.5 & 0; $-$0.3 & 19 & from $-$0.8 down to $-$3.6 & 20 \\
2.0 & 0; $-$0.3 & 22; 25 & from $-$0.5 down to $-$3.6& 26\\
3.0 & 0; $-$0.3 & 25 & $-$0.5; from $-$0.8 down to $-$1.6  & 26; 35 \\
5.0 & 0; $-$0.3 & 24 & from $-$0.5 down to $-$1.6 & 24 \\
7.0 & 0; $-$0.3 & 24 & from $-$0.5 down to $-$1.6 & 24 \\
\hline                                         
\end{tabular}}
\end{table*}

%% file: table3.tex
\begin{table*}
\caption{The [ls/Fe], [hs/Fe], [Pb/Fe], [hs/ls] and [Pb/hs] predicted
ratios by varying the $^{13}$C-pocket efficiency, for AGB stellar 
models of $M$ = 1.3, 1.4, 1.5 and 2
 $M_{\odot}$, at [Fe/H] = $-$2.6.}
 \label{bab9ltHThssuls_z5m5_nro16eq_diffM}
\centering
\resizebox{12cm}{!}{\begin{tabular}{l|lrrrrrrrr} 
\hline
Mass & $^{13}$C-pocket&  ST$\times$2   & ST$\times$1.3 &  
ST     & ST/1.5     &  ST/2   &  ST/3   & ST/4.5   &  ST/6      \\ 
\hline                                                                                                                   
                      & {[ls/Fe]} &	0.63	&	0.49	&	0.41	&	0.35	&	0.34	&	0.41	&	0.51	&	0.58	\\
$M$ = 1.3 $M_{\odot}$ & {[hs/Fe]} &	1.04	&	0.86	&	0.77	&	0.72	&	0.76	&	0.91	&	1.08	&	1.29	\\
                      & {[Pb/Fe]} &	3.27	&	3.24	&	3.22	&	3.21	&	3.20	&	3.18	&	3.16	&	3.13	\\
                      & {[hs/ls]} &	0.41	&	0.37	&	0.36	&	0.38	&	0.42	&	0.51	&	0.58	&	0.72	\\
                      & {[Pb/hs]} &	2.23	&	2.38	&	2.45	&	2.49	&	2.44	&	2.27	&	2.08	&	1.84	\\
\hline                                                                                                                           
                      & {[ls/Fe]} &	1.42	&	1.22	&	1.11	&	1.04	&	1.05	&	1.16	&	1.43	&	1.65	\\
$M$ = 1.4 $M_{\odot}$ & {[hs/Fe]} &	1.77	&	1.54	&	1.47	&	1.52	&	1.69	&	2.09	&	2.30	&	2.41	\\
                      & {[Pb/Fe]} &	3.86	&	3.80	&	3.78	&	3.75	&	3.74	&	3.69	&	3.62	&	3.54	\\
                      & {[hs/ls]} &	0.35	&	0.33	&	0.37	&	0.48	&	0.64	&	0.93	&	0.88	&	0.76	\\
                      & {[Pb/hs]} &	2.09	&	2.26	&	2.31	&	2.23	&	2.05	&	1.60	&	1.32	&	1.13	\\
\hline                                                                                                                          
                      & {[ls/Fe]} &	1.69	&	1.51	&	1.46	&	1.51	&	1.66	&	2.00	&	2.25	&	2.35	\\
$M$ = 1.5 $M_{\odot}$ & {[hs/Fe]} &	2.11	&	1.96	&	2.04	&	2.42	&	2.60	&	2.73	&	2.83	&	2.80	\\
                      & {[Pb/Fe]} &	4.16	&	4.10	&	4.07	&	4.02	&	3.98	&	3.88	&	3.72	&	3.60	\\
                      & {[hs/ls]} &	0.42	&	0.45	&	0.59	&	0.91	&	0.95	&	0.73	&	0.59	&	0.45	\\
                      & {[Pb/hs]} &	2.05	&	2.14	&	2.03	&	1.60	&	1.38	&	1.15	&	0.89	&	0.80	\\
\hline                                                                                                                          
                      & {[ls/Fe]} &	1.93	&	1.68	&	1.52	&	1.41	&	1.40	&	1.51	&	1.78	&	2.00	\\
$M$ = 2.0 $M_{\odot}$ & {[hs/Fe]} &	2.37	&	2.08	&	1.93	&	1.88	&	2.02	&	2.41	&	2.62	&	2.73	\\
                      & {[Pb/Fe]} &	4.29	&	4.18	&	4.14	&	4.10	&	4.07	&	4.03	&	3.96	&	3.88	\\
                      & {[hs/ls]} &	0.44	&	0.41	&	0.41	&	0.48	&	0.63	&	0.90	&	0.84	&	0.74	\\					
                      & {[Pb/hs]} &	1.92	&	2.10	&	2.21	&	2.22	&	2.05	&	1.62	&	1.34	&	1.15	\\						
\hline                                                                                                                          
\\
\hline
Mass & $^{13}$C-pocket &  ST/9      &  ST/12  & ST/18 & ST/24  & ST/30  & ST/45 &  ST/75  & ST/150 \\               
\hline                                                                      
                      & {[ls/Fe]} &	0.71  &	0.89	&	1.22	&	1.40	&	1.49	&	1.55	&	1.46	&	0.49   \\
$M$ = 1.3 $M_{\odot}$ & {[hs/Fe]} &	1.68  &	1.88	&	2.01	&	2.04	&	2.02	&	1.83	&	1.05	&	0.13   \\
                      & {[Pb/Fe]} &	3.07  &	2.99	&	2.81	&	2.62	&	2.40	&	1.77	&	0.89	&	0.03   \\
                      & {[hs/ls]} &	0.97  &	0.99	&	0.79	&	0.64	&	0.53	&	0.28	&	-0.40	&	-0.36  \\
                      & {[Pb/hs]} &	1.39  &	1.11	&	0.80	&	0.58	&	0.38	&	-0.06	&	-0.16	&	-0.10  \\
\hline                                                                            
                      & {[ls/Fe]} &	1.88  &	1.99	&	2.08	&	2.09	&	2.06	&	1.95	&	1.70	&	0.82  \\
$M$ = 1.4 $M_{\odot}$ & {[hs/Fe]} &	2.51  &	2.49	&	2.42	&	2.33	&	2.25	&	1.99	&	1.29	&	0.21  \\
                      & {[Pb/Fe]} &	3.40  &	3.26	&	3.02	&	2.81	&	2.58	&	2.01	&	1.12	&	0.12  \\
                      & {[hs/ls]} &	0.63  &	0.50	&	0.34	&	0.24	&	0.19	&	0.05	&	-0.41	&	-0.61 \\
                      & {[Pb/hs]} &	0.89  &	0.77	&	0.60	&	0.48	&	0.33	&	0.02	&	-0.17	&	-0.09 \\
\hline                                   
                      & {[ls/Fe]} &	2.44  &	2.43	&	2.35	&	2.28	&	2.20	&	2.05	&	1.81	&	1.22  	\\ 
$M$ = 1.5 $M_{\odot}$ & {[hs/Fe]} &	2.68  &	2.58	&	2.45	&	2.33	&	2.23	&	1.97	&	1.29	&	0.26 	\\ 
                      & {[Pb/Fe]} &	3.41  &	3.26	&	3.00	&	2.78	&	2.56	&	1.98	&	1.12	&	0.21 	\\ 
                      & {[hs/ls]} &	0.25  &	0.15	&	0.10	&	0.06	&	0.03	&	-0.08	&	-0.52	&	-0.95 	\\
                      & {[Pb/hs]} &	0.73  &	0.68	&	0.55	&	0.45	&	0.33	&	0.01	&	-0.17	&	-0.05 	\\   
\hline                                   
                      & {[ls/Fe]} &	2.22  &	2.34	&	2.45	&	2.46	&	2.44	&	2.32	&	2.06	&	1.35 	\\						   
$M$ = 2.0 $M_{\odot}$ & {[hs/Fe]} &	2.85  &	2.85	&	2.75	&	2.67	&	2.58	&	2.17	&	1.46	&	0.30 	\\						   
                      & {[Pb/Fe]} &	3.73  &	3.58	&	3.32	&	3.06	&	2.79	&	2.20	&	1.29	&	0.18 	\\						   
                      & {[hs/ls]} &	0.63  &	0.51	&	0.30	&	0.21	&	0.14	&	-0.14	&	-0.60	&	-1.05	 \\						
                      & {[Pb/hs]} &	0.88  &	0.73	&	0.57	&	0.39	&	0.21	&	0.03	&	-0.17	&	-0.12	 \\
                      \hline    
\end{tabular}}
\end{table*}

%% file: AppendixA.tex
\clearpage

\section{Solar $s$-process contribution} 

In Table~\ref{bab9ltHT}, we show the best representation 
of the solar main component with theoretical predictions  
in percentage for elements from Sr to Bi for an average between
$M$ = 1.5 and 3 $M_{\odot}$ at half solar metallicity and 
a case ST. 
Here, the main-$s$ percentages presented by \citet{arlandini99}, 
stellar model (reported as comparison in Col.~3), 
are updated with new solar abundances and a network upgraded
to 2009 (Col.~4). The results by \citet{bisterzo06NIC9}
 have been further updated with the recent cross sections 
 measurement of $^{62}$Ni \citep{alpizar08}, 
$^{90,91}$Zr \citep{tagliente08a,tagliente08b}, 
$^{186,187,188}$Os isotopes \citep{mosconi08}, 
$^{204,206,207}$Pb \citep{domingopardo06,domingopardo07a,domingopardo07b}, 
$^{209}$Bi(n, $\gamma$)$^{210}$Bi$^{\rm g}$ \citep{bisterzo07}
(see also KADoNiS, Karlsruhe Astrophysical Database of Nucleosynthesis in Stars,
web address http://nuclear-astrophysics.fzk.de/kadonis/.). 
The case ST at [Fe/H] = $-$0.3 gives a [hs/ls] = $-$0.25 
 (Col.~4, and Fig.~\ref{AA_hssuls_m1p3m1p5m2_noobs}, middle panel).
 In Col.~5 we report the normalization of 
the main-$s$ percentages to europium in logarithmic scale, [El/Eu].
As shown in Col.~4, only $\sim$ 6\% of solar europium is produced 
by the $s$-process, and it is considered a typical $r$-process element.
The normalization to europium highlights the amount of a pure $s$-process
contribution to each element [El/Eu]$_{\rm s}$.
This [El/Eu] ratio is useful to compare our theoretical predictions
with spectroscopic observations, especially in CEMP-$s$ (and CEMP-$s+r$) 
stars, to understand if there is competition between $r$- and $s$-process 
(see Table~\ref{bab9ltHTZlow} and Paper II).
A pure $s$-process contribution predicts [La/Eu]$_{\rm s}$ = 1.08 
at [Fe/H] = $-$0.3, and 0.8 $\leq$ [La/Eu]$_{\rm s}$ $\leq$ 1.1 at 
[Fe/H] = $-$2.6 (see Table~\ref{bab9ltHTZlow}, Cols.~7 to~10).
If lower [La/Eu]$_{\rm s}$ values are observed, this indicate stars that 
experienced an important $r$-process contribution in addition to the $s$-process 
enhancements.

\begin{figure}
   \centering
\includegraphics[angle=-90,width=12cm]{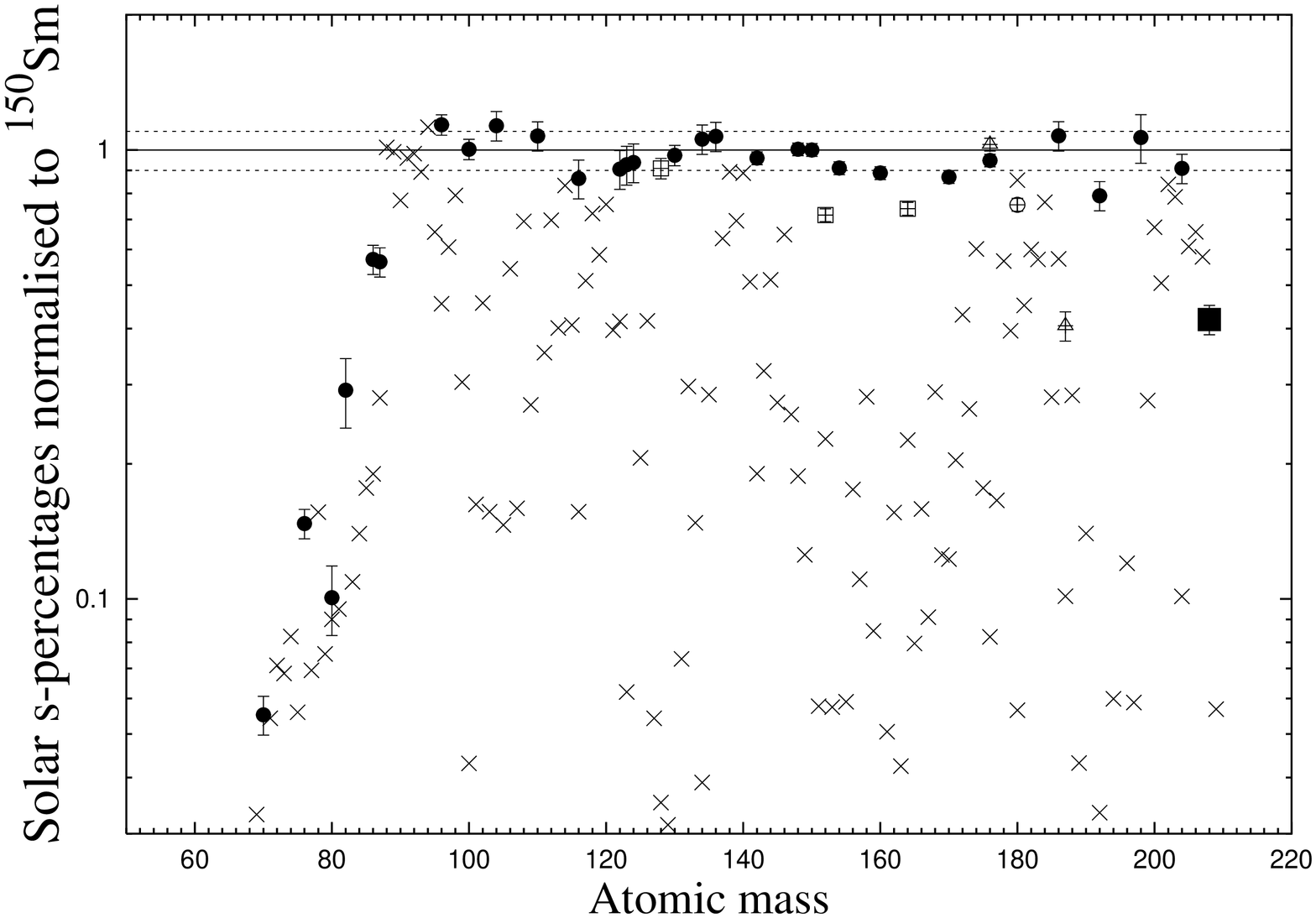}
\caption{Solar s\% normalized to $^{150}$Sm versus atomic mass
for the solar main component as in \citet{arlandini99}, updated to 2009.}
\label{maincomponent}
\end{figure}

The same model presented in Table~\ref{bab9ltHT}, is shown in Fig.~\ref{maincomponent} for isotopes from Sr to Bi normalised to the $s$-only nucleus $^{150}$Sm. 
The full circles are the $s$-only nuclei. We adopted different symbols for $^{128}$Xe, $^{152}$Gd, and $^{164}$Er, which have a not negligible p contribution (10\% for Xe), 
for $^{176}$Lu, a long-lived isotope (3.8 $\times$ 10$^{10}$ y) which decays into $^{176}$Hf, for $^{187}$Os, which is affected by the long-lived decay of $^{187}$Re (5 $\times$ 10$^{10}$ y), and for $^{180}$Ta, which receives also contributions from the $p$-process and from $\nu$-interactions in massive stars. 
The black full square corresponds to $^{208}$Pb, which receives a contribution of about 50\%  by the strong-$s$ component \citep{travaglio01,travaglio04,serminato09}.

\input{tableA1.tex}

%% file: tableA1.tex
\begin{table*}
\caption{Theoretical predictions in percentage for elements 
from Sr to Bi, (the label `El' stands for a generic elements) adopted to
reproduce the main component obtained by an average of $M$ = 1.5 and 3.0
$M_{\odot}$ models ($M_{\rm aver}$) at [Fe/H] = $-$0.3 
(case ST as in \citealt{arlandini99}, stellar model, Col.~3), improved
with cross section measurements and solar abundances upgraded to 2009
(Col.~4; see text of this Appendix).
In Col.~5 we report the normalization of the updated main-$s$ 
percentages to europium in logarithmic scale, [El/Eu].}
\label{bab9ltHT}
\centering
\resizebox{8cm}{!}{\begin{tabular}{|lc|ccc|} 
\hline
   &  & Arlandini ($M_{\rm aver}$)& Updated & Updated \\[0.5ex]
   &  &   [Fe/H] = $-$0.3 & [Fe/H] = $-$0.3 &  \\[0.5ex]
\hline
El& Z &   	\%	  &  \% &  [El/Eu]  \\                                  
 &     &      ST	& ST  &      ST     \\  
 (1)&(2)&(3)&(4)&(5)\\        
\hline                                                           
Sr & 38 &	85.0 &    93.1     & 1.21      \\  
Y  & 39 &	92.0 &    99.0     & 1.23      \\  
Zr & 40 &	83.0 &    88.1     & 1.18      \\  
Nb & 41 &	85.0 &    89.3     & 1.19      \\  
Mo & 42 &	50.0 &    54.8     & 0.98      \\  
Ru & 44 &	32.0 &    34.1     & 0.77      \\  
Rh & 45 &	14.0 &    15.6     & 0.43      \\  
Pd & 46 &	46.0 &    49.4     & 0.93      \\  
Ag & 47 &	20.0 &    21.3     & 0.56      \\  
Cd & 48 &	52.0 &    64.0     & 1.04      \\  
In & 49 &	35.0 &    39.0     & 0.83      \\  
Sn & 50 &	65.0 &    65.6     & 1.05      \\  
Sb & 51 &	25.0 &    25.3     & 0.64      \\  
Te & 52 &	17.0 &    18.2     & 0.50      \\  
I  & 53 &	5.3  &    5.4      & -0.03     \\  
Xe & 54 &	17.0 &    17.0     & 0.47      \\  
Cs & 55 &	15.0 &    14.8     & 0.41      \\  
Ba & 56 &	81.0 &    84.1     & 1.16      \\  
La & 57 &	62.0 &    69.5     & 1.08      \\  
Ce & 58 &	77.0 &    80.7     & 1.14      \\  
Pr & 59 &	49.0 &    50.8     & 0.94      \\  
Nd & 60 &	56.0 &    56.7     & 0.99      \\  
Sm & 62 &	29.0 &    30.9     & 0.73      \\  
Eu & 63 &	5.8  &    5.8      & 0.00      \\  
Gd & 64 &	15.0 &    11.0     & 0.28      \\  
Tb & 65 &	7.2  &    8.5      & 0.17      \\  
Dy & 66 &	15.0 &    14.4     & 0.39      \\  
Ho & 67 &	7.8  &    8.0     &  0.14     \\   
Er & 68 &	17.0 &    18.2     & 0.50      \\  
Tm & 69 &	13.0 &    12.5     & 0.33      \\  
Yb & 70 &	33.0 &    39.4     & 0.83      \\  
Lu & 71 &	20.0 &    19.9     & 0.54      \\  
Hf & 72 &	56.0 &    58.9     & 1.01      \\  
Ta & 73 &	41.0 &    45.0     & 0.89      \\  
W  & 74 &	56.0 &    63.8     & 1.04      \\  
Re & 75 &	8.9  &    16.9     & 0.46      \\  
Os & 76 &	9.4  &    11.9     & 0.31      \\  
Ir & 77 &	1.4  &    1.5      & -0.59     \\  
Pt & 78 &	5.1  &    6.4      & 0.04      \\  
Au & 79 &	5.8  &    5.9      & 0.01      \\  
Hg & 80 &	61.0 &    63.2     & 1.04      \\  
Tl & 81 &	76.0 &    66.1     & 1.06      \\  
Pb & 82 &	46.0 &    49.9     & 0.93      \\  
Bi & 83 &	4.9  &    5.7      & -0.01     \\  
\hline                                                                               
[hs/ls]& & -0.27 &-0.25  &         \\
\hline                                         
\end{tabular}}
\end{table*}

%% file: AppendixB.tex
\clearpage

\section{Data Tables} 

\subsection{[El/Fe] predictions as the $^{13}$C-pocket changes \label{dataappendixa}}

 In the following Tables we list the theoretical surface predictions [El/Fe]
for four initial AGB masses at [Fe/H] = $-$2.6, for elements from helium to bismuth: 
$M^{\rm AGB}_{\rm ini}$ = 1.3 $M_{\odot}$ (n5) (Table~\ref{datatablebab9ltHTm1p3z5m5a}),
$M^{\rm AGB}_{\rm ini}$ = 1.4 $M_{\odot}$ (n10) (Table~\ref{datatablebab9ltHTm1p4z5m5a}),
$M^{\rm AGB}_{\rm ini}$ = 1.5 $M_{\odot}$ (n20) (Tables~\ref{datatablebab9ltHTm1p5z5m5a}),
$M^{\rm AGB}_{\rm ini}$ = 2.0 $M_{\odot}$ (n26) (Tables~\ref{datatablebab9ltHTm2z5m5a}), 
where `$ni$' is the number of TPs with TDU.
Columns correspond to the different results obtained using a wide range 
of $^{13}$C-pockets (ST$\times$2 down to ST/150).
The case ST/150 mainly shows the $^{22}$Ne($\alpha$, n)$^{25}$Mg neutron source contribution.

\input{tableB1.tex}

\input{tableB2.tex}

\input{tableB3.tex}

\input{tableB4.tex}

\clearpage
\newpage

\subsection{[El/Fe] predictions as the metallicity changes \label{dataappendixb}}

In Tables~\ref{datatablebab9ltHTp1p5d8m1p3a} and~\ref{datatablebab9ltHTp1p5d8m1p5a} 
we list the theoretical surface predictions [El/Fe] for $M^{\rm AGB}_{\rm ini}$ = 
1.3 and 1.5 $M_{\odot}$ models for two $^{13}$C-pockets (ST and ST/12), by changing the metallicity.

\input{tableB5.tex}

\input{tableB6.tex}

%% file: tableB1.tex
\begin{table*}
\caption{Theoretical [El/Fe] predictions from He to Bi for $M^{\rm AGB}_{\rm ini}$ = 1.3 $M_{\odot}$ (which undergoes 5 TPs with TDU, n5) and [Fe/H] = $-$2.6 models, as the $^{13}$C-pocket changes (ST$\times$2 down to ST/150).}
 \label{datatablebab9ltHTm1p3z5m5a}
\centering
\resizebox{15cm}{!}{\begin{tabular}{cccccccccccccccccc} 
\hline
\\
El  &  Z  &  ST*2  & ST*1.3 &   ST  &  ST/1.5 &  ST/2  &  ST/3  & ST/4.5 &  ST/6  & ST/9  & ST/12  & ST/18 & ST/24 & ST/30 & ST/45 &  ST/75  & ST/150 \\
\hline
He  &  4  &  0.06  &  0.06  &  0.06 &   0.06  &  0.06  &  0.06  &  0.06  &  0.05  & 0.06  &  0.06  &  0.06 &  0.06 &  0.06 &  0.06 &   0.06  &  0.05  \\
C   &  6  &  3.44  &  3.44  &  3.44 &   3.44  &  3.43  &  3.43  &  3.43  &  3.43  & 3.43  &  3.43  &  3.43 &  3.43 &  3.43 &  3.43 &   3.43  &  3.43  \\
N   &  7  &  0.68  &  0.68  &  0.68 &   0.68  &  0.68  &  0.68  &  0.68  &  0.68  & 0.68  &  0.68  &  0.68 &  0.68 &  0.68 &  0.68 &   0.68  &  0.68  \\
O   &  8  &  1.54  &  1.51  &  1.49 &   1.46  &  1.44  &  1.42  &  1.40  &  1.38  & 1.37  &  1.36  &  1.35 &  1.34 &  1.33 &  1.33 &   1.32  &  1.32  \\
F   &  9  &  3.63  &  3.60  &  3.57 &   3.52  &  3.49  &  3.43  &  3.38  &  3.33  & 3.25  &  3.19  &  3.10 &  3.03 &  2.97 &  2.87 &   2.72  &  2.45  \\
Ne  & 10  &  1.38  &  1.35  &  1.33 &   1.31  &  1.30  &  1.29  &  1.28  &  1.28  & 1.27  &  1.27  &  1.26 &  1.26 &  1.26 &  1.26 &   1.26  &  1.25  \\
Na  & 11  &  0.98  &  0.94  &  0.90 &   0.86  &  0.84  &  0.80  &  0.77  &  0.75  & 0.73  &  0.71  &  0.69 &  0.68 &  0.67 &  0.65 &   0.64  &  0.61  \\
Mg  & 12  &  0.50  &  0.48  &  0.46 &   0.44  &  0.42  &  0.40  &  0.39  &  0.38  & 0.36  &  0.35  &  0.35 &  0.34 &  0.34 &  0.33 &   0.33  &  0.33  \\
Al  & 13  &  0.06  &  0.05  &  0.04 &   0.03  &  0.03  &  0.02  &  0.02  &  0.02  & 0.01  &  0.01  &  0.01 &  0.01 &  0.01 &  0.01 &   0.01  &  0.01  \\
Si  & 14  &  0.32  &  0.31  &  0.31 &   0.30  &  0.30  &  0.29  &  0.29  &  0.29  & 0.29  &  0.29  &  0.29 &  0.29 &  0.29 &  0.29 &   0.29  &  0.29  \\
P   & 15  &  0.81  &  0.65  &  0.55 &   0.42  &  0.36  &  0.29  &  0.24  &  0.22  & 0.19  &  0.18  &  0.16 &  0.14 &  0.13 &  0.11 &   0.08  &  0.04  \\
S   & 16  &  0.31  &  0.30  &  0.30 &   0.30  &  0.30  &  0.29  &  0.29  &  0.29  & 0.29  &  0.29  &  0.29 &  0.29 &  0.29 &  0.29 &   0.29  &  0.29  \\
Cl  & 17  &  0.07  &  0.06  &  0.06 &   0.05  &  0.05  &  0.05  &  0.04  &  0.04  & 0.04  &  0.04  &  0.04 &  0.05 &  0.05 &  0.05 &   0.06  &  0.05  \\
Ar  & 18  &  0.27  &  0.26  &  0.26 &   0.26  &  0.26  &  0.26  &  0.26  &  0.26  & 0.26  &  0.26  &  0.26 &  0.26 &  0.26 &  0.26 &   0.26  &  0.26  \\
K   & 19  &  0.33  &  0.33  &  0.32 &   0.32  &  0.32  &  0.32  &  0.31  &  0.31  & 0.31  &  0.31  &  0.31 &  0.31 &  0.31 &  0.31 &   0.31  &  0.31  \\
Ca  & 20  &  0.29  &  0.29  &  0.29 &   0.29  &  0.29  &  0.29  &  0.29  &  0.29  & 0.29  &  0.29  &  0.29 &  0.29 &  0.29 &  0.29 &   0.29  &  0.29  \\
Sc  & 21  &  0.39  &  0.36  &  0.34 &   0.31  &  0.29  &  0.26  &  0.24  &  0.23  & 0.22  &  0.22  &  0.23 &  0.24 &  0.24 &  0.24 &   0.22  &  0.16  \\
Ti  & 22  &  0.28  &  0.27  &  0.27 &   0.26  &  0.26  &  0.26  &  0.26  &  0.26  & 0.26  &  0.27  &  0.27 &  0.27 &  0.27 &  0.27 &   0.26  &  0.25  \\
V   & 23  &  0.02  &  0.01  &  0.01 &   0.01  &  0.00  &  0.00  &  0.00  &  0.00  & 0.01  &  0.01  &  0.01 &  0.01 &  0.01 &  0.00 &   0.00  &  0.00  \\
Cr  & 24  & -0.20  & -0.20  & -0.20 &  -0.20  & -0.20  & -0.20  & -0.20  & -0.20  &-0.20  & -0.20  & -0.20 & -0.20 & -0.20 & -0.20 &  -0.20  & -0.20  \\
Mn  & 25  & -0.40  & -0.40  & -0.40 &  -0.40  & -0.40  & -0.40  & -0.40  & -0.40  &-0.40  & -0.40  & -0.40 & -0.40 & -0.40 & -0.40 &  -0.40  & -0.40  \\
Fe  & 26  &  0.00  &  0.00  &  0.00 &   0.00  &  0.00  &  0.00  &  0.00  &  0.00  & 0.00  &  0.00  &  0.00 &  0.00 &  0.00 &  0.00 &   0.00  &  0.00  \\
Co  & 27  &  0.43  &  0.43  &  0.43 &   0.42  &  0.42  &  0.41  &  0.39  &  0.38  & 0.36  &  0.35  &  0.33 &  0.31 &  0.30 &  0.28 &   0.27  &  0.25  \\
Ni  & 28  &  0.10  &  0.09  &  0.08 &   0.08  &  0.07  &  0.06  &  0.05  &  0.04  & 0.04  &  0.03  &  0.03 &  0.02 &  0.02 &  0.02 &   0.02  &  0.01  \\
Cu  & 29  &  0.86  &  0.83  &  0.80 &   0.76  &  0.73  &  0.67  &  0.61  &  0.56  & 0.50  &  0.45  &  0.39 &  0.36 &  0.34 &  0.32 &   0.33  &  0.31  \\
Zn  & 30  &  0.38  &  0.34  &  0.31 &   0.26  &  0.23  &  0.18  &  0.14  &  0.12  & 0.09  &  0.07  &  0.06 &  0.05 &  0.06 &  0.08 &   0.12  &  0.11  \\
Ga  & 31  &  0.83  &  0.76  &  0.70 &   0.60  &  0.54  &  0.43  &  0.34  &  0.29  & 0.22  &  0.19  &  0.16 &  0.17 &  0.20 &  0.32 &   0.41  &  0.38  \\
Ge  & 32  &  0.81  &  0.72  &  0.65 &   0.54  &  0.47  &  0.36  &  0.28  &  0.23  & 0.18  &  0.16  &  0.15 &  0.18 &  0.24 &  0.41 &   0.49  &  0.43  \\
As  & 33  &  0.64  &  0.55  &  0.48 &   0.37  &  0.31  &  0.22  &  0.16  &  0.13  & 0.10  &  0.09  &  0.10 &  0.14 &  0.21 &  0.37 &   0.43  &  0.35  \\
Se  & 34  &  0.72  &  0.61  &  0.52 &   0.40  &  0.32  &  0.23  &  0.17  &  0.15  & 0.13  &  0.13  &  0.15 &  0.24 &  0.35 &  0.55 &   0.60  &  0.45  \\
Br  & 35  &  0.62  &  0.51  &  0.43 &   0.32  &  0.25  &  0.17  &  0.13  &  0.11  & 0.10  &  0.10  &  0.13 &  0.21 &  0.31 &  0.50 &   0.54  &  0.38  \\
Kr  & 36  &  0.70  &  0.56  &  0.46 &   0.33  &  0.27  &  0.22  &  0.21  &  0.22  & 0.24  &  0.26  &  0.36 &  0.50 &  0.63 &  0.79 &   0.79  &  0.47  \\
Rb  & 37  &  0.83  &  0.67  &  0.55 &   0.41  &  0.35  &  0.31  &  0.32  &  0.34  & 0.37  &  0.41  &  0.54 &  0.71 &  0.85 &  1.01 &   0.99  &  0.57  \\
Sr  & 38  &  0.63  &  0.49  &  0.40 &   0.31  &  0.29  &  0.32  &  0.39  &  0.45  & 0.55  &  0.69  &  1.00 &  1.20 &  1.32 &  1.42 &   1.38  &  0.63  \\
Y   & 39  &  0.63  &  0.49  &  0.41 &   0.34  &  0.33  &  0.39  &  0.49  &  0.56  & 0.68  &  0.84  &  1.17 &  1.37 &  1.47 &  1.55 &   1.48  &  0.57  \\
Zr  & 40  &  0.63  &  0.49  &  0.41 &   0.35  &  0.35  &  0.42  &  0.52  &  0.59  & 0.74  &  0.94  &  1.27 &  1.43 &  1.51 &  1.55 &   1.43  &  0.41  \\
Nb  & 41  &  0.66  &  0.51  &  0.43 &   0.37  &  0.37  &  0.44  &  0.55  &  0.62  & 0.78  &  0.98  &  1.32 &  1.48 &  1.55 &  1.59 &   1.46  &  0.39  \\
Mo  & 42  &  0.55  &  0.42  &  0.34 &   0.29  &  0.29  &  0.36  &  0.46  &  0.52  & 0.68  &  0.89  &  1.21 &  1.35 &  1.41 &  1.42 &   1.26  &  0.23  \\
Ru  & 44  &  0.40  &  0.29  &  0.23 &   0.18  &  0.18  &  0.22  &  0.29  &  0.34  & 0.46  &  0.65  &  0.94 &  1.07 &  1.13 &  1.15 &   1.00  &  0.12  \\
Rh  & 45  &  0.21  &  0.14  &  0.10 &   0.08  &  0.07  &  0.10  &  0.13  &  0.16  & 0.25  &  0.39  &  0.63 &  0.75 &  0.82 &  0.83 &   0.70  &  0.05  \\
Pd  & 46  &  0.43  &  0.31  &  0.24 &   0.19  &  0.19  &  0.24  &  0.32  &  0.39  & 0.54  &  0.75  &  1.08 &  1.22 &  1.29 &  1.31 &   1.15  &  0.17  \\
Ag  & 47  &  0.22  &  0.14  &  0.11 &   0.08  &  0.08  &  0.11  &  0.16  &  0.20  & 0.32  &  0.49  &  0.77 &  0.90 &  0.96 &  0.98 &   0.83  &  0.07  \\
Cd  & 48  &  0.44  &  0.33  &  0.27 &   0.24  &  0.25  &  0.33  &  0.44  &  0.52  & 0.72  &  0.96  &  1.28 &  1.41 &  1.47 &  1.48 &   1.29  &  0.22  \\
In  & 49  &  0.31  &  0.22  &  0.19 &   0.17  &  0.18  &  0.24  &  0.33  &  0.40  & 0.57  &  0.80  &  1.10 &  1.22 &  1.28 &  1.28 &   1.09  &  0.14  \\
Sn  & 50  &  0.60  &  0.47  &  0.40 &   0.35  &  0.37  &  0.46  &  0.57  &  0.66  & 0.89  &  1.15  &  1.45 &  1.56 &  1.61 &  1.60 &   1.35  &  0.24  \\
Sb  & 51  &  0.39  &  0.28  &  0.23 &   0.19  &  0.20  &  0.27  &  0.35  &  0.42  & 0.62  &  0.86  &  1.13 &  1.23 &  1.27 &  1.25 &   0.98  &  0.10  \\
Te  & 52  &  0.32  &  0.23  &  0.18 &   0.15  &  0.16  &  0.21  &  0.29  &  0.35  & 0.53  &  0.76  &  1.02 &  1.12 &  1.16 &  1.13 &   0.86  &  0.07  \\
I   & 53  &  0.12  &  0.08  &  0.05 &   0.04  &  0.05  &  0.07  &  0.10  &  0.13  & 0.23  &  0.38  &  0.59 &  0.67 &  0.70 &  0.68 &   0.45  &  0.01  \\
Xe  & 54  &  0.38  &  0.27  &  0.22 &   0.19  &  0.19  &  0.25  &  0.34  &  0.40  & 0.61  &  0.86  &  1.11 &  1.20 &  1.24 &  1.20 &   0.89  &  0.08  \\
Cs  & 55  &  0.28  &  0.20  &  0.16 &   0.13  &  0.14  &  0.19  &  0.26  &  0.33  & 0.52  &  0.76  &  1.01 &  1.10 &  1.13 &  1.10 &   0.78  &  0.06  \\
Ba  & 56  &  1.09  &  0.92  &  0.82 &   0.76  &  0.80  &  0.93  &  1.09  &  1.26  & 1.63  &  1.87  &  2.05 &  2.10 &  2.11 &  2.00 &   1.42  &  0.27  \\
La  & 57  &  1.09  &  0.92  &  0.82 &   0.77  &  0.80  &  0.95  &  1.11  &  1.29  & 1.68  &  1.90  &  2.06 &  2.11 &  2.10 &  1.97 &   1.29  &  0.21  \\
Ce  & 58  &  1.21  &  1.03  &  0.93 &   0.88  &  0.93  &  1.08  &  1.25  &  1.48  & 1.88  &  2.06  &  2.20 &  2.23 &  2.21 &  2.02 &   1.24  &  0.18  \\
Pr  & 59  &  1.00  &  0.83  &  0.74 &   0.70  &  0.74  &  0.89  &  1.06  &  1.29  & 1.68  &  1.87  &  2.01 &  2.03 &  2.02 &  1.83 &   1.05  &  0.11  \\
Nd  & 60  &  1.09  &  0.91  &  0.82 &   0.77  &  0.82  &  0.97  &  1.15  &  1.38  & 1.77  &  1.96  &  2.08 &  2.10 &  2.09 &  1.87 &   1.05  &  0.11  \\
Sm  & 62  &  0.94  &  0.76  &  0.67 &   0.63  &  0.67  &  0.81  &  0.98  &  1.21  & 1.59  &  1.77  &  1.88 &  1.90 &  1.88 &  1.64 &   0.82  &  0.06  \\
Eu  & 63  &  0.38  &  0.28  &  0.22 &   0.20  &  0.22  &  0.30  &  0.41  &  0.58  & 0.92  &  1.07  &  1.18 &  1.20 &  1.18 &  0.95 &   0.30  &  0.00  \\
Gd  & 64  &  0.69  &  0.54  &  0.46 &   0.43  &  0.46  &  0.58  &  0.73  &  0.95  & 1.32  &  1.49  &  1.60 &  1.61 &  1.59 &  1.35 &   0.57  &  0.02  \\
Tb  & 65  &  0.50  &  0.37  &  0.31 &   0.28  &  0.31  &  0.41  &  0.54  &  0.74  & 1.09  &  1.25  &  1.36 &  1.37 &  1.34 &  1.11 &   0.39  &  0.00  \\
Dy  & 66  &  0.68  &  0.53  &  0.45 &   0.42  &  0.45  &  0.57  &  0.72  &  0.94  & 1.31  &  1.47  &  1.57 &  1.58 &  1.56 &  1.31 &   0.53  &  0.02  \\
Ho  & 67  &  0.48  &  0.36  &  0.30 &   0.27  &  0.30  &  0.40  &  0.52  &  0.72  & 1.07  &  1.22  &  1.32 &  1.33 &  1.31 &  1.06 &   0.36  &  0.00  \\
Er  & 68  &  0.77  &  0.62  &  0.54 &   0.50  &  0.54  &  0.67  &  0.82  &  1.05  & 1.42  &  1.58  &  1.68 &  1.69 &  1.66 &  1.41 &   0.61  &  0.03  \\
Tm  & 69  &  0.67  &  0.52  &  0.45 &   0.41  &  0.45  &  0.56  &  0.71  &  0.93  & 1.29  &  1.44  &  1.54 &  1.54 &  1.51 &  1.26 &   0.49  &  0.01  \\
Yb  & 70  &  1.08  &  0.90  &  0.80 &   0.76  &  0.80  &  0.94  &  1.11  &  1.35  & 1.73  &  1.88  &  1.98 &  1.99 &  1.97 &  1.71 &   0.85  &  0.08  \\
Lu  & 71  &  0.83  &  0.66  &  0.58 &   0.53  &  0.57  &  0.69  &  0.84  &  1.07  & 1.43  &  1.59  &  1.69 &  1.70 &  1.68 &  1.42 &   0.60  &  0.03  \\
Hf  & 72  &  1.27  &  1.08  &  0.97 &   0.92  &  0.95  &  1.10  &  1.27  &  1.52  & 1.91  &  2.07  &  2.18 &  2.18 &  2.16 &  1.89 &   1.00  &  0.12  \\
Ta  & 73  &  1.15  &  0.96  &  0.86 &   0.80  &  0.84  &  0.98  &  1.16  &  1.41  & 1.79  &  1.95  &  2.06 &  2.07 &  2.04 &  1.76 &   0.89  &  0.08  \\
W   & 74  &  1.27  &  1.08  &  0.97 &   0.91  &  0.95  &  1.10  &  1.29  &  1.55  & 1.94  &  2.10  &  2.20 &  2.21 &  2.18 &  1.89 &   1.00  &  0.11  \\
Re  & 75  &  0.78  &  0.61  &  0.52 &   0.48  &  0.52  &  0.64  &  0.87  &  1.13  & 1.43  &  1.56  &  1.64 &  1.62 &  1.58 &  1.29 &   0.51  &  0.01  \\
Os  & 76  &  0.64  &  0.49  &  0.41 &   0.38  &  0.42  &  0.53  &  0.70  &  0.93  & 1.29  &  1.43  &  1.53 &  1.53 &  1.50 &  1.21 &   0.44  &  0.01  \\
Ir  & 77  &  0.16  &  0.10  &  0.07 &   0.07  &  0.07  &  0.11  &  0.17  &  0.29  & 0.53  &  0.64  &  0.73 &  0.73 &  0.70 &  0.47 &   0.07  &  0.00  \\
Pt  & 78  &  0.46  &  0.34  &  0.28 &   0.26  &  0.28  &  0.38  &  0.51  &  0.72  & 1.06  &  1.20  &  1.29 &  1.29 &  1.26 &  0.97 &   0.28  &  0.00  \\
Au  & 79  &  0.44  &  0.32  &  0.27 &   0.25  &  0.27  &  0.36  &  0.49  &  0.70  & 1.04  &  1.17  &  1.26 &  1.26 &  1.23 &  0.93 &   0.26  &  0.00  \\
Hg  & 80  &  1.31  &  1.12  &  1.03 &   0.98  &  1.03  &  1.19  &  1.39  &  1.66  & 2.04  &  2.18  &  2.27 &  2.27 &  2.24 &  1.90 &   1.00  &  0.14  \\
Tl  & 81  &  1.35  &  1.16  &  1.06 &   1.04  &  1.10  &  1.26  &  1.57  &  1.86  & 2.11  &  2.23  &  2.29 &  2.25 &  2.19 &  1.81 &   0.95  &  0.10  \\
Pb  & 82  &  3.27  &  3.24  &  3.22 &   3.21  &  3.20  &  3.18  &  3.16  &  3.13  & 3.07  &  2.99  &  2.81 &  2.62 &  2.40 &  1.77 &   0.89  &  0.03  \\
Bi  & 83  &  3.17  &  3.12  &  3.09 &   3.05  &  3.02  &  2.98  &  2.93  &  2.88  & 2.77  &  2.64  &  2.39 &  2.11 &  1.76 &  1.06 &   0.28  &  0.00  \\
 \hline    
\end{tabular}}
\end{table*}

%% file: tableB2.tex
\begin{table*}
\caption{The same as Table~\ref{datatablebab9ltHTm1p3z5m5a}, but for
$M^{\rm AGB}_{\rm ini}$ = 1.4 $M_{\odot}$ (n10).}
 \label{datatablebab9ltHTm1p4z5m5a}
\centering
\resizebox{15cm}{!}{\begin{tabular}{cccccccccccccccccc} 
\hline
\\
El &  Z  &  ST*2  & ST*1.3 &   ST   & ST/1.5 &  ST/2  &  ST/3 & ST/4.5&  ST/6  &  ST/9 & ST/12 &  ST/18 & ST/24 & ST/30 & ST/45 & ST/75 & ST/150  \\
\hline
He &  4  &  0.16  &  0.16  &  0.16  &  0.16  &  0.16  &  0.16 &  0.16 &  0.15  &  0.16 &  0.16 &   0.16 &  0.16 &  0.16 &  0.16 &  0.16 &  0.15   \\
C  &  6  &  3.90  &  3.90  &  3.90  &  3.90  &  3.90  &  3.90 &  3.90 &  3.90  &  3.89 &  3.89 &   3.89 &  3.89 &  3.89 &  3.89 &  3.89 &  3.89   \\
N  &  7  &  1.36  &  1.36  &  1.36  &  1.36  &  1.36  &  1.35 &  1.35 &  1.35  &  1.35 &  1.35 &   1.35 &  1.35 &  1.35 &  1.35 &  1.35 &  1.35   \\
O  &  8  &  1.95  &  1.91  &  1.88  &  1.84  &  1.81  &  1.78 &  1.75 &  1.74  &  1.72 &  1.71 &   1.70 &  1.69 &  1.68 &  1.68 &  1.67 &  1.66   \\
F  &  9  &  4.00  &  3.94  &  3.90  &  3.83  &  3.79  &  3.71 &  3.64 &  3.59  &  3.52 &  3.46 &   3.39 &  3.33 &  3.29 &  3.21 &  3.12 &  2.99   \\
Ne & 10  &  2.36  &  2.35  &  2.35  &  2.34  &  2.34  &  2.34 &  2.33 &  2.33  &  2.33 &  2.33 &   2.33 &  2.33 &  2.33 &  2.33 &  2.33 &  2.32   \\
Na & 11  &  2.03  &  2.01  &  1.99  &  1.98  &  1.98  &  1.97 &  1.96 &  1.96  &  1.96 &  1.95 &   1.95 &  1.94 &  1.94 &  1.93 &  1.92 &  1.91   \\
Mg & 12  &  1.45  &  1.41  &  1.37  &  1.32  &  1.29  &  1.24 &  1.20 &  1.17  &  1.14 &  1.12 &   1.11 &  1.10 &  1.09 &  1.08 &  1.07 &  1.07   \\
Al & 13  &  0.51  &  0.44  &  0.38  &  0.31  &  0.27  &  0.22 &  0.18 &  0.17  &  0.15 &  0.14 &   0.13 &  0.13 &  0.13 &  0.12 &  0.12 &  0.12   \\
Si & 14  &  0.51  &  0.45  &  0.42  &  0.38  &  0.36  &  0.34 &  0.33 &  0.32  &  0.32 &  0.32 &   0.31 &  0.31 &  0.31 &  0.31 &  0.31 &  0.31   \\
P  & 15  &  1.70  &  1.47  &  1.29  &  1.05  &  0.90  &  0.72 &  0.60 &  0.55  &  0.50 &  0.47 &   0.43 &  0.41 &  0.39 &  0.35 &  0.31 &  0.26   \\
S  & 16  &  0.43  &  0.38  &  0.35  &  0.32  &  0.32  &  0.31 &  0.30 &  0.30  &  0.30 &  0.30 &   0.30 &  0.30 &  0.30 &  0.30 &  0.30 &  0.30   \\
Cl & 17  &  0.25  &  0.19  &  0.16  &  0.14  &  0.14  &  0.13 &  0.13 &  0.13  &  0.14 &  0.14 &   0.15 &  0.16 &  0.16 &  0.16 &  0.16 &  0.14   \\
Ar & 18  &  0.27  &  0.27  &  0.26  &  0.26  &  0.26  &  0.26 &  0.26 &  0.26  &  0.26 &  0.26 &   0.26 &  0.26 &  0.26 &  0.26 &  0.26 &  0.26   \\
K  & 19  &  0.41  &  0.40  &  0.39  &  0.39  &  0.38  &  0.38 &  0.38 &  0.38  &  0.38 &  0.38 &   0.38 &  0.38 &  0.38 &  0.38 &  0.38 &  0.37   \\
Ca & 20  &  0.29  &  0.28  &  0.28  &  0.28  &  0.28  &  0.28 &  0.28 &  0.28  &  0.28 &  0.28 &   0.28 &  0.28 &  0.28 &  0.28 &  0.28 &  0.28   \\
Sc & 21  &  0.81  &  0.73  &  0.69  &  0.65  &  0.63  &  0.62 &  0.62 &  0.62  &  0.63 &  0.64 &   0.65 &  0.65 &  0.65 &  0.64 &  0.61 &  0.56   \\
Ti & 22  &  0.42  &  0.36  &  0.33  &  0.32  &  0.32  &  0.32 &  0.33 &  0.34  &  0.34 &  0.34 &   0.34 &  0.34 &  0.33 &  0.32 &  0.30 &  0.28   \\
V  & 23  &  0.12  &  0.06  &  0.04  &  0.03  &  0.03  &  0.03 &  0.04 &  0.05  &  0.05 &  0.05 &   0.04 &  0.03 &  0.03 &  0.01 &  0.00 &  0.00   \\
Cr & 24  & -0.20  & -0.20  & -0.20  & -0.20  & -0.20  & -0.20 & -0.20 & -0.20  & -0.20 & -0.20 &  -0.20 & -0.20 & -0.20 & -0.20 & -0.20 & -0.20   \\
Mn & 25  & -0.40  & -0.40  & -0.40  & -0.40  & -0.40  & -0.40 & -0.40 & -0.40  & -0.40 & -0.40 &  -0.40 & -0.40 & -0.40 & -0.40 & -0.40 & -0.40   \\
Fe & 26  &  0.00  &  0.00  &  0.00  &  0.00  &  0.00  &  0.00 &  0.00 &  0.00  &  0.00 &  0.00 &   0.00 &  0.00 &  0.00 &  0.00 &  0.00 &  0.00   \\
Co & 27  &  0.79  &  0.79  &  0.78  &  0.78  &  0.77  &  0.77 &  0.76 &  0.76  &  0.75 &  0.75 &   0.74 &  0.74 &  0.73 &  0.73 &  0.73 &  0.73   \\
Ni & 28  &  0.23  &  0.21  &  0.21  &  0.19  &  0.19  &  0.18 &  0.17 &  0.16  &  0.16 &  0.15 &   0.15 &  0.15 &  0.15 &  0.15 &  0.15 &  0.15   \\
Cu & 29  &  1.24  &  1.22  &  1.20  &  1.17  &  1.16  &  1.13 &  1.10 &  1.08  &  1.06 &  1.05 &   1.04 &  1.03 &  1.04 &  1.05 &  1.06 &  1.06   \\
Zn & 30  &  0.68  &  0.65  &  0.62  &  0.58  &  0.56  &  0.52 &  0.48 &  0.46  &  0.44 &  0.44 &   0.45 &  0.47 &  0.49 &  0.52 &  0.54 &  0.50   \\
Ga & 31  &  1.25  &  1.19  &  1.15  &  1.09  &  1.05  &  0.99 &  0.94 &  0.91  &  0.89 &  0.89 &   0.93 &  0.98 &  1.01 &  1.07 &  1.08 &  0.99   \\
Ge & 32  &  1.26  &  1.19  &  1.14  &  1.07  &  1.03  &  0.96 &  0.91 &  0.88  &  0.87 &  0.89 &   0.96 &  1.02 &  1.07 &  1.13 &  1.12 &  0.98   \\
As & 33  &  1.10  &  1.03  &  0.97  &  0.90  &  0.85  &  0.78 &  0.73 &  0.70  &  0.70 &  0.74 &   0.84 &  0.91 &  0.96 &  1.02 &  0.99 &  0.82   \\
Se & 34  &  1.24  &  1.15  &  1.09  &  1.01  &  0.95  &  0.88 &  0.82 &  0.80  &  0.83 &  0.90 &   1.04 &  1.12 &  1.17 &  1.22 &  1.16 &  0.93   \\
Br & 35  &  1.14  &  1.05  &  0.99  &  0.90  &  0.85  &  0.77 &  0.72 &  0.70  &  0.74 &  0.83 &   0.96 &  1.05 &  1.10 &  1.14 &  1.08 &  0.83   \\
Kr & 36  &  1.39  &  1.26  &  1.18  &  1.07  &  1.01  &  0.94 &  0.91 &  0.94  &  1.07 &  1.20 &   1.35 &  1.43 &  1.47 &  1.47 &  1.34 &  0.96   \\
Rb & 37  &  1.64  &  1.51  &  1.41  &  1.29  &  1.23  &  1.16 &  1.14 &  1.18  &  1.35 &  1.50 &   1.65 &  1.73 &  1.77 &  1.75 &  1.60 &  1.15   \\
Sr & 38  &  1.41  &  1.25  &  1.15  &  1.06  &  1.02  &  1.04 &  1.20 &  1.39  &  1.63 &  1.75 &   1.87 &  1.91 &  1.91 &  1.84 &  1.65 &  0.98   \\
Y  & 39  &  1.44  &  1.25  &  1.14  &  1.06  &  1.05  &  1.13 &  1.36 &  1.58  &  1.83 &  1.95 &   2.05 &  2.07 &  2.05 &  1.96 &  1.73 &  0.91   \\
Zr & 40  &  1.40  &  1.18  &  1.07  &  1.02  &  1.04  &  1.18 &  1.49 &  1.71  &  1.93 &  2.03 &   2.11 &  2.10 &  2.06 &  1.93 &  1.67 &  0.73   \\
Nb & 41  &  1.42  &  1.20  &  1.09  &  1.04  &  1.06  &  1.21 &  1.52 &  1.75  &  1.97 &  2.07 &   2.15 &  2.14 &  2.10 &  1.97 &  1.70 &  0.73   \\
Mo & 42  &  1.26  &  1.03  &  0.93  &  0.89  &  0.93  &  1.11 &  1.44 &  1.66  &  1.85 &  1.95 &   2.01 &  1.98 &  1.93 &  1.78 &  1.50 &  0.50   \\
Ru & 44  &  1.01  &  0.79  &  0.69  &  0.66  &  0.69  &  0.86 &  1.18 &  1.39  &  1.58 &  1.68 &   1.72 &  1.69 &  1.64 &  1.49 &  1.22 &  0.29   \\
Rh & 45  &  0.67  &  0.48  &  0.40  &  0.37  &  0.39  &  0.54 &  0.83 &  1.02  &  1.20 &  1.29 &   1.34 &  1.31 &  1.27 &  1.13 &  0.88 &  0.12   \\
Pd & 46  &  1.03  &  0.80  &  0.70  &  0.67  &  0.70  &  0.88 &  1.21 &  1.42  &  1.60 &  1.71 &   1.77 &  1.75 &  1.71 &  1.58 &  1.32 &  0.33   \\
Ag & 47  &  0.68  &  0.48  &  0.40  &  0.38  &  0.41  &  0.56 &  0.85 &  1.04  &  1.23 &  1.33 &   1.39 &  1.38 &  1.35 &  1.23 &  0.98 &  0.14   \\
Cd & 48  &  1.08  &  0.87  &  0.78  &  0.76  &  0.81  &  1.01 &  1.35 &  1.56  &  1.75 &  1.85 &   1.91 &  1.90 &  1.87 &  1.74 &  1.46 &  0.40   \\
In & 49  &  0.89  &  0.69  &  0.62  &  0.61  &  0.66  &  0.85 &  1.18 &  1.38  &  1.57 &  1.67 &   1.73 &  1.71 &  1.68 &  1.54 &  1.25 &  0.27   \\
Sn & 50  &  1.30  &  1.07  &  0.98  &  0.98  &  1.03  &  1.28 &  1.64 &  1.83  &  1.99 &  2.08 &   2.10 &  2.06 &  2.01 &  1.85 &  1.52 &  0.42   \\
Sb & 51  &  1.00  &  0.79  &  0.70  &  0.70  &  0.75  &  1.00 &  1.35 &  1.53  &  1.68 &  1.76 &   1.76 &  1.71 &  1.66 &  1.49 &  1.15 &  0.19   \\
Te & 52  &  0.88  &  0.67  &  0.59  &  0.59  &  0.64  &  0.88 &  1.24 &  1.40  &  1.55 &  1.63 &   1.63 &  1.58 &  1.52 &  1.36 &  1.02 &  0.14   \\
I  & 53  &  0.47  &  0.31  &  0.26  &  0.26  &  0.29  &  0.48 &  0.78 &  0.93  &  1.07 &  1.14 &   1.13 &  1.09 &  1.03 &  0.87 &  0.57 &  0.00   \\
Xe & 54  &  0.97  &  0.76  &  0.68  &  0.68  &  0.73  &  1.00 &  1.35 &  1.51  &  1.66 &  1.73 &   1.71 &  1.66 &  1.60 &  1.42 &  1.06 &  0.15   \\
Cs & 55  &  0.80  &  0.60  &  0.53  &  0.53  &  0.58  &  0.83 &  1.18 &  1.33  &  1.49 &  1.56 &   1.55 &  1.51 &  1.45 &  1.29 &  0.93 &  0.09   \\
Ba & 56  &  1.82  &  1.59  &  1.51  &  1.54  &  1.66  &  2.04 &  2.32 &  2.44  &  2.57 &  2.59 &   2.54 &  2.47 &  2.39 &  2.18 &  1.63 &  0.44   \\
La & 57  &  1.83  &  1.60  &  1.52  &  1.55  &  1.69  &  2.09 &  2.34 &  2.45  &  2.58 &  2.58 &   2.52 &  2.44 &  2.36 &  2.14 &  1.52 &  0.34   \\
Ce & 58  &  1.95  &  1.73  &  1.65  &  1.70  &  1.88  &  2.28 &  2.49 &  2.60  &  2.70 &  2.68 &   2.61 &  2.52 &  2.43 &  2.18 &  1.48 &  0.30   \\
Pr & 59  &  1.72  &  1.51  &  1.44  &  1.49  &  1.66  &  2.06 &  2.27 &  2.38  &  2.48 &  2.47 &   2.40 &  2.31 &  2.23 &  1.98 &  1.28 &  0.19   \\
Nd & 60  &  1.81  &  1.59  &  1.52  &  1.58  &  1.76  &  2.16 &  2.36 &  2.47  &  2.56 &  2.54 &   2.46 &  2.37 &  2.29 &  2.03 &  1.29 &  0.19   \\
Sm & 62  &  1.66  &  1.44  &  1.37  &  1.42  &  1.61  &  2.01 &  2.20 &  2.30  &  2.38 &  2.36 &   2.28 &  2.17 &  2.09 &  1.81 &  1.07 &  0.10   \\
Eu & 63  &  0.98  &  0.78  &  0.71  &  0.76  &  0.93  &  1.30 &  1.49 &  1.59  &  1.67 &  1.65 &   1.56 &  1.46 &  1.38 &  1.11 &  0.46 &  0.00   \\
Gd & 64  &  1.38  &  1.17  &  1.10  &  1.16  &  1.34  &  1.73 &  1.92 &  2.03  &  2.10 &  2.08 &   1.99 &  1.89 &  1.80 &  1.52 &  0.80 &  0.03   \\
Tb & 65  &  1.15  &  0.94  &  0.87  &  0.93  &  1.10  &  1.49 &  1.67 &  1.78  &  1.86 &  1.83 &   1.74 &  1.64 &  1.55 &  1.27 &  0.59 &  0.00   \\
Dy & 66  &  1.37  &  1.16  &  1.09  &  1.15  &  1.33  &  1.72 &  1.91 &  2.01  &  2.09 &  2.06 &   1.97 &  1.86 &  1.77 &  1.48 &  0.77 &  0.02   \\
Ho & 67  &  1.13  &  0.93  &  0.86  &  0.92  &  1.10  &  1.48 &  1.66 &  1.77  &  1.84 &  1.81 &   1.72 &  1.61 &  1.52 &  1.24 &  0.56 &  0.00   \\
Er & 68  &  1.50  &  1.28  &  1.21  &  1.27  &  1.46  &  1.86 &  2.04 &  2.14  &  2.21 &  2.18 &   2.09 &  1.98 &  1.88 &  1.59 &  0.86 &  0.05   \\
Tm & 69  &  1.36  &  1.15  &  1.08  &  1.14  &  1.32  &  1.72 &  1.89 &  2.00  &  2.07 &  2.04 &   1.94 &  1.83 &  1.73 &  1.44 &  0.73 &  0.02   \\
Yb & 70  &  1.83  &  1.61  &  1.54  &  1.60  &  1.79  &  2.19 &  2.37 &  2.47  &  2.54 &  2.50 &   2.40 &  2.29 &  2.19 &  1.90 &  1.13 &  0.14   \\
Lu & 71  &  1.55  &  1.33  &  1.26  &  1.32  &  1.51  &  1.90 &  2.08 &  2.18  &  2.24 &  2.20 &   2.10 &  1.99 &  1.90 &  1.60 &  0.86 &  0.05   \\
Hf & 72  &  2.02  &  1.79  &  1.71  &  1.77  &  1.97  &  2.37 &  2.54 &  2.65  &  2.71 &  2.67 &   2.57 &  2.46 &  2.37 &  2.07 &  1.28 &  0.21   \\
Ta & 73  &  1.89  &  1.67  &  1.59  &  1.65  &  1.85  &  2.24 &  2.41 &  2.52  &  2.58 &  2.54 &   2.44 &  2.34 &  2.25 &  1.94 &  1.16 &  0.16   \\
W  & 74  &  2.01  &  1.79  &  1.71  &  1.77  &  1.97  &  2.37 &  2.54 &  2.65  &  2.70 &  2.67 &   2.58 &  2.47 &  2.38 &  2.07 &  1.28 &  0.22   \\
Re & 75  &  1.44  &  1.22  &  1.15  &  1.25  &  1.48  &  1.83 &  1.98 &  2.08  &  2.11 &  2.06 &   1.96 &  1.84 &  1.75 &  1.44 &  0.71 &  0.02   \\
Os & 76  &  1.32  &  1.10  &  1.03  &  1.10  &  1.31  &  1.68 &  1.85 &  1.96  &  2.01 &  1.98 &   1.88 &  1.77 &  1.69 &  1.38 &  0.64 &  0.01   \\
Ir & 77  &  0.55  &  0.39  &  0.35  &  0.39  &  0.54  &  0.86 &  1.00 &  1.11  &  1.16 &  1.13 &   1.04 &  0.94 &  0.86 &  0.59 &  0.12 &  0.00   \\
Pt & 78  &  1.08  &  0.88  &  0.82  &  0.88  &  1.08  &  1.45 &  1.61 &  1.72  &  1.77 &  1.74 &   1.64 &  1.54 &  1.45 &  1.13 &  0.44 &  0.00   \\
Au & 79  &  1.05  &  0.85  &  0.79  &  0.86  &  1.05  &  1.42 &  1.58 &  1.69  &  1.74 &  1.70 &   1.61 &  1.50 &  1.41 &  1.09 &  0.41 &  0.00   \\
Hg & 80  &  2.04  &  1.83  &  1.76  &  1.83  &  2.05  &  2.44 &  2.59 &  2.71  &  2.75 &  2.71 &   2.62 &  2.51 &  2.42 &  2.07 &  1.26 &  0.25   \\
Tl & 81  &  2.05  &  1.84  &  1.79  &  1.94  &  2.20  &  2.51 &  2.64 &  2.74  &  2.72 &  2.67 &   2.55 &  2.43 &  2.34 &  1.96 &  1.17 &  0.23   \\
Pb & 82  &  3.86  &  3.80  &  3.78  &  3.75  &  3.74  &  3.69 &  3.62 &  3.54  &  3.40 &  3.26 &   3.02 &  2.81 &  2.58 &  2.01 &  1.12 &  0.12   \\
Bi & 83  &  3.75  &  3.69  &  3.65  &  3.61  &  3.58  &  3.51 &  3.41 &  3.31  &  3.13 &  2.96 &   2.67 &  2.39 &  2.09 &  1.45 &  0.56 &  0.03   \\
\hline
\end{tabular}}
\end{table*}

%% file: tableB3.tex
\begin{table*}
\caption{The same as Table~\ref{datatablebab9ltHTm1p3z5m5a}, but for
$M^{\rm AGB}_{\rm ini}$ = 1.5 $M_{\odot}$ (n20).}
 \label{datatablebab9ltHTm1p5z5m5a}
\centering
\resizebox{15cm}{!}{\begin{tabular}{cccccccccccccccccc} 
\hline
\\
El &  Z &  ST*2  &ST*1.3 &   ST  & ST/1.5 & ST/2  & ST/3 & ST/4.5& ST/6 & ST/9 & ST/12 &  ST/18 & ST/24  & ST/30  & ST/45 & ST/75 & ST/150 \\
\hline
He &  4 &  0.26  & 0.26  &  0.26 &  0.26  & 0.26  & 0.26 &  0.26 & 0.26 & 0.26 &  0.26 &   0.26 &  0.26  &  0.26  &  0.26 &  0.26 &  0.26  \\
C  &  6 &  4.13  & 4.13  &  4.13 &  4.13  & 4.13  & 4.13 &  4.13 & 4.13 & 4.13 &  4.13 &   4.13 &  4.13  &  4.13  &  4.13 &  4.13 &  4.12  \\
N  &  7 &  2.17  & 2.16  &  2.16 &  2.16  & 2.16  & 2.16 &  2.16 & 2.16 & 2.16 &  2.16 &   2.16 &  2.16  &  2.16  &  2.15 &  2.15 &  2.15  \\
O  &  8 &  2.18  & 2.12  &  2.09 &  2.04  & 2.02  & 1.98 &  1.96 & 1.94 & 1.93 &  1.92 &   1.90 &  1.90  &  1.89  &  1.89 &  1.88 &  1.88  \\
F  &  9 &  4.15  & 4.07  &  4.02 &  3.94  & 3.89  & 3.82 &  3.75 & 3.71 & 3.64 &  3.60 &   3.54 &  3.51  &  3.48  &  3.43 &  3.38 &  3.31  \\
Ne & 10 &  2.88  & 2.88  &  2.87 &  2.87  & 2.87  & 2.87 &  2.87 & 2.87 & 2.87 &  2.87 &   2.87 &  2.87  &  2.87  &  2.86 &  2.86 &  2.86  \\
Na & 11 &  2.57  & 2.57  &  2.57 &  2.57  & 2.57  & 2.58 &  2.58 & 2.58 & 2.58 &  2.58 &   2.58 &  2.57  &  2.57  &  2.56 &  2.56 &  2.55  \\
Mg & 12 &  2.07  & 2.02  &  1.98 &  1.93  & 1.90  & 1.85 &  1.81 & 1.79 & 1.77 &  1.76 &   1.74 &  1.74  &  1.73  &  1.73 &  1.72 &  1.72  \\
Al & 13 &  0.98  & 0.88  &  0.81 &  0.72  & 0.67  & 0.62 &  0.58 & 0.56 & 0.54 &  0.53 &   0.51 &  0.51  &  0.50  &  0.49 &  0.48 &  0.48  \\
Si & 14 &  0.77  & 0.66  &  0.59 &  0.50  & 0.46  & 0.42 &  0.39 & 0.38 & 0.37 &  0.36 &   0.36 &  0.36  &  0.35  &  0.35 &  0.35 &  0.35  \\
P  & 15 &  2.12  & 1.87  &  1.66 &  1.39  & 1.22  & 1.00 &  0.86 & 0.79 & 0.73 &  0.69 &   0.65 &  0.63  &  0.61  &  0.58 &  0.54 &  0.50  \\
S  & 16 &  0.59  & 0.47  &  0.41 &  0.36  & 0.34  & 0.32 &  0.32 & 0.32 & 0.31 &  0.31 &   0.31 &  0.31  &  0.31  &  0.31 &  0.31 &  0.31  \\
Cl & 17 &  0.40  & 0.30  &  0.26 &  0.23  & 0.22  & 0.22 &  0.23 & 0.24 & 0.25 &  0.26 &   0.26 &  0.27  &  0.27  &  0.27 &  0.26 &  0.23  \\
Ar & 18 &  0.27  & 0.26  &  0.26 &  0.26  & 0.26  & 0.26 &  0.26 & 0.26 & 0.26 &  0.26 &   0.26 &  0.26  &  0.26  &  0.26 &  0.26 &  0.26  \\
K  & 19 &  0.49  & 0.47  &  0.46 &  0.45  & 0.45  & 0.45 &  0.45 & 0.46 & 0.46 &  0.46 &   0.46 &  0.46  &  0.46  &  0.46 &  0.46 &  0.46  \\
Ca & 20 &  0.28  & 0.28  &  0.28 &  0.27  & 0.27  & 0.27 &  0.28 & 0.28 & 0.28 &  0.28 &   0.28 &  0.28  &  0.28  &  0.28 &  0.27 &  0.27  \\
Sc & 21 &  1.04  & 0.94  &  0.89 &  0.88  & 0.88  & 0.89 &  0.91 & 0.92 & 0.93 &  0.93 &   0.93 &  0.93  &  0.92  &  0.91 &  0.88 &  0.85  \\
Ti & 22 &  0.54  & 0.44  &  0.41 &  0.41  & 0.41  & 0.43 &  0.43 & 0.43 & 0.43 &  0.42 &   0.41 &  0.40  &  0.38  &  0.37 &  0.35 &  0.33  \\
V  & 23 &  0.21  & 0.12  &  0.09 &  0.09  & 0.09  & 0.10 &  0.10 & 0.10 & 0.09 &  0.08 &   0.06 &  0.05  &  0.04  &  0.02 &  0.01 &  0.00  \\
Cr & 24 & -0.20  &-0.20  & -0.20 & -0.20  &-0.20  &-0.20 & -0.20 &-0.20 &-0.20 & -0.20 &  -0.20 & -0.20  & -0.20  & -0.20 & -0.20 & -0.20  \\
Mn & 25 & -0.40  &-0.40  & -0.40 & -0.40  &-0.40  &-0.40 & -0.40 &-0.40 &-0.40 & -0.40 &  -0.40 & -0.40  & -0.40  & -0.40 & -0.40 & -0.40  \\
Fe & 26 &  0.00  & 0.00  &  0.00 &  0.00  & 0.00  & 0.00 &  0.00 & 0.00 & 0.00 &  0.00 &   0.00 &  0.00  &  0.00  &  0.00 &  0.00 &  0.00  \\
Co & 27 &  1.01  & 1.01  &  1.01 &  1.01  & 1.01  & 1.00 &  1.00 & 1.00 & 1.00 &  1.00 &   1.00 &  1.00  &  1.00  &  1.00 &  1.00 &  1.00  \\
Ni & 28 &  0.33  & 0.31  &  0.31 &  0.30  & 0.30  & 0.30 &  0.29 & 0.29 & 0.29 &  0.29 &   0.29 &  0.30  &  0.30  &  0.30 &  0.31 &  0.31  \\
Cu & 29 &  1.44  & 1.42  &  1.41 &  1.40  & 1.39  & 1.38 &  1.38 & 1.37 & 1.37 &  1.38 &   1.39 &  1.39  &  1.40  &  1.42 &  1.43 &  1.43  \\
Zn & 30 &  0.82  & 0.80  &  0.78 &  0.76  & 0.74  & 0.73 &  0.72 & 0.73 & 0.75 &  0.77 &   0.81 &  0.83  &  0.85  &  0.86 &  0.86 &  0.82  \\
Ga & 31 &  1.40  & 1.36  &  1.33 &  1.30  & 1.28  & 1.27 &  1.26 & 1.28 & 1.32 &  1.36 &   1.42 &  1.46  &  1.47  &  1.48 &  1.46 &  1.40  \\
Ge & 32 &  1.41  & 1.35  &  1.32 &  1.29  & 1.27  & 1.25 &  1.25 & 1.28 & 1.36 &  1.41 &   1.48 &  1.52  &  1.53  &  1.53 &  1.49 &  1.40  \\
As & 33 &  1.24  & 1.18  &  1.14 &  1.10  & 1.08  & 1.06 &  1.08 & 1.13 & 1.22 &  1.29 &   1.36 &  1.40  &  1.41  &  1.40 &  1.34 &  1.23  \\
Se & 34 &  1.38  & 1.31  &  1.27 &  1.22  & 1.20  & 1.18 &  1.23 & 1.30 & 1.43 &  1.50 &   1.58 &  1.61  &  1.61  &  1.58 &  1.50 &  1.36  \\
Br & 35 &  1.28  & 1.21  &  1.16 &  1.11  & 1.09  & 1.08 &  1.13 & 1.21 & 1.35 &  1.43 &   1.51 &  1.53  &  1.53  &  1.49 &  1.41 &  1.26  \\
Kr & 36 &  1.53  & 1.43  &  1.37 &  1.33  & 1.32  & 1.36 &  1.49 & 1.62 & 1.76 &  1.84 &   1.90 &  1.89  &  1.86  &  1.79 &  1.66 &  1.45  \\
Rb & 37 &  1.80  & 1.69  &  1.63 &  1.58  & 1.57  & 1.62 &  1.79 & 1.93 & 2.08 &  2.16 &   2.21 &  2.19  &  2.16  &  2.07 &  1.92 &  1.69  \\
Sr & 38 &  1.62  & 1.48  &  1.42 &  1.41  & 1.48  & 1.74 &  1.99 & 2.11 & 2.22 &  2.25 &   2.21 &  2.16  &  2.11  &  2.00 &  1.81 &  1.40  \\
Y  & 39 &  1.69  & 1.52  &  1.46 &  1.49  & 1.61  & 1.93 &  2.19 & 2.30 & 2.40 &  2.41 &   2.35 &  2.28  &  2.21  &  2.08 &  1.85 &  1.32  \\
Zr & 40 &  1.69  & 1.50  &  1.45 &  1.53  & 1.70  & 2.07 &  2.30 & 2.40 & 2.47 &  2.45 &   2.35 &  2.27  &  2.19  &  2.02 &  1.76 &  1.11  \\
Nb & 41 &  1.71  & 1.53  &  1.47 &  1.55  & 1.73  & 2.11 &  2.34 & 2.43 & 2.51 &  2.49 &   2.39 &  2.30  &  2.22  &  2.05 &  1.78 &  1.10  \\
Mo & 42 &  1.57  & 1.38  &  1.33 &  1.44  & 1.64  & 2.02 &  2.22 & 2.31 & 2.37 &  2.33 &   2.21 &  2.12  &  2.04  &  1.85 &  1.55 &  0.81  \\
Ru & 44 &  1.30  & 1.11  &  1.07 &  1.18  & 1.39  & 1.77 &  1.96 & 2.06 & 2.11 &  2.06 &   1.93 &  1.83  &  1.75  &  1.56 &  1.27 &  0.55  \\
Rh & 45 &  0.93  & 0.75  &  0.71 &  0.82  & 1.02  & 1.39 &  1.58 & 1.67 & 1.72 &  1.67 &   1.54 &  1.45  &  1.37  &  1.19 &  0.92 &  0.27  \\
Pd & 46 &  1.31  & 1.12  &  1.08 &  1.20  & 1.42  & 1.79 &  1.98 & 2.07 & 2.12 &  2.07 &   1.96 &  1.87  &  1.80  &  1.63 &  1.35 &  0.56  \\
Ag & 47 &  0.94  & 0.76  &  0.72 &  0.83  & 1.04  & 1.40 &  1.59 & 1.68 & 1.73 &  1.68 &   1.57 &  1.49  &  1.42  &  1.26 &  1.00 &  0.28  \\
Cd & 48 &  1.41  & 1.22  &  1.18 &  1.31  & 1.53  & 1.90 &  2.07 & 2.16 & 2.20 &  2.16 &   2.06 &  1.99  &  1.93  &  1.76 &  1.47 &  0.59  \\
In & 49 &  1.22  & 1.04  &  1.01 &  1.14  & 1.35  & 1.71 &  1.88 & 1.97 & 2.01 &  1.96 &   1.87 &  1.79  &  1.73  &  1.56 &  1.27 &  0.42  \\
Sn & 50 &  1.63  & 1.45  &  1.42 &  1.58  & 1.83  & 2.18 &  2.33 & 2.42 & 2.43 &  2.36 &   2.24 &  2.14  &  2.06  &  1.87 &  1.54 &  0.62  \\
Sb & 51 &  1.32  & 1.14  &  1.11 &  1.29  & 1.54  & 1.89 &  2.02 & 2.11 & 2.10 &  2.02 &   1.89 &  1.79  &  1.71  &  1.51 &  1.17 &  0.32  \\
Te & 52 &  1.19  & 1.01  &  0.99 &  1.17  & 1.42  & 1.76 &  1.89 & 1.97 & 1.96 &  1.88 &   1.75 &  1.65  &  1.57  &  1.37 &  1.03 &  0.22  \\
I  & 53 &  0.73  & 0.57  &  0.55 &  0.71  & 0.95  & 1.27 &  1.39 & 1.47 & 1.45 &  1.37 &   1.24 &  1.15  &  1.07  &  0.88 &  0.57 &  0.01  \\
Xe & 54 &  1.29  & 1.11  &  1.09 &  1.29  & 1.56  & 1.87 &  2.00 & 2.08 & 2.04 &  1.95 &   1.82 &  1.72  &  1.63  &  1.43 &  1.06 &  0.23  \\
Cs & 55 &  1.10  & 0.93  &  0.91 &  1.11  & 1.37  & 1.69 &  1.81 & 1.89 & 1.85 &  1.77 &   1.65 &  1.56  &  1.48  &  1.29 &  0.92 &  0.14  \\
Ba & 56 &  2.17  & 2.00  &  2.03 &  2.37  & 2.59  & 2.78 &  2.90 & 2.92 & 2.82 &  2.73 &   2.60 &  2.49  &  2.39  &  2.17 &  1.63 &  0.56  \\
La & 57 &  2.17  & 2.01  &  2.06 &  2.42  & 2.63  & 2.79 &  2.90 & 2.90 & 2.79 &  2.69 &   2.56 &  2.45  &  2.35  &  2.12 &  1.52 &  0.43  \\
Ce & 58 &  2.30  & 2.15  &  2.23 &  2.62  & 2.80  & 2.92 &  3.03 & 2.99 & 2.86 &  2.77 &   2.64 &  2.52  &  2.42  &  2.16 &  1.47 &  0.37  \\
Pr & 59 &  2.08  & 1.92  &  2.01 &  2.40  & 2.57  & 2.70 &  2.80 & 2.77 & 2.64 &  2.55 &   2.42 &  2.31  &  2.21  &  1.96 &  1.27 &  0.23  \\
Nd & 60 &  2.16  & 2.01  &  2.11 &  2.50  & 2.67  & 2.78 &  2.88 & 2.84 & 2.71 &  2.62 &   2.49 &  2.37  &  2.27  &  2.00 &  1.28 &  0.23  \\
Sm & 62 &  2.01  & 1.86  &  1.96 &  2.35  & 2.51  & 2.63 &  2.72 & 2.67 & 2.54 &  2.44 &   2.30 &  2.17  &  2.07  &  1.79 &  1.06 &  0.13  \\
Eu & 63 &  1.30  & 1.16  &  1.26 &  1.64  & 1.80  & 1.91 &  2.01 & 1.95 & 1.82 &  1.72 &   1.58 &  1.46  &  1.36  &  1.08 &  0.45 &  0.00  \\
Gd & 64 &  1.73  & 1.58  &  1.68 &  2.07  & 2.23  & 2.35 &  2.44 & 2.39 & 2.25 &  2.16 &   2.01 &  1.89  &  1.78  &  1.49 &  0.79 &  0.03  \\
Tb & 65 &  1.49  & 1.34  &  1.44 &  1.83  & 1.98  & 2.10 &  2.19 & 2.14 & 2.00 &  1.91 &   1.76 &  1.64  &  1.53  &  1.25 &  0.57 &  0.00  \\
Dy & 66 &  1.72  & 1.57  &  1.68 &  2.07  & 2.22  & 2.34 &  2.43 & 2.37 & 2.23 &  2.14 &   1.99 &  1.86  &  1.76  &  1.46 &  0.76 &  0.02  \\
Ho & 67 &  1.48  & 1.33  &  1.44 &  1.82  & 1.98  & 2.09 &  2.18 & 2.12 & 1.98 &  1.89 &   1.74 &  1.61  &  1.51  &  1.21 &  0.55 &  0.00  \\
Er & 68 &  1.85  & 1.70  &  1.81 &  2.20  & 2.35  & 2.46 &  2.56 & 2.49 & 2.36 &  2.26 &   2.11 &  1.97  &  1.87  &  1.57 &  0.85 &  0.05  \\
Tm & 69 &  1.71  & 1.56  &  1.67 &  2.06  & 2.21  & 2.32 &  2.41 & 2.35 & 2.21 &  2.11 &   1.96 &  1.83  &  1.72  &  1.42 &  0.72 &  0.01  \\
Yb & 70 &  2.18  & 2.04  &  2.15 &  2.54  & 2.69  & 2.80 &  2.89 & 2.82 & 2.68 &  2.58 &   2.42 &  2.28  &  2.18  &  1.87 &  1.13 &  0.18  \\
Lu & 71 &  1.90  & 1.75  &  1.87 &  2.26  & 2.41  & 2.52 &  2.60 & 2.53 & 2.39 &  2.28 &   2.12 &  1.99  &  1.88  &  1.58 &  0.85 &  0.06  \\
Hf & 72 &  2.36  & 2.22  &  2.33 &  2.73  & 2.87  & 2.99 &  3.07 & 3.00 & 2.85 &  2.75 &   2.59 &  2.46  &  2.36  &  2.05 &  1.28 &  0.27  \\
Ta & 73 &  2.24  & 2.09  &  2.21 &  2.60  & 2.75  & 2.86 &  2.94 & 2.87 & 2.72 &  2.62 &   2.46 &  2.33  &  2.23  &  1.92 &  1.16 &  0.20  \\
W  & 74 &  2.36  & 2.21  &  2.33 &  2.73  & 2.87  & 2.98 &  3.06 & 2.99 & 2.84 &  2.75 &   2.60 &  2.47  &  2.36  &  2.05 &  1.27 &  0.28  \\
Re & 75 &  1.77  & 1.63  &  1.77 &  2.17  & 2.29  & 2.40 &  2.47 & 2.39 & 2.24 &  2.13 &   1.97 &  1.84  &  1.73  &  1.41 &  0.70 &  0.03  \\
Os & 76 &  1.66  & 1.52  &  1.64 &  2.04  & 2.17  & 2.28 &  2.35 & 2.28 & 2.14 &  2.04 &   1.90 &  1.77  &  1.67  &  1.35 &  0.63 &  0.00  \\
Ir & 77 &  0.83  & 0.71  &  0.82 &  1.18  & 1.30  & 1.41 &  1.49 & 1.41 & 1.28 &  1.19 &   1.05 &  0.93  &  0.84  &  0.56 &  0.09 &  0.00  \\
Pt & 78 &  1.42  & 1.28  &  1.41 &  1.80  & 1.93  & 2.04 &  2.11 & 2.03 & 1.89 &  1.80 &   1.65 &  1.53  &  1.43  &  1.11 &  0.43 &  0.00  \\
Au & 79 &  1.39  & 1.25  &  1.38 &  1.77  & 1.89  & 2.00 &  2.07 & 1.99 & 1.86 &  1.76 &   1.62 &  1.49  &  1.39  &  1.07 &  0.40 &  0.00  \\
Hg & 80 &  2.39  & 2.26  &  2.40 &  2.80  & 2.91  & 3.02 &  3.09 & 3.00 & 2.87 &  2.77 &   2.63 &  2.50  &  2.40  &  2.05 &  1.26 &  0.32  \\
Tl & 81 &  2.40  & 2.28  &  2.46 &  2.85  & 2.93  & 3.04 &  3.07 & 2.97 & 2.82 &  2.72 &   2.55 &  2.41  &  2.31  &  1.93 &  1.16 &  0.32  \\
Pb & 82 &  4.16  & 4.10  &  4.07 &  4.02  & 3.98  & 3.88 &  3.72 & 3.60 & 3.41 &  3.26 &   3.00 &  2.78  &  2.56  &  1.98 &  1.12 &  0.21  \\
Bi & 83 &  4.03  & 3.97  &  3.93 &  3.86  & 3.80  & 3.67 &  3.49 & 3.35 & 3.14 &  2.96 &   2.65 &  2.37  &  2.08  &  1.45 &  0.58 &  0.08  \\
\hline
\end{tabular}}
\end{table*}

%% file: tableB4.tex
\begin{table*}
\caption{The same as Table~\ref{datatablebab9ltHTm1p3z5m5a}, but for
$M^{\rm AGB}_{\rm ini}$ = 2.0 $M_{\odot}$ (n26).}
 \label{datatablebab9ltHTm2z5m5a}
\centering
\resizebox{15cm}{!}{\begin{tabular}{cccccccccccccccccc} 
\hline
\\
El &  Z  &  ST*2 &  ST*1.3 &   ST   & ST/1.5 &  ST/2  &  ST/3 & ST/4.5 &  ST/6 & ST/9  & ST/12 &  ST/18 &  ST/24 & ST/30  & ST/45  & ST/75 & ST/150  \\
\hline
He &  4  &  0.25 &   0.25  &  0.24  &  0.24  &  0.24  &  0.24 &  0.24  &  0.24 & 0.25  &  0.25 &   0.24 &   0.24 &  0.24  &  0.24  &  0.24 &  0.24   \\
C  &  6  &  4.07 &   4.07  &  4.07  &  4.06  &  4.06  &  4.06 &  4.06  &  4.06 & 4.06  &  4.06 &   4.06 &   4.06 &  4.06  &  4.06  &  4.05 &  4.05   \\
N  &  7  &  1.93 &   1.93  &  1.93  &  1.92  &  1.92  &  1.92 &  1.92  &  1.92 & 1.92  &  1.92 &   1.91 &   1.91 &  1.91  &  1.91  &  1.91 &  1.91   \\
O  &  8  &  2.18 &   2.13  &  2.09  &  2.04  &  2.01  &  1.96 &  1.93  &  1.91 & 1.88  &  1.87 &   1.85 &   1.85 &  1.84  &  1.83  &  1.82 &  1.81   \\
F  &  9  &  4.27 &   4.22  &  4.18  &  4.11  &  4.07  &  4.01 &  3.94  &  3.89 & 3.82  &  3.76 &   3.68 &   3.62 &  3.58  &  3.49  &  3.38 &  3.17   \\
Ne & 10  &  2.46 &   2.45  &  2.44  &  2.43  &  2.42  &  2.42 &  2.41  &  2.41 & 2.41  &  2.41 &   2.41 &   2.41 &  2.40  &  2.40  &  2.40 &  2.40   \\
Na & 11  &  2.22 &   2.20  &  2.19  &  2.19  &  2.18  &  2.18 &  2.18  &  2.18 & 2.18  &  2.18 &   2.18 &   2.17 &  2.17  &  2.16  &  2.15 &  2.14   \\
Mg & 12  &  1.80 &   1.75  &  1.72  &  1.67  &  1.64  &  1.59 &  1.55  &  1.52 & 1.49  &  1.47 &   1.45 &   1.44 &  1.43  &  1.42  &  1.41 &  1.41   \\
Al & 13  &  0.85 &   0.77  &  0.71  &  0.62  &  0.57  &  0.51 &  0.46  &  0.44 & 0.41  &  0.40 &   0.38 &   0.38 &  0.37  &  0.36  &  0.35 &  0.34   \\
Si & 14  &  0.74 &   0.65  &  0.59  &  0.51  &  0.47  &  0.42 &  0.39  &  0.38 & 0.36  &  0.35 &   0.35 &   0.35 &  0.34  &  0.34  &  0.34 &  0.34   \\
P  & 15  &  2.17 &   1.96  &  1.79  &  1.54  &  1.38  &  1.15 &  0.97  &  0.87 & 0.78  &  0.74 &   0.68 &   0.65 &  0.63  &  0.58  &  0.53 &  0.47   \\
S  & 16  &  0.63 &   0.52  &  0.46  &  0.39  &  0.36  &  0.34 &  0.32  &  0.32 & 0.31  &  0.31 &   0.31 &   0.31 &  0.31  &  0.31  &  0.31 &  0.31   \\
Cl & 17  &  0.49 &   0.37  &  0.30  &  0.24  &  0.21  &  0.20 &  0.19  &  0.19 & 0.20  &  0.21 &   0.22 &   0.23 &  0.24  &  0.25  &  0.25 &  0.22   \\
Ar & 18  &  0.28 &   0.27  &  0.27  &  0.26  &  0.26  &  0.26 &  0.26  &  0.26 & 0.26  &  0.26 &   0.26 &   0.26 &  0.26  &  0.26  &  0.26 &  0.26   \\
K  & 19  &  0.50 &   0.47  &  0.46  &  0.44  &  0.44  &  0.43 &  0.43  &  0.43 & 0.43  &  0.43 &   0.44 &   0.44 &  0.44  &  0.44  &  0.44 &  0.43   \\
Ca & 20  &  0.29 &   0.28  &  0.28  &  0.28  &  0.28  &  0.28 &  0.28  &  0.28 & 0.28  &  0.28 &   0.28 &   0.28 &  0.28  &  0.28  &  0.28 &  0.28   \\
Sc & 21  &  1.19 &   1.04  &  0.96  &  0.89  &  0.86  &  0.85 &  0.85  &  0.86 & 0.88  &  0.89 &   0.91 &   0.91 &  0.91  &  0.90  &  0.87 &  0.82   \\
Ti & 22  &  0.66 &   0.52  &  0.45  &  0.39  &  0.38  &  0.39 &  0.40  &  0.41 & 0.43  &  0.43 &   0.43 &   0.42 &  0.41  &  0.38  &  0.35 &  0.32   \\
V  & 23  &  0.34 &   0.20  &  0.13  &  0.08  &  0.07  &  0.08 &  0.09  &  0.10 & 0.11  &  0.11 &   0.10 &   0.09 &  0.07  &  0.04  &  0.02 &  0.00   \\
Cr & 24  & -0.20 &  -0.20  & -0.20  & -0.20  & -0.20  & -0.20 & -0.20  & -0.20 &-0.20  & -0.20 &  -0.20 &  -0.20 & -0.20  & -0.20  & -0.20 & -0.20   \\
Mn & 25  & -0.40 &  -0.40  & -0.40  & -0.40  & -0.40  & -0.40 & -0.40  & -0.40 &-0.40  & -0.40 &  -0.40 &  -0.40 & -0.40  & -0.40  & -0.40 & -0.40   \\
Fe & 26  &  0.00 &   0.00  &  0.00  &  0.00  &  0.00  &  0.00 &  0.00  &  0.00 & 0.00  &  0.00 &   0.00 &   0.00 &  0.00  &  0.00  &  0.00 &  0.00   \\
Co & 27  &  0.99 &   0.99  &  0.98  &  0.98  &  0.97  &  0.96 &  0.95  &  0.95 & 0.94  &  0.93 &   0.92 &   0.92 &  0.92  &  0.91  &  0.91 &  0.92   \\
Ni & 28  &  0.34 &   0.32  &  0.31  &  0.30  &  0.29  &  0.28 &  0.27  &  0.26 & 0.26  &  0.25 &   0.25 &   0.25 &  0.25  &  0.26  &  0.26 &  0.27   \\
Cu & 29  &  1.44 &   1.41  &  1.39  &  1.37  &  1.35  &  1.33 &  1.30  &  1.29 & 1.27  &  1.26 &   1.26 &   1.26 &  1.27  &  1.29  &  1.32 &  1.33   \\
Zn & 30  &  0.84 &   0.80  &  0.77  &  0.73  &  0.71  &  0.68 &  0.65  &  0.64 & 0.63  &  0.63 &   0.65 &   0.68 &  0.71  &  0.77  &  0.80 &  0.77   \\
Ga & 31  &  1.43 &   1.37  &  1.33  &  1.28  &  1.25  &  1.21 &  1.17  &  1.16 & 1.15  &  1.17 &   1.22 &   1.27 &  1.32  &  1.40  &  1.42 &  1.36   \\
Ge & 32  &  1.46 &   1.39  &  1.34  &  1.28  &  1.25  &  1.20 &  1.17  &  1.16 & 1.17  &  1.20 &   1.28 &   1.35 &  1.41  &  1.49  &  1.50 &  1.39   \\
As & 33  &  1.30 &   1.22  &  1.17  &  1.10  &  1.07  &  1.02 &  0.99  &  0.99 & 1.01  &  1.06 &   1.16 &   1.25 &  1.31  &  1.39  &  1.38 &  1.24   \\
Se & 34  &  1.45 &   1.36  &  1.30  &  1.22  &  1.18  &  1.14 &  1.11  &  1.11 & 1.16  &  1.24 &   1.38 &   1.47 &  1.54  &  1.61  &  1.57 &  1.37   \\
Br & 35  &  1.35 &   1.26  &  1.19  &  1.12  &  1.08  &  1.03 &  1.01  &  1.01 & 1.07  &  1.15 &   1.30 &   1.40 &  1.46  &  1.53  &  1.48 &  1.27   \\
Kr & 36  &  1.66 &   1.52  &  1.43  &  1.34  &  1.30  &  1.26 &  1.27  &  1.32 & 1.48  &  1.61 &   1.77 &   1.86 &  1.90  &  1.90  &  1.80 &  1.50   \\
Rb & 37  &  1.96 &   1.80  &  1.70  &  1.60  &  1.56  &  1.53 &  1.55  &  1.62 & 1.81  &  1.95 &   2.11 &   2.20 &  2.23  &  2.21  &  2.08 &  1.77   \\
Sr & 38  &  1.79 &   1.59  &  1.46  &  1.35  &  1.33  &  1.36 &  1.52  &  1.70 & 1.94  &  2.07 &   2.21 &   2.27 &  2.28  &  2.21  &  2.06 &  1.47   \\
Y  & 39  &  1.89 &   1.66  &  1.51  &  1.40  &  1.38  &  1.46 &  1.69  &  1.90 & 2.14  &  2.27 &   2.40 &   2.43 &  2.42  &  2.32  &  2.11 &  1.43   \\
Zr & 40  &  1.96 &   1.69  &  1.53  &  1.41  &  1.41  &  1.56 &  1.87  &  2.09 & 2.30  &  2.41 &   2.50 &   2.49 &  2.45  &  2.31  &  2.01 &  1.27   \\
Nb & 41  &  1.96 &   1.69  &  1.53  &  1.41  &  1.41  &  1.56 &  1.87  &  2.09 & 2.32  &  2.43 &   2.52 &   2.51 &  2.47  &  2.33  &  2.02 &  1.25   \\
Mo & 42  &  1.84 &   1.56  &  1.39  &  1.28  &  1.28  &  1.47 &  1.81  &  2.02 & 2.22  &  2.32 &   2.39 &   2.36 &  2.30  &  2.15  &  1.77 &  0.96   \\
Ru & 44  &  1.57 &   1.29  &  1.12  &  1.00  &  1.01  &  1.20 &  1.55  &  1.75 & 1.94  &  2.04 &   2.10 &   2.06 &  2.00  &  1.85  &  1.48 &  0.68   \\
Rh & 45  &  1.20 &   0.93  &  0.77  &  0.66  &  0.67  &  0.85 &  1.18  &  1.38 & 1.56  &  1.66 &   1.72 &   1.69 &  1.63  &  1.49  &  1.12 &  0.39   \\
Pd & 46  &  1.61 &   1.32  &  1.15  &  1.03  &  1.05  &  1.25 &  1.60  &  1.81 & 1.99  &  2.10 &   2.16 &   2.14 &  2.08  &  1.96  &  1.57 &  0.72   \\
Ag & 47  &  1.23 &   0.95  &  0.79  &  0.69  &  0.70  &  0.89 &  1.23  &  1.42 & 1.61  &  1.71 &   1.78 &   1.76 &  1.70  &  1.59  &  1.20 &  0.42   \\
Cd & 48  &  1.65 &   1.37  &  1.20  &  1.10  &  1.12  &  1.33 &  1.67  &  1.87 & 2.07  &  2.19 &   2.26 &   2.25 &  2.20  &  2.09  &  1.66 &  0.76   \\
In & 49  &  1.40 &   1.14  &  0.98  &  0.89  &  0.91  &  1.10 &  1.44  &  1.64 & 1.84  &  1.96 &   2.04 &   2.03 &  1.98  &  1.88  &  1.44 &  0.55   \\
Sn & 50  &  1.87 &   1.59  &  1.42  &  1.32  &  1.34  &  1.58 &  1.94  &  2.12 & 2.30  &  2.40 &   2.45 &   2.41 &  2.34  &  2.20  &  1.70 &  0.76   \\
Sb & 51  &  1.59 &   1.31  &  1.15  &  1.05  &  1.07  &  1.32 &  1.69  &  1.86 & 2.03  &  2.12 &   2.14 &   2.08 &  2.01  &  1.85  &  1.32 &  0.46   \\
Te & 52  &  1.45 &   1.17  &  1.01  &  0.91  &  0.94  &  1.20 &  1.56  &  1.72 & 1.89  &  1.98 &   2.00 &   1.93 &  1.86  &  1.71  &  1.16 &  0.34   \\
I  & 53  &  0.97 &   0.71  &  0.57  &  0.49  &  0.52  &  0.74 &  1.07  &  1.23 & 1.39  &  1.48 &   1.49 &   1.43 &  1.35  &  1.20  &  0.69 &  0.08   \\
Xe & 54  &  1.57 &   1.28  &  1.12  &  1.02  &  1.06  &  1.35 &  1.71  &  1.86 & 2.02  &  2.09 &   2.09 &   2.01 &  1.94  &  1.77  &  1.18 &  0.34   \\
Cs & 55  &  1.36 &   1.08  &  0.92  &  0.83  &  0.87  &  1.14 &  1.50  &  1.65 & 1.81  &  1.90 &   1.91 &   1.85 &  1.78  &  1.63  &  1.03 &  0.23   \\
Ba & 56  &  2.43 &   2.14  &  1.98  &  1.91  &  2.01  &  2.37 &  2.65  &  2.77 & 2.91  &  2.94 &   2.88 &   2.80 &  2.72  &  2.45  &  1.75 &  0.65   \\
La & 57  &  2.42 &   2.14  &  1.98  &  1.92  &  2.03  &  2.41 &  2.66  &  2.77 & 2.91  &  2.93 &   2.85 &   2.77 &  2.69  &  2.37  &  1.65 &  0.48   \\
Ce & 58  &  2.55 &   2.27  &  2.11  &  2.07  &  2.22  &  2.60 &  2.81  &  2.92 & 3.04  &  3.04 &   2.94 &   2.85 &  2.77  &  2.37  &  1.65 &  0.40   \\
Pr & 59  &  2.31 &   2.04  &  1.88  &  1.84  &  1.99  &  2.37 &  2.58  &  2.69 & 2.81  &  2.82 &   2.73 &   2.65 &  2.57  &  2.17  &  1.45 &  0.26   \\
Nd & 60  &  2.40 &   2.12  &  1.97  &  1.93  &  2.08  &  2.46 &  2.66  &  2.78 & 2.89  &  2.89 &   2.79 &   2.71 &  2.62  &  2.19  &  1.48 &  0.26   \\
Sm & 62  &  2.28 &   1.99  &  1.84  &  1.80  &  1.96  &  2.35 &  2.54  &  2.65 & 2.75  &  2.73 &   2.61 &   2.52 &  2.42  &  1.96  &  1.26 &  0.15   \\
Eu & 63  &  1.57 &   1.29  &  1.14  &  1.11  &  1.27  &  1.64 &  1.83  &  1.94 & 2.03  &  2.01 &   1.89 &   1.80 &  1.70  &  1.25  &  0.61 &  0.00   \\
Gd & 64  &  2.01 &   1.73  &  1.57  &  1.54  &  1.71  &  2.09 &  2.27  &  2.38 & 2.47  &  2.45 &   2.33 &   2.23 &  2.13  &  1.66  &  0.98 &  0.05   \\
Tb & 65  &  1.76 &   1.48  &  1.33  &  1.30  &  1.46  &  1.84 &  2.03  &  2.13 & 2.23  &  2.20 &   2.08 &   1.98 &  1.87  &  1.41  &  0.75 &  0.00   \\
Dy & 66  &  2.00 &   1.72  &  1.56  &  1.53  &  1.70  &  2.08 &  2.26  &  2.37 & 2.46  &  2.43 &   2.30 &   2.21 &  2.10  &  1.62  &  0.94 &  0.04   \\
Ho & 67  &  1.75 &   1.47  &  1.32  &  1.29  &  1.46  &  1.83 &  2.02  &  2.12 & 2.21  &  2.18 &   2.05 &   1.96 &  1.84  &  1.37  &  0.72 &  0.00   \\
Er & 68  &  2.12 &   1.84  &  1.68  &  1.65  &  1.82  &  2.20 &  2.38  &  2.49 & 2.58  &  2.55 &   2.42 &   2.32 &  2.20  &  1.73  &  1.04 &  0.07   \\
Tm & 69  &  1.96 &   1.68  &  1.52  &  1.49  &  1.66  &  2.04 &  2.22  &  2.33 & 2.42  &  2.39 &   2.26 &   2.16 &  2.04  &  1.57  &  0.89 &  0.03   \\
Yb & 70  &  2.45 &   2.17  &  2.01  &  1.98  &  2.15  &  2.54 &  2.71  &  2.82 & 2.90  &  2.87 &   2.73 &   2.63 &  2.52  &  2.03  &  1.33 &  0.20   \\
Lu & 71  &  2.15 &   1.87  &  1.72  &  1.69  &  1.86  &  2.24 &  2.42  &  2.52 & 2.60  &  2.57 &   2.43 &   2.33 &  2.22  &  1.73  &  1.05 &  0.08   \\
Hf & 72  &  2.64 &   2.35  &  2.19  &  2.17  &  2.34  &  2.72 &  2.90  &  3.00 & 3.08  &  3.04 &   2.90 &   2.81 &  2.70  &  2.19  &  1.49 &  0.29   \\
Ta & 73  &  2.52 &   2.23  &  2.07  &  2.05  &  2.23  &  2.61 &  2.78  &  2.88 & 2.96  &  2.92 &   2.78 &   2.69 &  2.57  &  2.07  &  1.37 &  0.22   \\
W  & 74  &  2.66 &   2.37  &  2.21  &  2.19  &  2.38  &  2.76 &  2.92  &  3.02 & 3.10  &  3.06 &   2.92 &   2.83 &  2.71  &  2.20  &  1.50 &  0.28   \\
Re & 75  &  2.08 &   1.79  &  1.63  &  1.63  &  1.83  &  2.19 &  2.34  &  2.45 & 2.51  &  2.46 &   2.31 &   2.22 &  2.10  &  1.59  &  0.93 &  0.04   \\
Os & 76  &  1.95 &   1.66  &  1.50  &  1.49  &  1.68  &  2.05 &  2.21  &  2.32 & 2.39  &  2.35 &   2.21 &   2.13 &  2.01  &  1.49  &  0.83 &  0.01   \\
Ir & 77  &  1.08 &   0.82  &  0.69  &  0.67  &  0.84  &  1.18 &  1.33  &  1.44 & 1.51  &  1.48 &   1.34 &   1.27 &  1.15  &  0.67  &  0.20 &  0.00   \\
Pt & 78  &  1.67 &   1.39  &  1.24  &  1.23  &  1.42  &  1.78 &  1.94  &  2.05 & 2.13  &  2.09 &   1.96 &   1.88 &  1.75  &  1.23  &  0.60 &  0.00   \\
Au & 79  &  1.63 &   1.35  &  1.21  &  1.19  &  1.38  &  1.74 &  1.90  &  2.01 & 2.08  &  2.05 &   1.92 &   1.84 &  1.71  &  1.18  &  0.57 &  0.00   \\
Hg & 80  &  2.64 &   2.36  &  2.21  &  2.20  &  2.39  &  2.75 &  2.91  &  3.03 & 3.10  &  3.07 &   2.93 &   2.86 &  2.73  &  2.17  &  1.48 &  0.33   \\
Tl & 81  &  2.66 &   2.38  &  2.23  &  2.25  &  2.47  &  2.81 &  2.95  &  3.06 & 3.10  &  3.05 &   2.91 &   2.82 &  2.67  &  2.10  &  1.42 &  0.29   \\
Pb & 82  &  4.29 &   4.18  &  4.14  &  4.10  &  4.07  &  4.03 &  3.96  &  3.88 & 3.73  &  3.58 &   3.32 &   3.06 &  2.79  &  2.20  &  1.29 &  0.18   \\
Bi & 83  &  4.20 &   4.09  &  4.03  &  3.98  &  3.94  &  3.87 &  3.76  &  3.65 & 3.43  &  3.24 &   2.90 &   2.55 &  2.24  &  1.63  &  0.63 &  0.07   \\
\hline    
\end{tabular}}
\end{table*}

%% file: tableB5.tex
\begin{table*}
\caption{Theoretical [El/Fe] predictions from He to Bi for $M^{\rm AGB}_{\rm ini}$ = 1.3 $M_{\odot}$ and two $^{13}$C-pockets (ST and ST/12), as the metallicity changes.}
 \label{datatablebab9ltHTp1p5d8m1p3a}
\centering
\resizebox{14cm}{!}{\begin{tabular}{cc|cccccccc|cc|cccccccc} 
\hline
    &    & \multicolumn{7}{c}{Case ST} & & & &   \multicolumn{7}{c}{case ST/12} & \\
    &    & \multicolumn{7}{c}{[Fe/H]}  & & & &   \multicolumn{7}{c}{[Fe/H]} & \\
El  &  Z &  -3.6 &  -3.0 &  -2.6 &  -2.3 &  -2.0  &  -1.6   &  -1.3 & -1.0 & & &
-3.6 &  -3.0 &  -2.6 &  -2.3 &  -2.0  &  -1.6   &  -1.3 & -1.0    \\
\hline
He  &  4  &  0.06  &  0.06  &  0.06  &  0.06  &  0.06  &  0.03 &  0.01  &  0.01  & & &     0.06  &  0.05 &  0.05 &  0.05 &   0.05  &  0.03  &  0.01  &  0.01 \\
C   &  6  &  4.44  &  3.83  &  3.44  &  3.13  &  2.83  &  2.13 &  1.54  &  1.24  & & &     4.44  &  3.83 &  3.43 &  3.13 &   2.83  &  2.13  &  1.54  &  1.24 \\
N   &  7  &  1.41  &  0.92  &  0.68  &  0.56  &  0.48  &  0.42 &  0.40  &  0.40  & & &     1.41  &  0.92 &  0.68 &  0.56 &   0.48  &  0.41  &  0.40  &  0.40 \\
O   &  8  &  2.37  &  1.82  &  1.49  &  1.25  &  1.03  &  0.70 &  0.53  &  0.40  & & &     2.21  &  1.68 &  1.36 &  1.14 &   0.94  &  0.67  &  0.52  &  0.40 \\
F   &  9  &  4.57  &  3.97  &  3.57  &  3.25  &  2.92  &  2.10 &  1.31  &  0.74  & & &     4.23  &  3.63 &  3.19 &  2.79 &   2.30  &  1.34  &  0.64  &  0.35 \\
Ne  & 10  &  2.28  &  1.69  &  1.33  &  1.09  &  0.88  &  0.53 &  0.39  &  0.37  & & &     2.20  &  1.62 &  1.27 &  1.03 &   0.84  &  0.52  &  0.39  &  0.37 \\
Na  & 11  &  1.85  &  1.26  &  0.90  &  0.66  &  0.46  &  0.15 &  0.05  &  0.04  & & &     1.65  &  1.06 &  0.71 &  0.48 &   0.31  &  0.10  &  0.04  &  0.03 \\
Mg  & 12  &  1.06  &  0.63  &  0.46  &  0.38  &  0.33  &  0.28 &  0.26  &  0.26  & & &     0.79  &  0.47 &  0.35 &  0.31 &   0.28  &  0.26  &  0.25  &  0.25 \\
Al  & 13  &  0.25  &  0.09  &  0.04  &  0.03  &  0.02  &  0.01 &  0.00  &  0.00  & & &     0.04  &  0.02 &  0.01 &  0.01 &   0.01  &  0.01  &  0.00  &  0.00 \\
Si  & 14  &  0.38  &  0.32  &  0.31  &  0.30  &  0.29  &  0.29 &  0.28  &  0.28  & & &     0.29  &  0.29 &  0.29 &  0.29 &   0.29  &  0.28  &  0.28  &  0.28 \\
P   & 15  &  1.21  &  0.76  &  0.55  &  0.43  &  0.34  &  0.14 &  0.05  &  0.04  & & &     0.25  &  0.20 &  0.18 &  0.15 &   0.12  &  0.04  &  0.01  &  0.01 \\
S   & 16  &  0.34  &  0.31  &  0.30  &  0.30  &  0.29  &  0.29 &  0.29  &  0.29  & & &     0.29  &  0.29 &  0.29 &  0.29 &   0.29  &  0.29  &  0.29  &  0.29 \\
Cl  & 17  &  0.08  &  0.06  &  0.06  &  0.05  &  0.05  &  0.02 &  0.01  &  0.01  & & &     0.05  &  0.04 &  0.04 &  0.04 &   0.04  &  0.03  &  0.01  &  0.01 \\
Ar  & 18  &  0.27  &  0.26  &  0.26  &  0.26  &  0.26  &  0.26 &  0.26  &  0.26  & & &     0.26  &  0.26 &  0.26 &  0.26 &   0.26  &  0.26  &  0.26  &  0.26 \\
K   & 19  &  0.33  &  0.33  &  0.32  &  0.32  &  0.31  &  0.30 &  0.29  &  0.29  & & &     0.32  &  0.31 &  0.31 &  0.31 &   0.30  &  0.29  &  0.29  &  0.29 \\
Ca  & 20  &  0.29  &  0.29  &  0.29  &  0.29  &  0.29  &  0.29 &  0.29  &  0.29  & & &     0.29  &  0.29 &  0.29 &  0.29 &   0.29  &  0.29  &  0.29  &  0.29 \\
Sc  & 21  &  0.41  &  0.36  &  0.34  &  0.30  &  0.26  &  0.09 &  0.04  &  0.04  & & &     0.28  &  0.25 &  0.22 &  0.21 &   0.21  &  0.11  &  0.04  &  0.03 \\
Ti  & 22  &  0.29  &  0.27  &  0.27  &  0.26  &  0.26  &  0.25 &  0.24  &  0.25  & & &     0.27  &  0.27 &  0.27 &  0.27 &   0.28  &  0.25  &  0.24  &  0.24 \\
V   & 23  &  0.03  &  0.01  &  0.01  &  0.01  &  0.00  &  0.00 &  0.00  &  0.00  & & &     0.01  &  0.01 &  0.01 &  0.01 &   0.01  &  0.00  &  0.00  &  0.00 \\
Cr  & 24  & -0.20  & -0.20  & -0.20  & -0.20  & -0.20  & -0.20 & -0.20  & -0.20  & & &    -0.20  & -0.20 & -0.20 & -0.20 &  -0.20  & -0.20  & -0.20  & -0.20 \\
Mn  & 25  & -0.40  & -0.40  & -0.40  & -0.40  & -0.40  & -0.40 & -0.40  & -0.40  & & &    -0.40  & -0.40 & -0.40 & -0.40 &  -0.40  & -0.40  & -0.40  & -0.40 \\
Fe  & 26  &  0.00  &  0.00  &  0.00  &  0.00  &  0.00  &  0.00 &  0.00  &  0.00  & & &     0.00  &  0.00 &  0.00 &  0.00 &   0.00  &  0.00  &  0.00  &  0.00 \\
Co  & 27  &  0.42  &  0.43  &  0.43  &  0.42  &  0.40  &  0.19 &  0.06  &  0.03  & & &     0.41  &  0.38 &  0.35 &  0.30 &   0.22  &  0.07  &  0.03  &  0.03 \\
Ni  & 28  &  0.10  &  0.09  &  0.08  &  0.07  &  0.05  &  0.01 &  0.00  &  0.00  & & &     0.07  &  0.05 &  0.03 &  0.02 &   0.01  &  0.00  &  0.00  &  0.00 \\
Cu  & 29  &  0.87  &  0.85  &  0.80  &  0.74  &  0.63  &  0.22 &  0.04  &  0.02  & & &     0.71  &  0.57 &  0.45 &  0.32 &   0.18  &  0.07  &  0.05  &  0.07 \\
Zn  & 30  &  0.40  &  0.36  &  0.31  &  0.24  &  0.16  &  0.02 &  0.00  &  0.00  & & &     0.23  &  0.13 &  0.07 &  0.04 &   0.02  &  0.04  &  0.04  &  0.04 \\
Ga  & 31  &  0.87  &  0.80  &  0.70  &  0.57  &  0.38  &  0.06 &  0.02  &  0.02  & & &     0.50  &  0.31 &  0.19 &  0.12 &   0.11  &  0.19  &  0.18  &  0.18 \\
Ge  & 32  &  0.85  &  0.77  &  0.65  &  0.50  &  0.31  &  0.05 &  0.02  &  0.02  & & &     0.44  &  0.25 &  0.16 &  0.12 &   0.15  &  0.29  &  0.25  &  0.22 \\
As  & 33  &  0.68  &  0.59  &  0.48  &  0.34  &  0.18  &  0.02 &  0.01  &  0.01  & & &     0.28  &  0.15 &  0.09 &  0.08 &   0.12  &  0.26  &  0.21  &  0.16 \\
Se  & 34  &  0.76  &  0.66  &  0.52  &  0.36  &  0.19  &  0.03 &  0.02  &  0.03  & & &     0.30  &  0.17 &  0.13 &  0.12 &   0.23  &  0.40  &  0.31  &  0.22 \\
Br  & 35  &  0.66  &  0.56  &  0.43  &  0.28  &  0.14  &  0.02 &  0.01  &  0.02  & & &     0.23  &  0.13 &  0.10 &  0.10 &   0.20  &  0.36  &  0.27  &  0.18 \\
Kr  & 36  &  0.74  &  0.61  &  0.46  &  0.31  &  0.18  &  0.07 &  0.05  &  0.12  & & &     0.31  &  0.26 &  0.26 &  0.32 &   0.49  &  0.56  &  0.38  &  0.22 \\
Rb  & 37  &  0.88  &  0.72  &  0.55  &  0.39  &  0.26  &  0.11 &  0.07  &  0.16  & & &     0.44  &  0.40 &  0.41 &  0.47 &   0.66  &  0.63  &  0.38  &  0.18 \\
Sr  & 38  &  0.67  &  0.51  &  0.40  &  0.32  &  0.26  &  0.18 &  0.15  &  0.48  & & &     0.53  &  0.57 &  0.69 &  0.92 &   1.29  &  1.26  &  0.89  &  0.42 \\
Y   & 39  &  0.66  &  0.49  &  0.41  &  0.36  &  0.31  &  0.23 &  0.18  &  0.56  & & &     0.65  &  0.71 &  0.84 &  1.09 &   1.44  &  1.31  &  0.84  &  0.31 \\
Zr  & 40  &  0.65  &  0.48  &  0.41  &  0.37  &  0.33  &  0.24 &  0.20  &  0.66  & & &     0.72  &  0.79 &  0.94 &  1.20 &   1.50  &  1.31  &  0.70  &  0.20 \\
Nb  & 41  &  0.68  &  0.50  &  0.43  &  0.39  &  0.35  &  0.26 &  0.21  &  0.68  & & &     0.76  &  0.83 &  0.98 &  1.25 &   1.54  &  1.32  &  0.67  &  0.18 \\
Mo  & 42  &  0.56  &  0.40  &  0.34  &  0.31  &  0.27  &  0.20 &  0.16  &  0.61  & & &     0.67  &  0.74 &  0.89 &  1.15 &   1.41  &  1.17  &  0.47  &  0.10 \\
Ru  & 44  &  0.40  &  0.27  &  0.23  &  0.19  &  0.16  &  0.11 &  0.10  &  0.50  & & &     0.47  &  0.52 &  0.65 &  0.90 &   1.18  &  0.99  &  0.34  &  0.07 \\
Rh  & 45  &  0.21  &  0.13  &  0.10  &  0.08  &  0.06  &  0.05 &  0.05  &  0.31  & & &     0.25  &  0.28 &  0.39 &  0.61 &   0.88  &  0.71  &  0.19  &  0.03 \\
Pd  & 46  &  0.42  &  0.29  &  0.24  &  0.21  &  0.18  &  0.15 &  0.16  &  0.64  & & &     0.52  &  0.58 &  0.75 &  1.06 &   1.37  &  1.16  &  0.43  &  0.09 \\
Ag  & 47  &  0.21  &  0.13  &  0.11  &  0.09  &  0.07  &  0.07 &  0.07  &  0.40  & & &     0.29  &  0.34 &  0.49 &  0.76 &   1.03  &  0.83  &  0.23  &  0.04 \\
Cd  & 48  &  0.44  &  0.31  &  0.27  &  0.25  &  0.25  &  0.23 &  0.22  &  0.78  & & &     0.66  &  0.76 &  0.96 &  1.26 &   1.53  &  1.29  &  0.48  &  0.11 \\
In  & 49  &  0.31  &  0.21  &  0.19  &  0.18  &  0.18  &  0.15 &  0.15  &  0.62  & & &     0.53  &  0.61 &  0.80 &  1.08 &   1.34  &  1.10  &  0.34  &  0.06 \\
Sn  & 50  &  0.61  &  0.45  &  0.40  &  0.37  &  0.35  &  0.30 &  0.27  &  0.88  & & &     0.84  &  0.95 &  1.15 &  1.43 &   1.65  &  1.35  &  0.44  &  0.09 \\
Sb  & 51  &  0.40  &  0.27  &  0.23  &  0.21  &  0.19  &  0.15 &  0.14  &  0.59  & & &     0.58  &  0.68 &  0.86 &  1.11 &   1.29  &  0.98  &  0.21  &  0.03 \\
Te  & 52  &  0.33  &  0.22  &  0.18  &  0.16  &  0.15  &  0.12 &  0.11  &  0.53  & & &     0.50  &  0.59 &  0.76 &  1.01 &   1.18  &  0.88  &  0.16  &  0.02 \\
I   & 53  &  0.12  &  0.07  &  0.05  &  0.05  &  0.04  &  0.03 &  0.03  &  0.25  & & &     0.21  &  0.27 &  0.38 &  0.58 &   0.73  &  0.48  &  0.05  &  0.00 \\
Xe  & 54  &  0.38  &  0.26  &  0.22  &  0.20  &  0.18  &  0.15 &  0.14  &  0.62  & & &     0.58  &  0.67 &  0.86 &  1.11 &   1.27  &  0.93  &  0.17  &  0.02 \\
Cs  & 55  &  0.29  &  0.19  &  0.16  &  0.14  &  0.13  &  0.12 &  0.12  &  0.56  & & &     0.48  &  0.57 &  0.76 &  1.01 &   1.17  &  0.84  &  0.13  &  0.02 \\
Ba  & 56  &  1.09  &  0.89  &  0.82  &  0.78  &  0.76  &  0.69 &  0.83  &  1.46  & & &     1.57  &  1.68 &  1.87 &  2.05 &   2.15  &  1.50  &  0.46  &  0.06 \\
La  & 57  &  1.09  &  0.89  &  0.82  &  0.79  &  0.76  &  0.70 &  0.88  &  1.46  & & &     1.61  &  1.72 &  1.90 &  2.07 &   2.15  &  1.38  &  0.40  &  0.03 \\
Ce  & 58  &  1.20  &  1.00  &  0.93  &  0.89  &  0.88  &  0.83 &  1.11  &  1.58  & & &     1.80  &  1.90 &  2.06 &  2.19 &   2.29  &  1.35  &  0.40  &  0.02 \\
Pr  & 59  &  1.00  &  0.80  &  0.74  &  0.71  &  0.70  &  0.68 &  0.95  &  1.38  & & &     1.60  &  1.70 &  1.87 &  2.00 &   2.10  &  1.15  &  0.28  &  0.01 \\
Nd  & 60  &  1.08  &  0.88  &  0.82  &  0.79  &  0.77  &  0.74 &  1.05  &  1.46  & & &     1.69  &  1.79 &  1.96 &  2.08 &   2.18  &  1.16  &  0.29  &  0.01 \\
Sm  & 62  &  0.93  &  0.74  &  0.67  &  0.64  &  0.62  &  0.58 &  0.86  &  1.21  & & &     1.51  &  1.61 &  1.77 &  1.87 &   1.95  &  0.89  &  0.17  &  0.00 \\
Eu  & 63  &  0.38  &  0.26  &  0.22  &  0.21  &  0.20  &  0.18 &  0.34  &  0.58  & & &     0.85  &  0.93 &  1.07 &  1.17 &   1.24  &  0.34  &  0.03  &  0.00 \\
Gd  & 64  &  0.68  &  0.51  &  0.46  &  0.43  &  0.42  &  0.39 &  0.63  &  0.93  & & &     1.25  &  1.33 &  1.49 &  1.58 &   1.66  &  0.62  &  0.09  &  0.00 \\
Tb  & 65  &  0.49  &  0.35  &  0.31  &  0.29  &  0.28  &  0.26 &  0.45  &  0.71  & & &     1.01  &  1.10 &  1.25 &  1.34 &   1.41  &  0.44  &  0.05  &  0.00 \\
Dy  & 66  &  0.67  &  0.50  &  0.45  &  0.43  &  0.41  &  0.38 &  0.60  &  0.90  & & &     1.23  &  1.32 &  1.47 &  1.55 &   1.62  &  0.59  &  0.08  &  0.00 \\
Ho  & 67  &  0.47  &  0.34  &  0.30  &  0.28  &  0.27  &  0.24 &  0.43  &  0.70  & & &     1.00  &  1.08 &  1.22 &  1.30 &   1.38  &  0.41  &  0.05  &  0.00 \\
Er  & 68  &  0.77  &  0.59  &  0.54  &  0.51  &  0.50  &  0.44 &  0.70  &  1.01  & & &     1.35  &  1.44 &  1.58 &  1.66 &   1.73  &  0.67  &  0.10  &  0.00 \\
Tm  & 69  &  0.66  &  0.50  &  0.45  &  0.42  &  0.41  &  0.35 &  0.57  &  0.87  & & &     1.22  &  1.30 &  1.44 &  1.50 &   1.58  &  0.55  &  0.07  &  0.00 \\
Yb  & 70  &  1.07  &  0.87  &  0.80  &  0.77  &  0.74  &  0.67 &  0.99  &  1.32  & & &     1.66  &  1.74 &  1.88 &  1.96 &   2.06  &  0.94  &  0.20  &  0.01 \\
Lu  & 71  &  0.82  &  0.63  &  0.58  &  0.54  &  0.51  &  0.45 &  0.75  &  1.05  & & &     1.37  &  1.45 &  1.59 &  1.67 &   1.77  &  0.69  &  0.11  &  0.00 \\
Hf  & 72  &  1.25  &  1.05  &  0.97  &  0.92  &  0.89  &  0.84 &  1.19  &  1.50  & & &     1.83  &  1.92 &  2.07 &  2.16 &   2.26  &  1.09  &  0.26  &  0.01 \\
Ta  & 73  &  1.13  &  0.93  &  0.86  &  0.81  &  0.78  &  0.74 &  1.09  &  1.39  & & &     1.71  &  1.80 &  1.95 &  2.04 &   2.14  &  0.97  &  0.20  &  0.01 \\
W   & 74  &  1.25  &  1.04  &  0.97  &  0.92  &  0.90  &  0.86 &  1.23  &  1.53  & & &     1.86  &  1.94 &  2.10 &  2.18 &   2.28  &  1.10  &  0.26  &  0.01 \\
Re  & 75  &  0.79  &  0.59  &  0.52  &  0.48  &  0.47  &  0.44 &  0.76  &  1.01  & & &     1.36  &  1.43 &  1.56 &  1.63 &   1.71  &  0.62  &  0.09  &  0.00 \\
Os  & 76  &  0.63  &  0.47  &  0.41  &  0.39  &  0.38  &  0.36 &  0.65  &  0.86  & & &     1.20  &  1.28 &  1.43 &  1.50 &   1.60  &  0.50  &  0.06  &  0.00 \\
Ir  & 77  &  0.15  &  0.09  &  0.07  &  0.07  &  0.06  &  0.06 &  0.16  &  0.25  & & &     0.48  &  0.53 &  0.64 &  0.70 &   0.78  &  0.10  &  0.00  &  0.00 \\
Pt  & 78  &  0.45  &  0.32  &  0.28  &  0.26  &  0.26  &  0.24 &  0.47  &  0.64  & & &     0.98  &  1.06 &  1.20 &  1.27 &   1.36  &  0.32  &  0.03  &  0.00 \\
Au  & 79  &  0.43  &  0.30  &  0.27  &  0.25  &  0.24  &  0.22 &  0.44  &  0.62  & & &     0.96  &  1.03 &  1.17 &  1.23 &   1.32  &  0.30  &  0.03  &  0.00 \\
Hg  & 80  &  1.29  &  1.09  &  1.03  &  0.99  &  0.98  &  0.94 &  1.35  &  1.55  & & &     1.95  &  2.03 &  2.18 &  2.24 &   2.35  &  1.07  &  0.24  &  0.03 \\
Tl  & 81  &  1.39  &  1.14  &  1.06  &  1.03  &  1.04  &  0.99 &  1.47  &  1.62  & & &     2.05  &  2.11 &  2.23 &  2.29 &   2.35  &  1.12  &  0.24  &  0.01 \\
Pb  & 82  &  3.24  &  3.23  &  3.22  &  3.22  &  3.21  &  2.89 &  2.53  &  2.26  & & &     3.08  &  3.06 &  2.99 &  2.85 &   2.43  &  1.06  &  0.10  &  0.00 \\
Bi  & 83  &  3.14  &  3.11  &  3.09  &  3.06  &  3.01  &  2.58 &  2.06  &  1.65  & & &     2.86  &  2.78 &  2.64 &  2.41 &   1.66  &  0.34  &  0.01  &  0.00 \\
\hline    
\end{tabular}}
\end{table*}

%% file: tableB6.tex
\begin{table*}
\caption{The same as Table~\ref{datatablebab9ltHTp1p5d8m1p3a}, 
but for $M^{\rm AGB}_{\rm ini}$ = 1.5 $M_{\odot}$.}
 \label{datatablebab9ltHTp1p5d8m1p5a}
\centering
\resizebox{14cm}{!}{\begin{tabular}{cc|cccccccc|cc|cccccccc} 
\hline
    &    & \multicolumn{7}{c}{Case ST} & & & & \multicolumn{7}{c}{case ST/12} & \\
    &    & \multicolumn{7}{c}{[Fe/H]}  & & & & \multicolumn{7}{c}{[Fe/H]} & \\
El  &  Z &  -3.6 &  -3.0 &  -2.6 &  -2.3 &  -2.0  &  -1.6   &  -1.3 & -1.0 & & &
-3.6 &  -3.0 &  -2.6 &  -2.3 &  -2.0  &  -1.6   &  -1.3 & -1.0    \\
\hline
He &  4  &  0.26  &  0.26  &  0.26  &  0.26  &  0.25  &  0.14  &  0.07 &  0.07  & & &   0.26 &   0.26 &  0.26 &  0.25 &  0.25 &   0.14  &  0.07  &  0.07  \\
C  &  6  &  5.14  &  4.53  &  4.13  &  3.83  &  3.52  &  2.83  &  2.24 &  1.94  & & &   5.13 &   4.53 &  4.13 &  3.83 &  3.52 &   2.83  &  2.24  &  1.94  \\
N  &  7  &  3.15  &  2.55  &  2.16  &  1.86  &  1.57  &  0.98  &  0.64 &  0.53  & & &   3.15 &   2.55 &  2.15 &  1.86 &  1.56 &   0.98  &  0.63  &  0.52  \\
O  &  8  &  3.07  &  2.48  &  2.09  &  1.79  &  1.50  &  0.95  &  0.63 &  0.46  & & &   2.89 &   2.30 &  1.92 &  1.63 &  1.35 &   0.86  &  0.59  &  0.43  \\
F  &  9  &  5.02  &  4.42  &  4.02  &  3.70  &  3.38  &  2.62  &  1.90 &  1.41  & & &   4.63 &   4.02 &  3.60 &  3.25 &  2.88 &   2.10  &  1.44  &  1.10  \\
Ne & 10  &  3.87  &  3.27  &  2.87  &  2.58  &  2.28  &  1.62  &  1.11 &  0.93  & & &   3.87 &   3.26 &  2.87 &  2.57 &  2.28 &   1.62  &  1.11  &  0.93  \\
Na & 11  &  3.57  &  2.97  &  2.57  &  2.26  &  1.96  &  1.27  &  0.73 &  0.52  & & &   3.59 &   2.98 &  2.58 &  2.27 &  1.96 &   1.25  &  0.71  &  0.49  \\
Mg & 12  &  2.98  &  2.38  &  1.98  &  1.69  &  1.40  &  0.80  &  0.46 &  0.37  & & &   2.75 &   2.15 &  1.76 &  1.47 &  1.19 &   0.67  &  0.41  &  0.35  \\
Al & 13  &  1.75  &  1.17  &  0.81  &  0.57  &  0.37  &  0.10  &  0.03 &  0.02  & & &   1.39 &   0.84 &  0.53 &  0.35 &  0.21 &   0.06  &  0.02  &  0.02  \\
Si & 14  &  1.29  &  0.81  &  0.59  &  0.48  &  0.41  &  0.32  &  0.29 &  0.29  & & &   0.52 &   0.40 &  0.36 &  0.35 &  0.34 &   0.31  &  0.29  &  0.29  \\
P  & 15  &  2.63  &  2.04  &  1.66  &  1.40  &  1.16  &  0.65  &  0.33 &  0.26  & & &   0.88 &   0.74 &  0.69 &  0.66 &  0.61 &   0.34  &  0.16  &  0.11  \\
S  & 16  &  0.87  &  0.53  &  0.41  &  0.37  &  0.34  &  0.30  &  0.29 &  0.29  & & &   0.32 &   0.31 &  0.31 &  0.31 &  0.31 &   0.30  &  0.29  &  0.29  \\
Cl & 17  &  0.53  &  0.32  &  0.26  &  0.23  &  0.22  &  0.11  &  0.06 &  0.06  & & &   0.26 &   0.26 &  0.26 &  0.26 &  0.26 &   0.16  &  0.09  &  0.09  \\
Ar & 18  &  0.28  &  0.27  &  0.26  &  0.26  &  0.26  &  0.26  &  0.26 &  0.26  & & &   0.26 &   0.26 &  0.26 &  0.26 &  0.26 &   0.26  &  0.26  &  0.26  \\
K  & 19  &  0.50  &  0.47  &  0.46  &  0.45  &  0.44  &  0.36  &  0.32 &  0.31  & & &   0.47 &   0.47 &  0.46 &  0.46 &  0.45 &   0.37  &  0.33  &  0.32  \\
Ca & 20  &  0.28  &  0.28  &  0.28  &  0.27  &  0.27  &  0.28  &  0.29 &  0.29  & & &   0.28 &   0.28 &  0.28 &  0.28 &  0.28 &   0.28  &  0.29  &  0.29  \\
Sc & 21  &  1.11  &  0.95  &  0.89  &  0.86  &  0.83  &  0.54  &  0.33 &  0.34  & & &   0.95 &   0.94 &  0.93 &  0.92 &  0.90 &   0.62  &  0.37  &  0.31  \\
Ti & 22  &  0.61  &  0.46  &  0.41  &  0.40  &  0.38  &  0.32  &  0.29 &  0.30  & & &   0.42 &   0.42 &  0.42 &  0.42 &  0.41 &   0.31  &  0.27  &  0.26  \\
V  & 23  &  0.28  &  0.13  &  0.09  &  0.07  &  0.06  &  0.03  &  0.02 &  0.03  & & &   0.09 &   0.08 &  0.08 &  0.08 &  0.06 &   0.01  &  0.00  &  0.00  \\
Cr & 24  & -0.20  & -0.20  & -0.20  & -0.20  & -0.20  & -0.20  & -0.20 & -0.20  & & &  -0.20 &  -0.20 & -0.20 & -0.20 & -0.20 &  -0.20  & -0.20  & -0.20  \\
Mn & 25  & -0.40  & -0.40  & -0.40  & -0.40  & -0.40  & -0.40  & -0.40 & -0.40  & & &  -0.40 &  -0.40 & -0.40 & -0.40 & -0.40 &  -0.40  & -0.40  & -0.40  \\
Fe & 26  &  0.00  &  0.00  &  0.00  &  0.00  &  0.00  &  0.00  &  0.00 &  0.00  & & &   0.00 &   0.00 &  0.00 &  0.00 &  0.00 &   0.00  &  0.00  &  0.00  \\
Co & 27  &  1.01  &  1.01  &  1.01  &  1.00  &  0.99  &  0.70  &  0.43 &  0.36  & & &   1.01 &   1.01 &  1.00 &  0.98 &  0.95 &   0.66  &  0.41  &  0.37  \\
Ni & 28  &  0.32  &  0.32  &  0.31  &  0.30  &  0.27  &  0.12  &  0.05 &  0.03  & & &   0.32 &   0.31 &  0.29 &  0.27 &  0.25 &   0.12  &  0.05  &  0.04  \\
Cu & 29  &  1.44  &  1.43  &  1.41  &  1.39  &  1.34  &  0.95  &  0.57 &  0.43  & & &   1.43 &   1.40 &  1.38 &  1.34 &  1.29 &   0.94  &  0.64  &  0.58  \\
Zn & 30  &  0.83  &  0.81  &  0.78  &  0.74  &  0.68  &  0.35  &  0.14 &  0.08  & & &   0.83 &   0.80 &  0.77 &  0.74 &  0.71 &   0.49  &  0.32  &  0.31  \\
Ga & 31  &  1.41  &  1.37  &  1.33  &  1.28  &  1.19  &  0.74  &  0.36 &  0.24  & & &   1.42 &   1.39 &  1.36 &  1.34 &  1.31 &   1.04  &  0.80  &  0.75  \\
Ge & 32  &  1.41  &  1.37  &  1.32  &  1.26  &  1.16  &  0.70  &  0.33 &  0.24  & & &   1.46 &   1.43 &  1.41 &  1.39 &  1.38 &   1.15  &  0.90  &  0.82  \\
As & 33  &  1.24  &  1.19  &  1.14  &  1.08  &  0.97  &  0.53  &  0.22 &  0.17  & & &   1.33 &   1.30 &  1.29 &  1.28 &  1.28 &   1.06  &  0.82  &  0.71  \\
Se & 34  &  1.38  &  1.33  &  1.27  &  1.19  &  1.07  &  0.60  &  0.28 &  0.26  & & &   1.53 &   1.52 &  1.50 &  1.50 &  1.52 &   1.30  &  1.02  &  0.85  \\
Br & 35  &  1.28  &  1.22  &  1.16  &  1.09  &  0.97  &  0.51  &  0.22 &  0.22  & & &   1.45 &   1.44 &  1.43 &  1.43 &  1.44 &   1.22  &  0.94  &  0.76  \\
Kr & 36  &  1.55  &  1.45  &  1.37  &  1.28  &  1.16  &  0.72  &  0.48 &  0.65  & & &   1.86 &   1.84 &  1.84 &  1.84 &  1.85 &   1.57  &  1.19  &  0.90  \\
Rb & 37  &  1.82  &  1.71  &  1.63  &  1.53  &  1.40  &  0.95  &  0.69 &  0.88  & & &   2.17 &   2.16 &  2.16 &  2.16 &  2.17 &   1.87  &  1.45  &  1.09  \\
Sr & 38  &  1.68  &  1.51  &  1.42  &  1.34  &  1.27  &  0.97  &  0.90 &  1.34  & & &   2.24 &   2.24 &  2.25 &  2.26 &  2.29 &   2.00  &  1.53  &  1.09  \\
Y  & 39  &  1.76  &  1.55  &  1.46  &  1.40  &  1.35  &  1.11  &  1.07 &  1.53  & & &   2.40 &   2.40 &  2.41 &  2.42 &  2.44 &   2.10  &  1.55  &  1.06  \\
Zr & 40  &  1.78  &  1.54  &  1.45  &  1.41  &  1.38  &  1.19  &  1.18 &  1.64  & & &   2.45 &   2.45 &  2.45 &  2.46 &  2.45 &   2.05  &  1.41  &  0.88  \\
Nb & 41  &  1.80  &  1.56  &  1.47  &  1.43  &  1.41  &  1.22  &  1.22 &  1.69  & & &   2.49 &   2.49 &  2.49 &  2.49 &  2.49 &   2.07  &  1.41  &  0.87  \\
Mo & 42  &  1.67  &  1.42  &  1.33  &  1.29  &  1.29  &  1.12  &  1.14 &  1.59  & & &   2.33 &   2.33 &  2.33 &  2.33 &  2.31 &   1.87  &  1.16  &  0.62  \\
Ru & 44  &  1.40  &  1.15  &  1.07  &  1.03  &  1.01  &  0.85  &  0.87 &  1.30  & & &   2.07 &   2.06 &  2.06 &  2.05 &  2.02 &   1.57  &  0.86  &  0.38  \\
Rh & 45  &  1.03  &  0.79  &  0.71  &  0.68  &  0.66  &  0.53  &  0.55 &  0.98  & & &   1.68 &   1.67 &  1.67 &  1.66 &  1.64 &   1.21  &  0.55  &  0.19  \\
Pd & 46  &  1.42  &  1.16  &  1.08  &  1.04  &  1.04  &  0.90  &  0.96 &  1.48  & & &   2.08 &   2.08 &  2.07 &  2.07 &  2.07 &   1.66  &  0.94  &  0.46  \\
Ag & 47  &  1.04  &  0.80  &  0.72  &  0.69  &  0.69  &  0.58  &  0.66 &  1.15  & & &   1.68 &   1.68 &  1.68 &  1.69 &  1.69 &   1.30  &  0.63  &  0.25  \\
Cd & 48  &  1.51  &  1.26  &  1.18  &  1.16  &  1.18  &  1.08  &  1.18 &  1.68  & & &   2.14 &   2.15 &  2.16 &  2.18 &  2.21 &   1.81  &  1.07  &  0.55  \\
In & 49  &  1.32  &  1.08  &  1.01  &  0.99  &  1.01  &  0.92  &  1.00 &  1.48  & & &   1.94 &   1.95 &  1.96 &  1.99 &  2.01 &   1.60  &  0.87  &  0.39  \\
Sn & 50  &  1.74  &  1.49  &  1.42  &  1.40  &  1.41  &  1.30  &  1.39 &  1.84  & & &   2.35 &   2.35 &  2.36 &  2.37 &  2.36 &   1.88  &  1.09  &  0.54  \\
Sb & 51  &  1.43  &  1.18  &  1.11  &  1.09  &  1.11  &  1.01  &  1.10 &  1.52  & & &   2.02 &   2.02 &  2.02 &  2.02 &  2.00 &   1.51  &  0.74  &  0.28  \\
Te & 52  &  1.30  &  1.05  &  0.99  &  0.97  &  0.99  &  0.90  &  0.99 &  1.40  & & &   1.87 &   1.87 &  1.88 &  1.88 &  1.86 &   1.36  &  0.62  &  0.20  \\
I  & 53  &  0.82  &  0.61  &  0.55  &  0.54  &  0.55  &  0.49  &  0.56 &  0.92  & & &   1.37 &   1.37 &  1.37 &  1.37 &  1.35 &   0.88  &  0.27  &  0.05  \\
Xe & 54  &  1.40  &  1.16  &  1.09  &  1.07  &  1.09  &  1.01  &  1.10 &  1.50  & & &   1.95 &   1.95 &  1.95 &  1.96 &  1.93 &   1.40  &  0.64  &  0.20  \\
Cs & 55  &  1.21  &  0.98  &  0.91  &  0.90  &  0.92  &  0.86  &  0.96 &  1.38  & & &   1.77 &   1.77 &  1.77 &  1.78 &  1.77 &   1.26  &  0.52  &  0.14  \\
Ba & 56  &  2.30  &  2.08  &  2.03  &  2.04  &  2.11  &  2.09  &  2.15 &  2.40  & & &   2.72 &   2.72 &  2.73 &  2.73 &  2.69 &   2.02  &  1.15  &  0.44  \\
La & 57  &  2.31  &  2.10  &  2.06  &  2.07  &  2.15  &  2.14  &  2.20 &  2.40  & & &   2.68 &   2.69 &  2.69 &  2.70 &  2.64 &   1.92  &  1.05  &  0.32  \\
Ce & 58  &  2.45  &  2.26  &  2.23  &  2.26  &  2.35  &  2.35  &  2.37 &  2.52  & & &   2.76 &   2.76 &  2.77 &  2.77 &  2.71 &   1.91  &  1.01  &  0.26  \\
Pr & 59  &  2.23  &  2.04  &  2.01  &  2.04  &  2.13  &  2.14  &  2.17 &  2.33  & & &   2.53 &   2.54 &  2.55 &  2.56 &  2.50 &   1.70  &  0.82  &  0.17  \\
Nd & 60  &  2.32  &  2.14  &  2.11  &  2.13  &  2.23  &  2.23  &  2.24 &  2.39  & & &   2.60 &   2.61 &  2.62 &  2.62 &  2.56 &   1.72  &  0.83  &  0.16  \\
Sm & 62  &  2.17  &  1.99  &  1.96  &  1.99  &  2.08  &  2.08  &  2.07 &  2.19  & & &   2.43 &   2.43 &  2.44 &  2.43 &  2.35 &   1.50  &  0.63  &  0.09  \\
Eu & 63  &  1.46  &  1.28  &  1.26  &  1.28  &  1.37  &  1.37  &  1.37 &  1.48  & & &   1.72 &   1.72 &  1.72 &  1.71 &  1.64 &   0.82  &  0.19  &  0.00  \\
Gd & 64  &  1.89  &  1.71  &  1.68  &  1.71  &  1.81  &  1.81  &  1.80 &  1.91  & & &   2.15 &   2.15 &  2.16 &  2.14 &  2.07 &   1.21  &  0.41  &  0.03  \\
Tb & 65  &  1.65  &  1.47  &  1.44  &  1.47  &  1.57  &  1.57  &  1.56 &  1.66  & & &   1.90 &   1.90 &  1.91 &  1.89 &  1.81 &   0.97  &  0.26  &  0.01  \\
Dy & 66  &  1.88  &  1.70  &  1.68  &  1.71  &  1.80  &  1.81  &  1.79 &  1.89  & & &   2.13 &   2.13 &  2.14 &  2.12 &  2.04 &   1.18  &  0.39  &  0.03  \\
Ho & 67  &  1.64  &  1.46  &  1.44  &  1.47  &  1.56  &  1.57  &  1.55 &  1.64  & & &   1.88 &   1.88 &  1.89 &  1.87 &  1.79 &   0.94  &  0.25  &  0.00  \\
Er & 68  &  2.01  &  1.83  &  1.81  &  1.84  &  1.94  &  1.94  &  1.91 &  1.99  & & &   2.25 &   2.25 &  2.26 &  2.24 &  2.15 &   1.28  &  0.45  &  0.04  \\
Tm & 69  &  1.87  &  1.69  &  1.67  &  1.70  &  1.79  &  1.79  &  1.75 &  1.83  & & &   2.11 &   2.11 &  2.11 &  2.09 &  2.00 &   1.13  &  0.35  &  0.02  \\
Yb & 70  &  2.35  &  2.17  &  2.15  &  2.18  &  2.28  &  2.26  &  2.21 &  2.30  & & &   2.58 &   2.58 &  2.58 &  2.55 &  2.46 &   1.57  &  0.67  &  0.11  \\
Lu & 71  &  2.07  &  1.89  &  1.87  &  1.89  &  1.99  &  1.97  &  1.91 &  2.00  & & &   2.29 &   2.29 &  2.28 &  2.26 &  2.17 &   1.28  &  0.44  &  0.04  \\
Hf & 72  &  2.53  &  2.36  &  2.33  &  2.36  &  2.45  &  2.43  &  2.38 &  2.48  & & &   2.75 &   2.75 &  2.75 &  2.72 &  2.64 &   1.73  &  0.79  &  0.16  \\
Ta & 73  &  2.41  &  2.23  &  2.21  &  2.23  &  2.33  &  2.31  &  2.26 &  2.37  & & &   2.62 &   2.62 &  2.62 &  2.59 &  2.51 &   1.60  &  0.68  &  0.12  \\
W  & 74  &  2.53  &  2.36  &  2.33  &  2.36  &  2.46  &  2.45  &  2.41 &  2.52  & & &   2.74 &   2.74 &  2.75 &  2.72 &  2.64 &   1.73  &  0.79  &  0.16  \\
Re & 75  &  1.95  &  1.79  &  1.77  &  1.80  &  1.90  &  1.89  &  1.84 &  1.95  & & &   2.13 &   2.13 &  2.13 &  2.11 &  2.02 &   1.14  &  0.35  &  0.03  \\
Os & 76  &  1.83  &  1.66  &  1.64  &  1.68  &  1.78  &  1.77  &  1.73 &  1.83  & & &   2.03 &   2.04 &  2.04 &  2.02 &  1.94 &   1.05  &  0.28  &  0.01  \\
Ir & 77  &  0.99  &  0.83  &  0.82  &  0.85  &  0.94  &  0.94  &  0.90 &  0.98  & & &   1.18 &   1.18 &  1.19 &  1.17 &  1.09 &   0.34  &  0.03  &  0.00  \\
Pt & 78  &  1.59  &  1.43  &  1.41  &  1.45  &  1.55  &  1.54  &  1.49 &  1.58  & & &   1.79 &   1.79 &  1.80 &  1.78 &  1.69 &   0.81  &  0.16  &  0.00  \\
Au & 79  &  1.56  &  1.40  &  1.38  &  1.42  &  1.52  &  1.51  &  1.45 &  1.54  & & &   1.76 &   1.76 &  1.76 &  1.74 &  1.65 &   0.77  &  0.15  &  0.00  \\
Hg & 80  &  2.58  &  2.42  &  2.40  &  2.44  &  2.54  &  2.53  &  2.47 &  2.57  & & &   2.76 &   2.76 &  2.77 &  2.75 &  2.67 &   1.71  &  0.74  &  0.19  \\
Tl & 81  &  2.61  &  2.47  &  2.46  &  2.51  &  2.62  &  2.59  &  2.49 &  2.61  & & &   2.71 &   2.71 &  2.72 &  2.69 &  2.60 &   1.66  &  0.68  &  0.14  \\
Pb & 82  &  4.12  &  4.08  &  4.07  &  4.06  &  4.05  &  3.72  &  3.33 &  3.07  & & &   3.37 &   3.33 &  3.26 &  3.11 &  2.75 &   1.65  &  0.50  &  0.06  \\
Bi & 83  &  3.98  &  3.94  &  3.93  &  3.91  &  3.87  &  3.48  &  3.00 &  2.56  & & &   3.15 &   3.08 &  2.96 &  2.74 &  2.21 &   1.00  &  0.11  &  0.01  \\
\hline    
\end{tabular}}
\end{table*}

%% file: AppendixC.tex
\section{Online-only material}

\clearpage
\newpage

\begin{figure}
   \centering
\includegraphics[angle=-90,width=9cm]{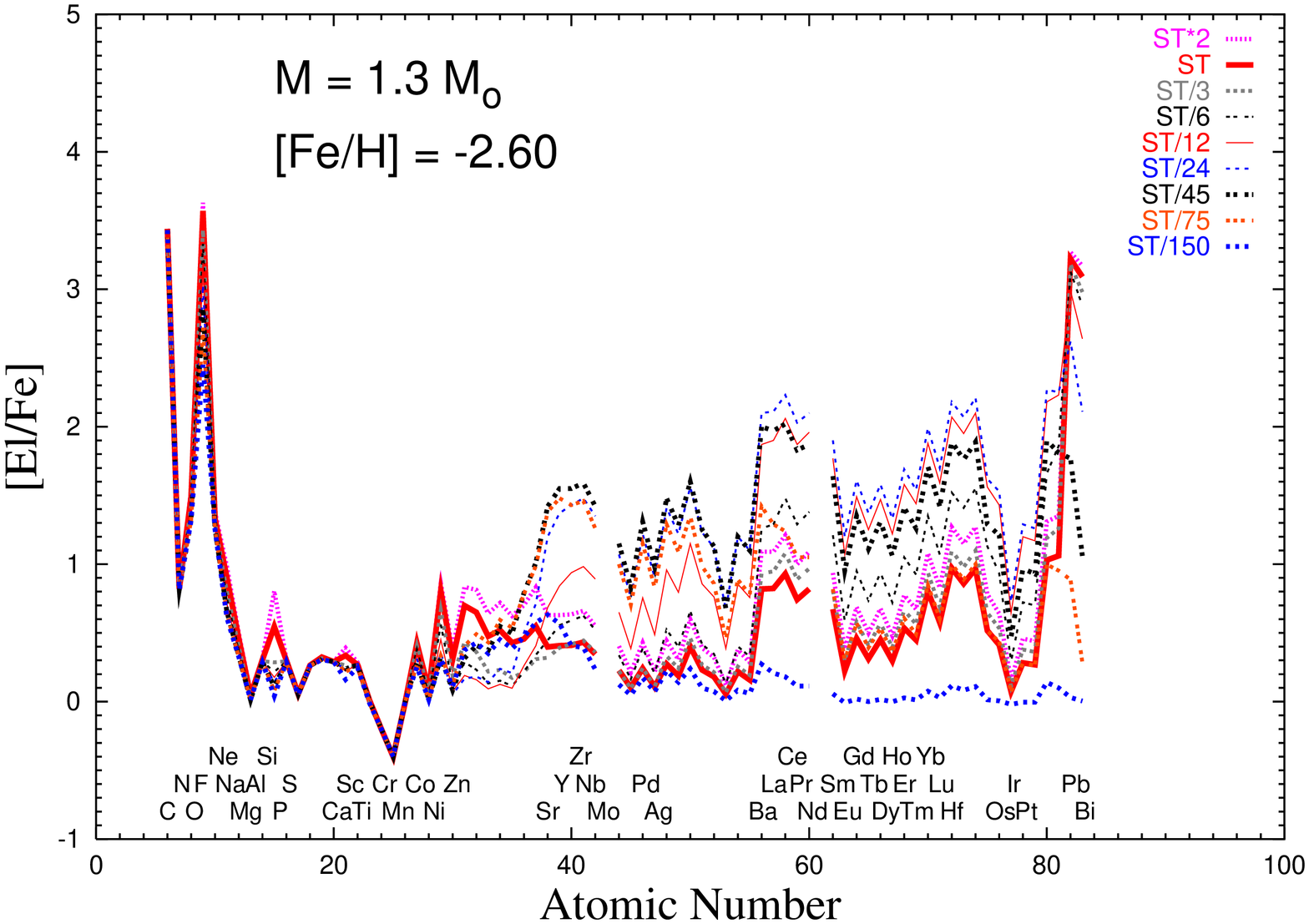}
\caption{Elemental composition in the envelope at the last TDU,
 for AGB models of initial mass $M$ = 1.3 $M_{\odot}$, 
initial metallicity [Fe/H] = $-$2.6, and different choices of 
$^{13}$C-pocket efficiency.}
\label{m1p5z5m5nro16eq_alcuniST_n5}
\end{figure}



\begin{figure}
   \centering
\includegraphics[angle=-90,width=8cm]{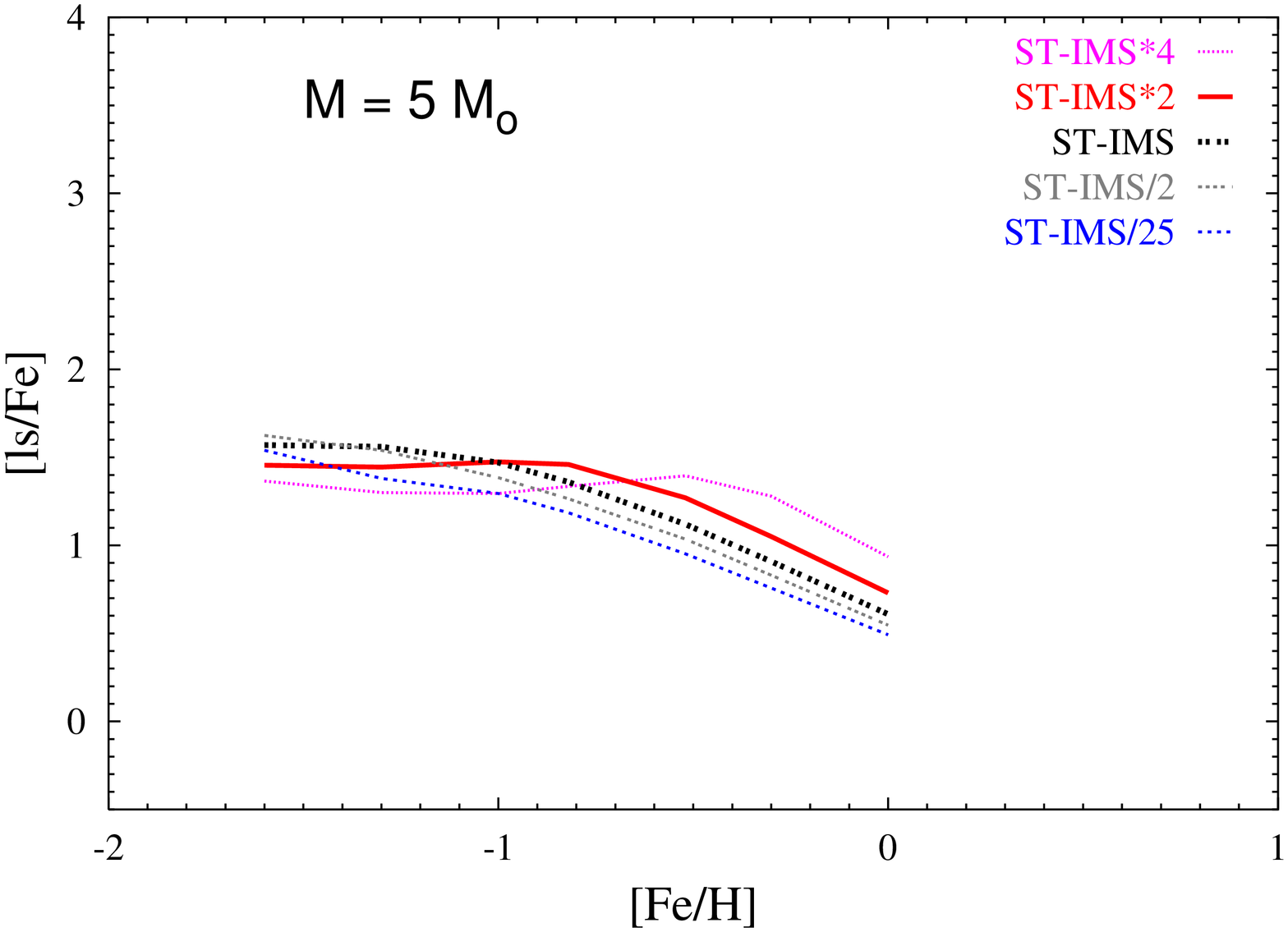}
\includegraphics[angle=-90,width=8cm]{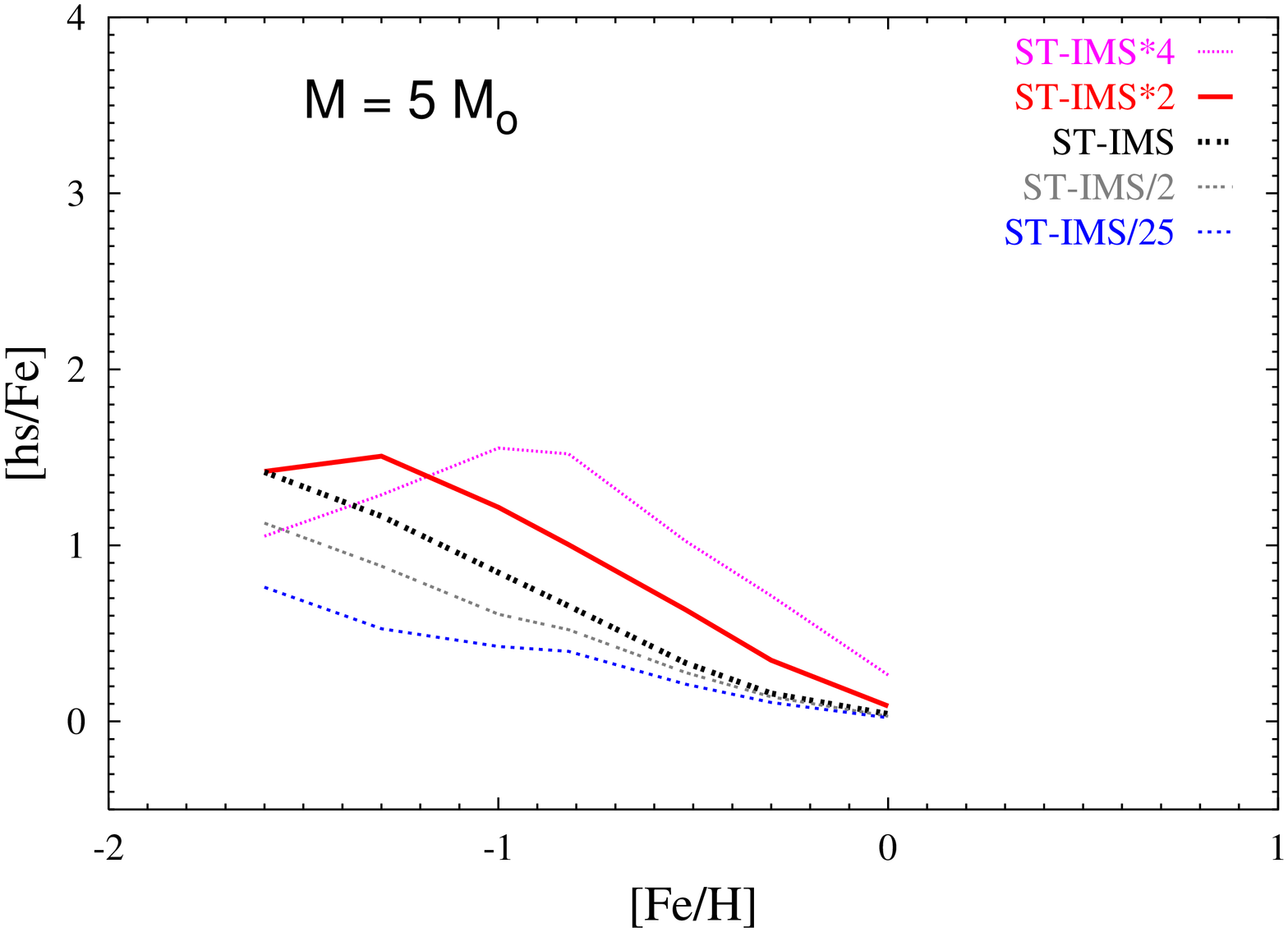}
\includegraphics[angle=-90,width=8cm]{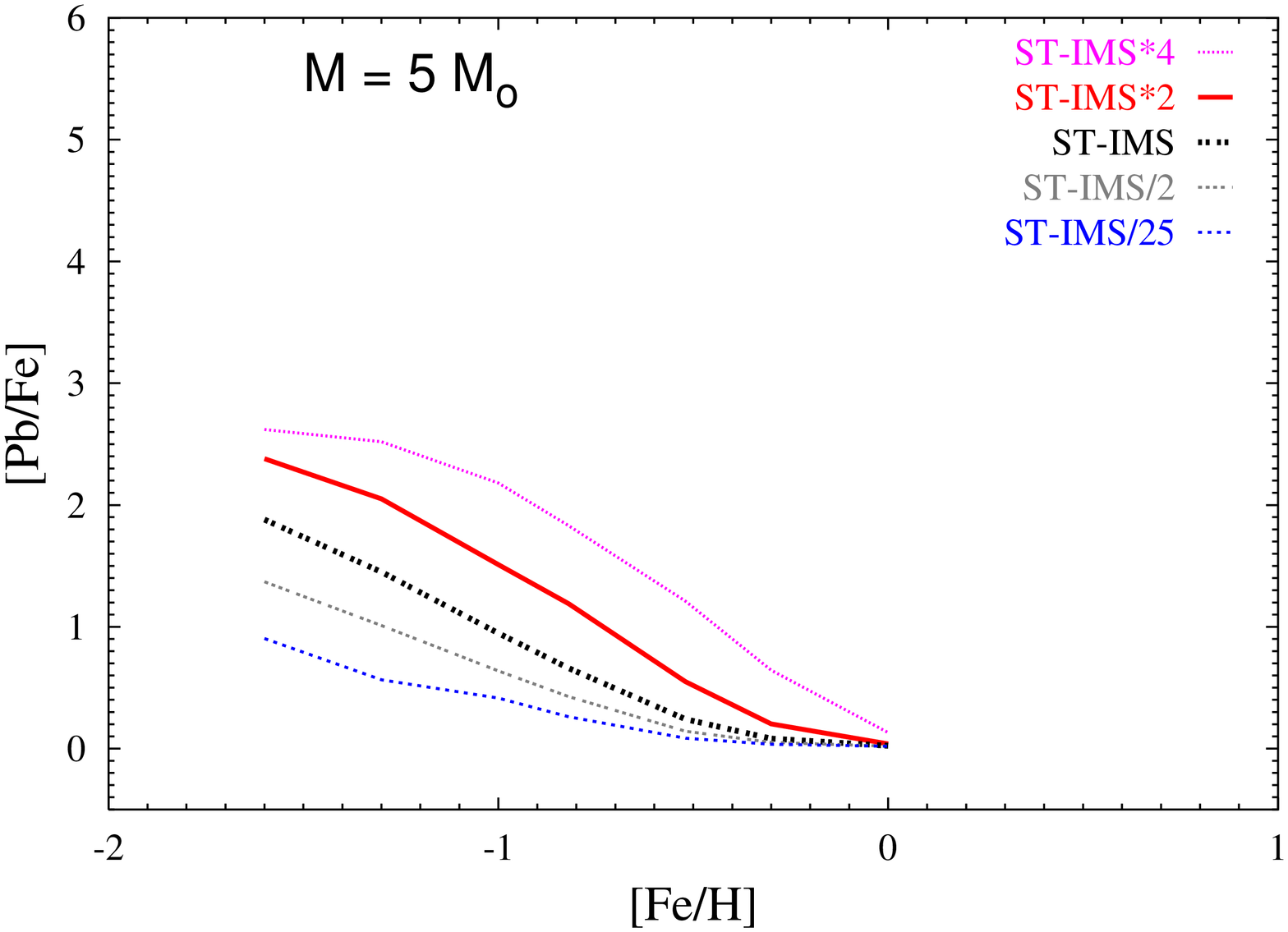}
\caption{Theoretical results of [ls/Fe], [hs/Fe], and [Pb/Fe] (\textit{top}, 
\textit{middle} and \textit{bottom panels}, respectively) 
for AGB models of initial mass $M$ = 5 $M_{\odot}$ 
and a range of $^{13}$C-pocket efficiencies (ST-IMS$\times$4 down to
ST-IMS/25) in the metallicity range $-$1.6 $\leq$ [Fe/H] $\leq$ 0.}
\label{AA_lssufe_m5_noobs}
\end{figure}

\begin{figure}
   \centering
\includegraphics[angle=-90,width=8cm]{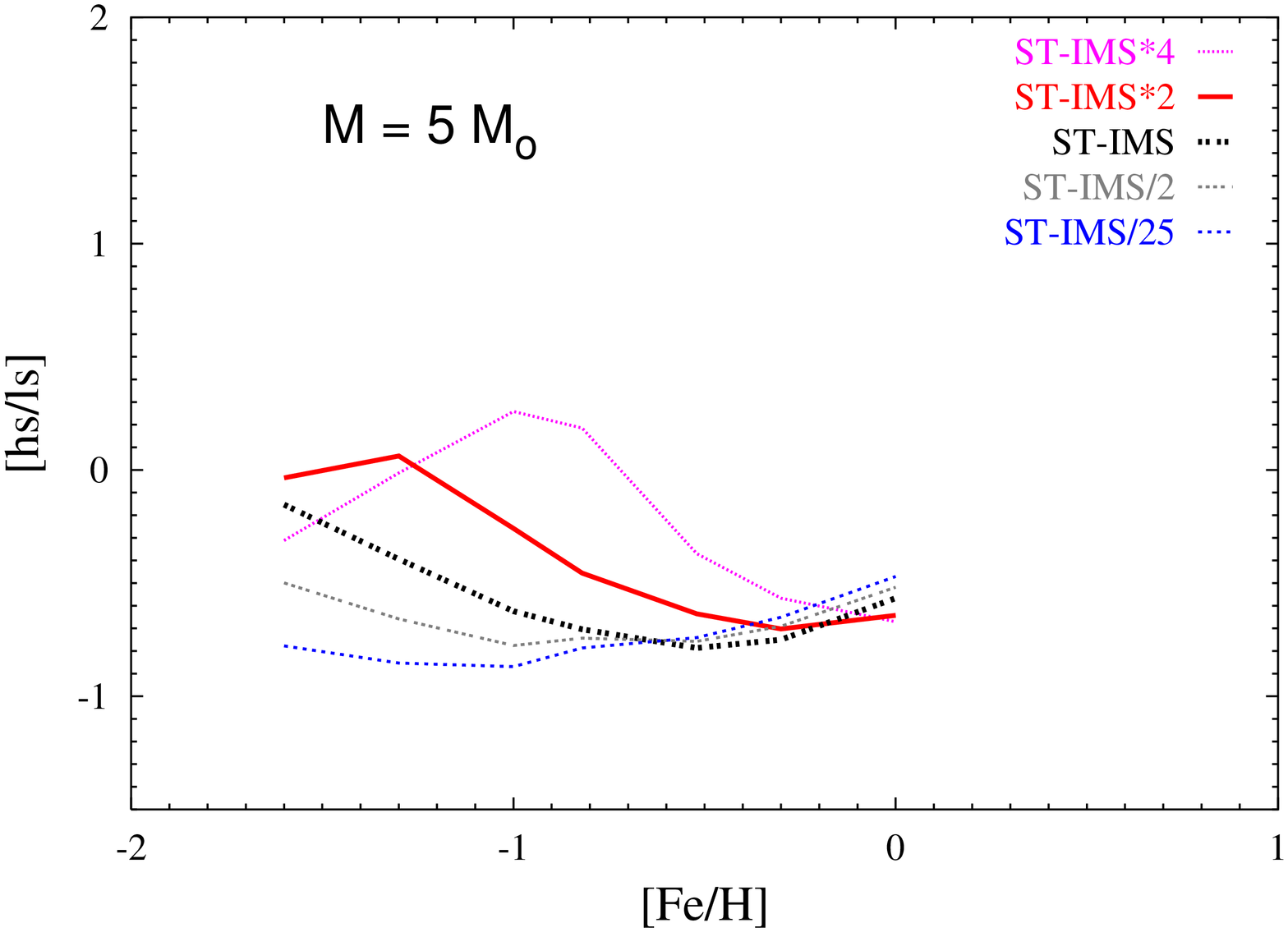}
\includegraphics[angle=-90,width=8cm]{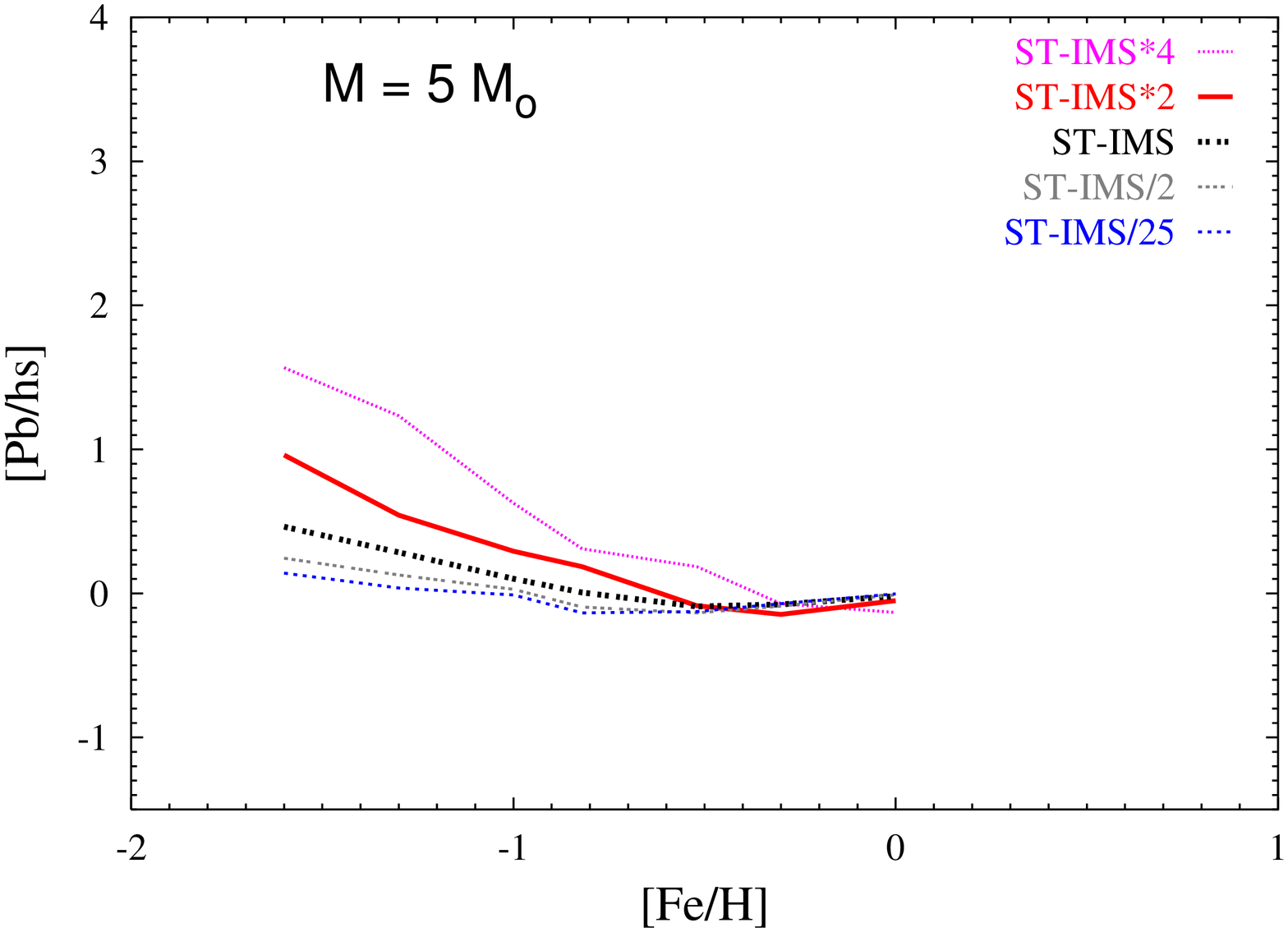}
\caption{The same as Fig.~\ref{AA_lssufe_m5_noobs}, but for the two
$s$-process indicators [hs/ls]
(\textit{top panel}) and [Pb/hs] (\textit{bottom panel}).}
\label{AA_pbsuhs_m5_noobs}
\end{figure}

\begin{figure}
   \centering
\includegraphics[angle=-90,width=8cm]{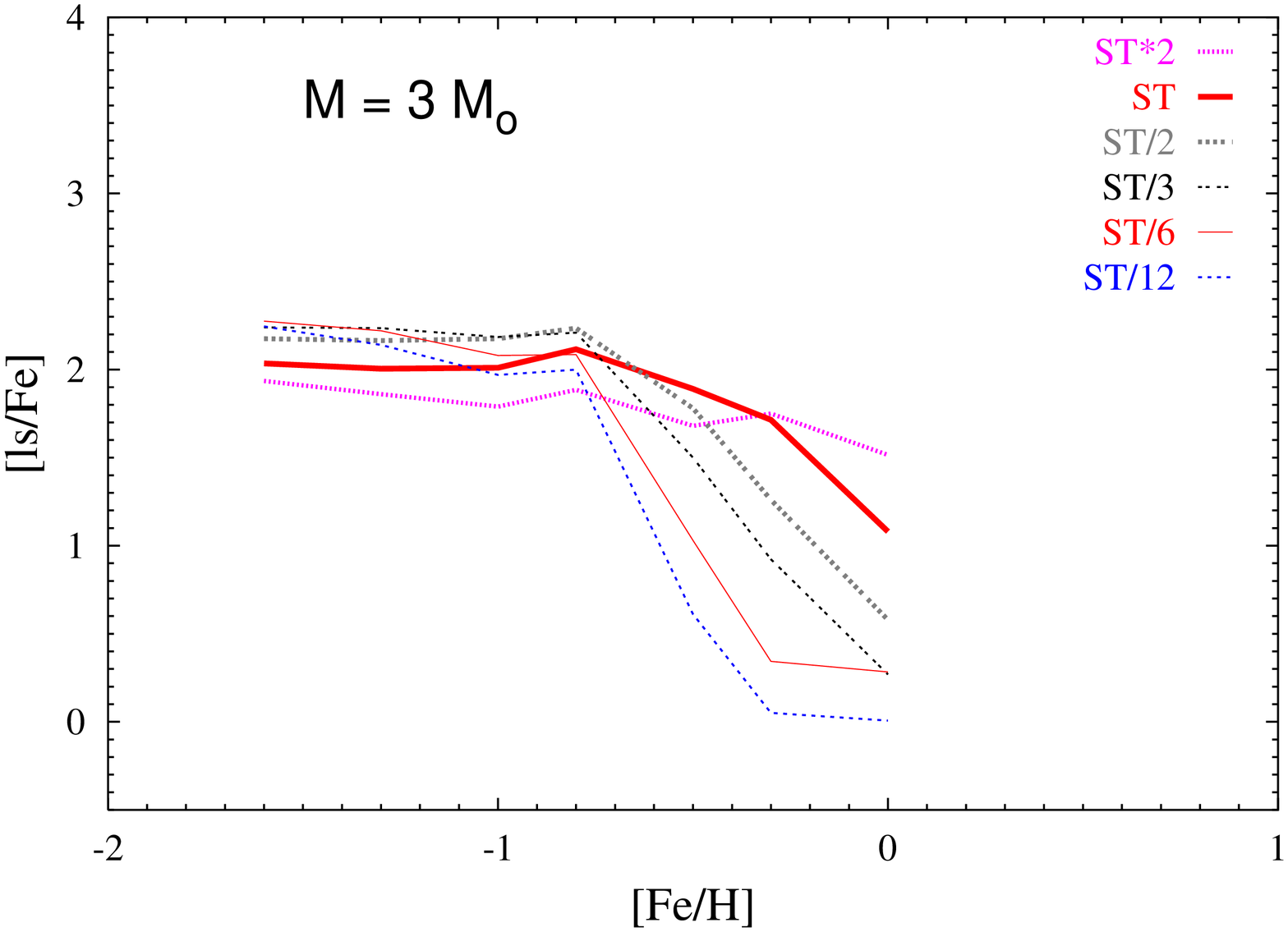}
\includegraphics[angle=-90,width=8cm]{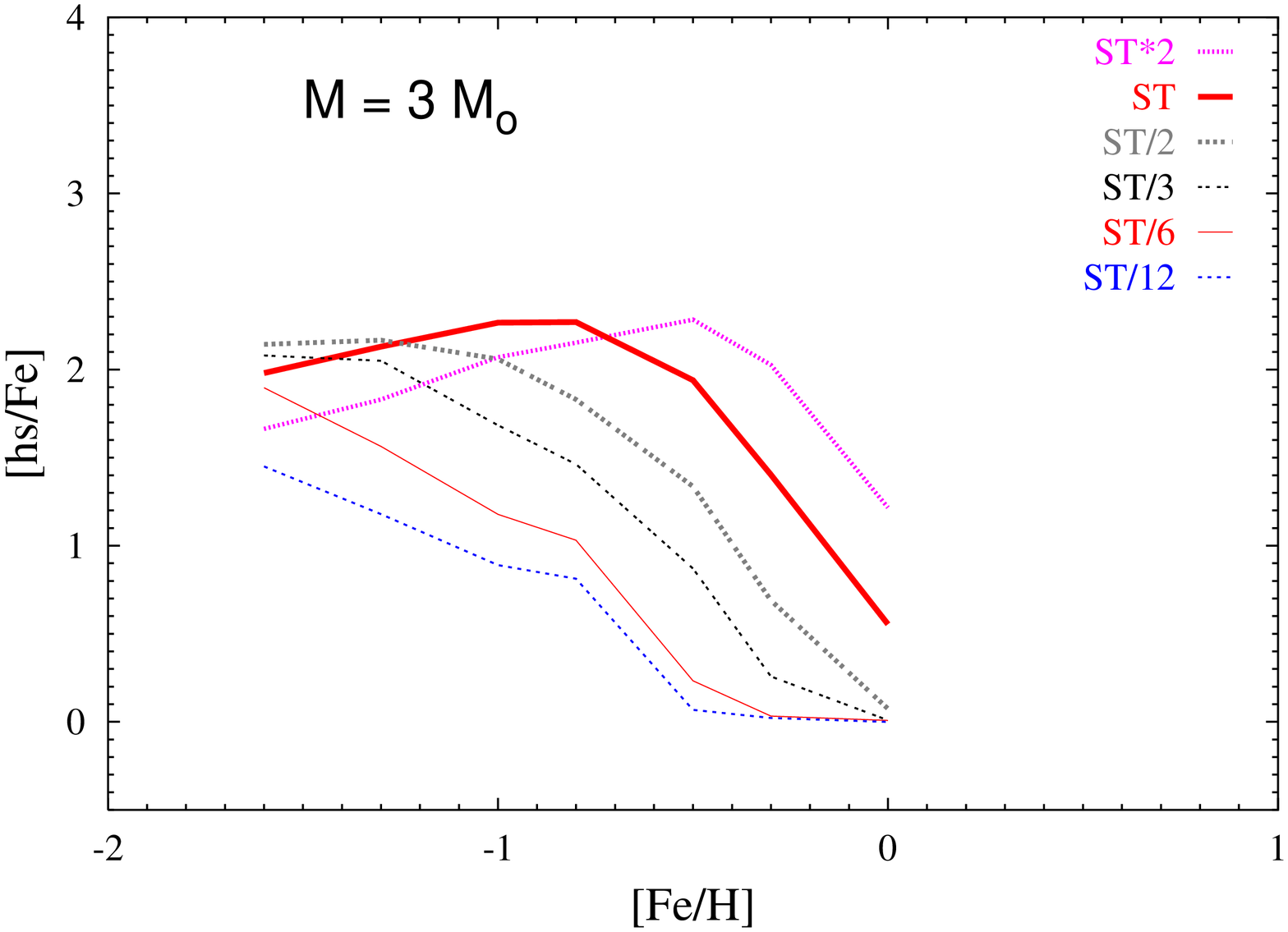}
\includegraphics[angle=-90,width=8cm]{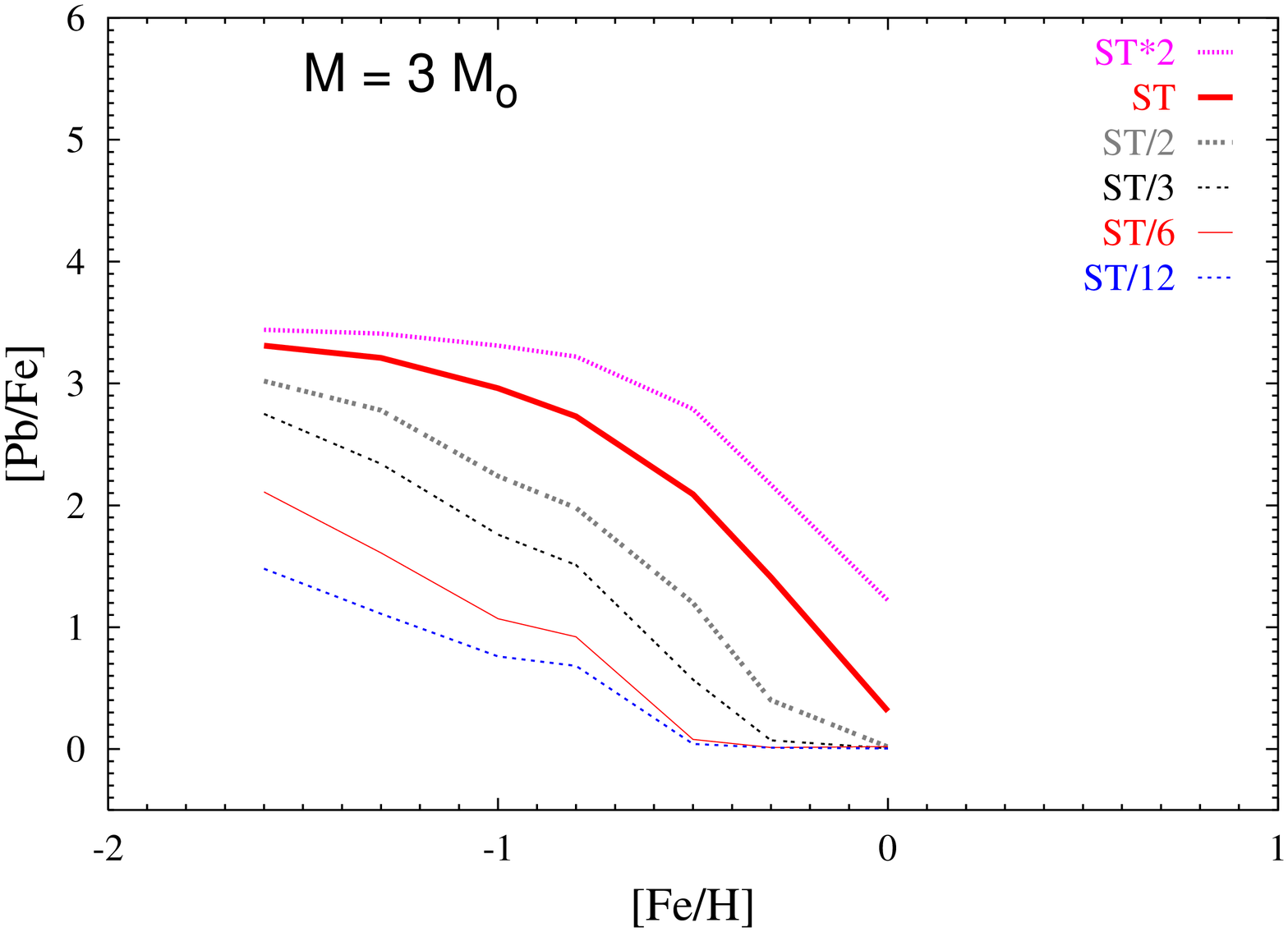}
\caption{Theoretical results of [ls/Fe], [hs/Fe], and [Pb/Fe] (\textit{top}, 
\textit{middle} and \textit{bottom panels}, respectively) 
for AGB models of initial mass $M$ = 3 $M_{\odot}$ 
and a range of $^{13}$C-pocket efficiencies (ST$\times$2 down to
ST/12) in the metallicity range $-$1.6 $\leq$ [Fe/H] $\leq$ 0.}
\label{AA_m3novlz_noobs_disk}
\end{figure}

\begin{figure}
   \centering
\includegraphics[angle=-90,width=8cm]{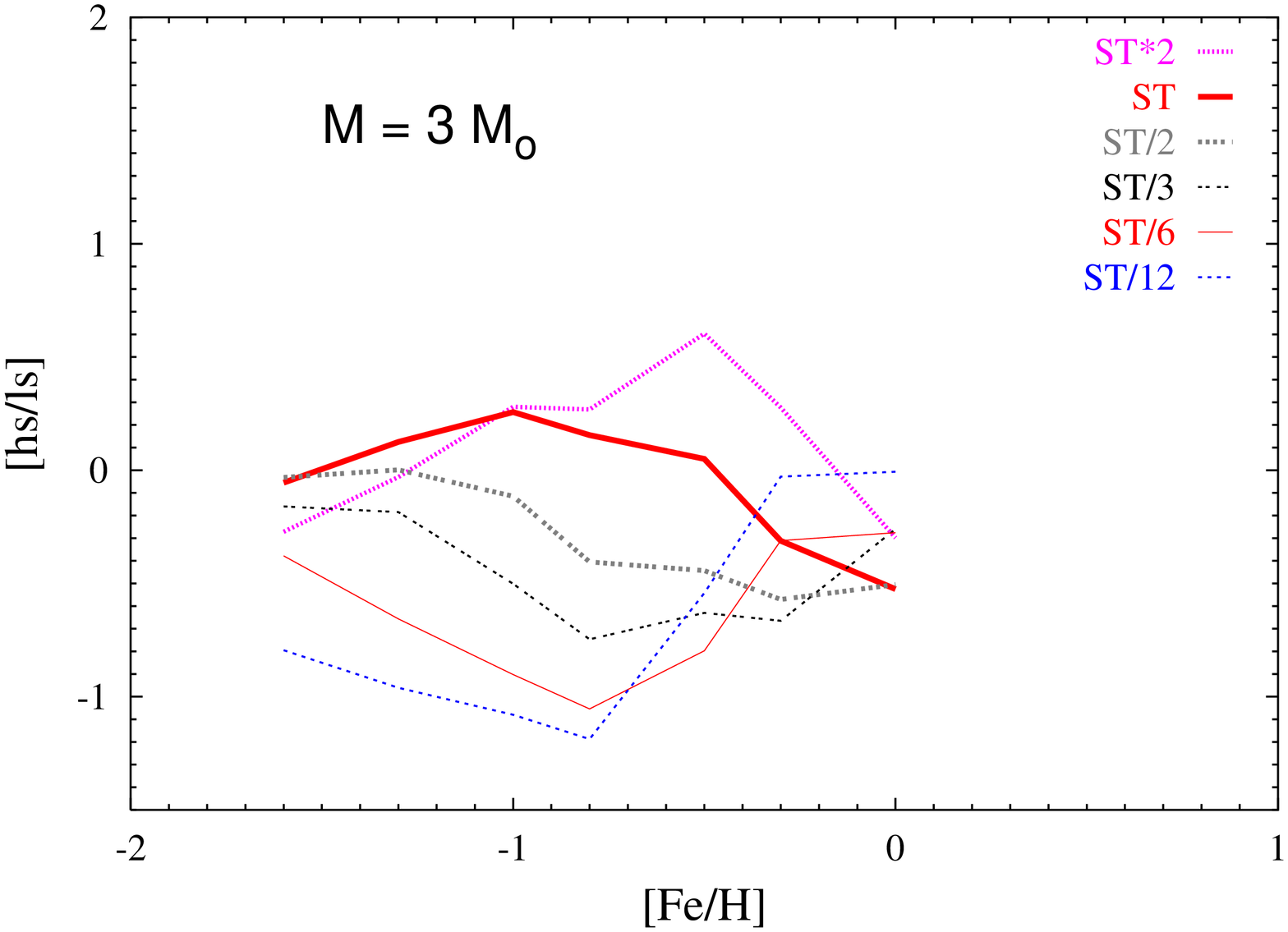}
\includegraphics[angle=-90,width=8cm]{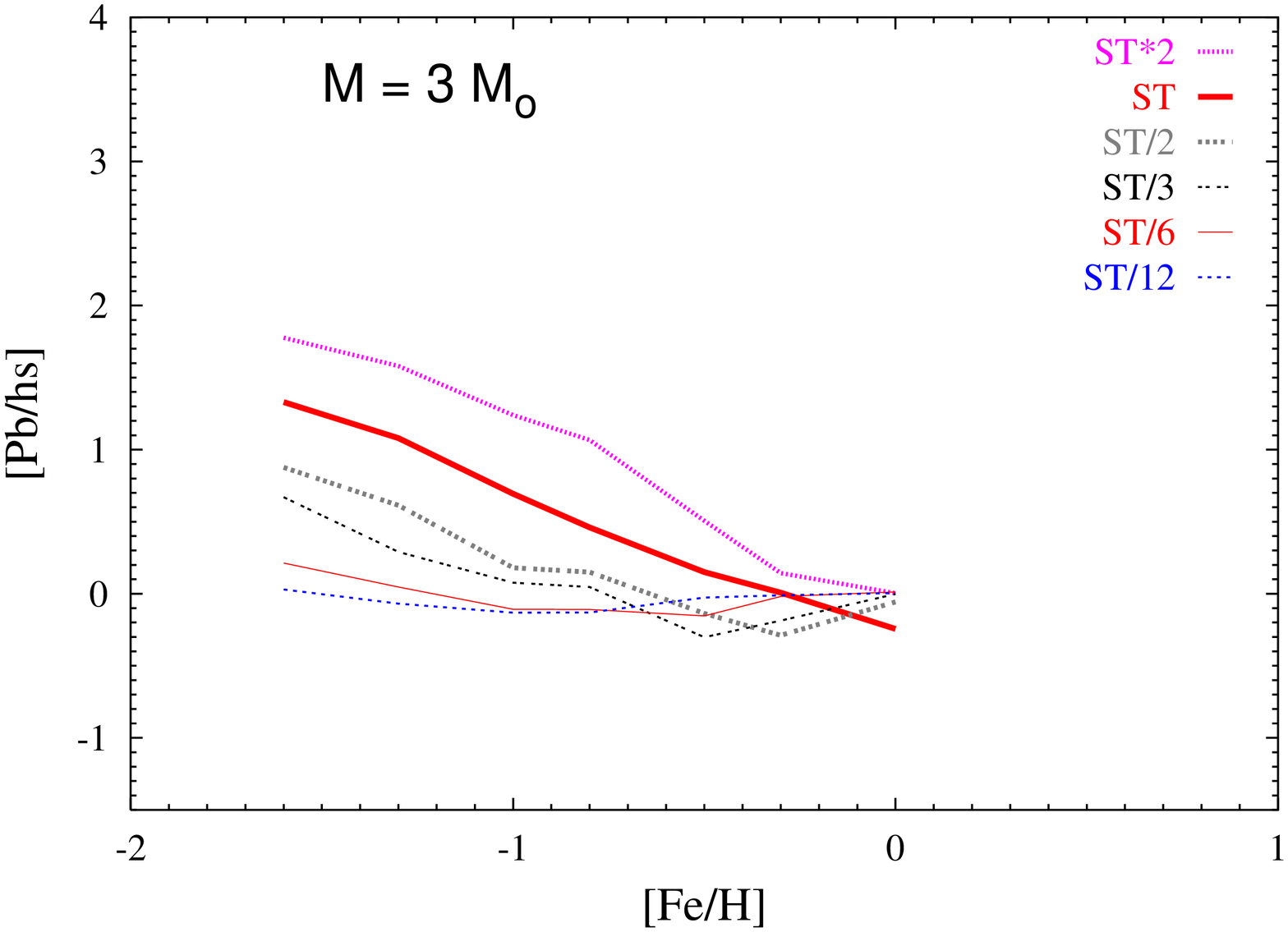}
\caption{The same as Fig.~\ref{AA_m3novlz_noobs_disk}, but for the two
$s$-process indicators [hs/ls] (\textit{top panel}) and [Pb/hs] 
(\textit{bottom panel}).}
\label{AA_m3novlz_noobs_disk1}
\end{figure}

\begin{figure}
   \centering
\includegraphics[angle=-90,width=9cm]{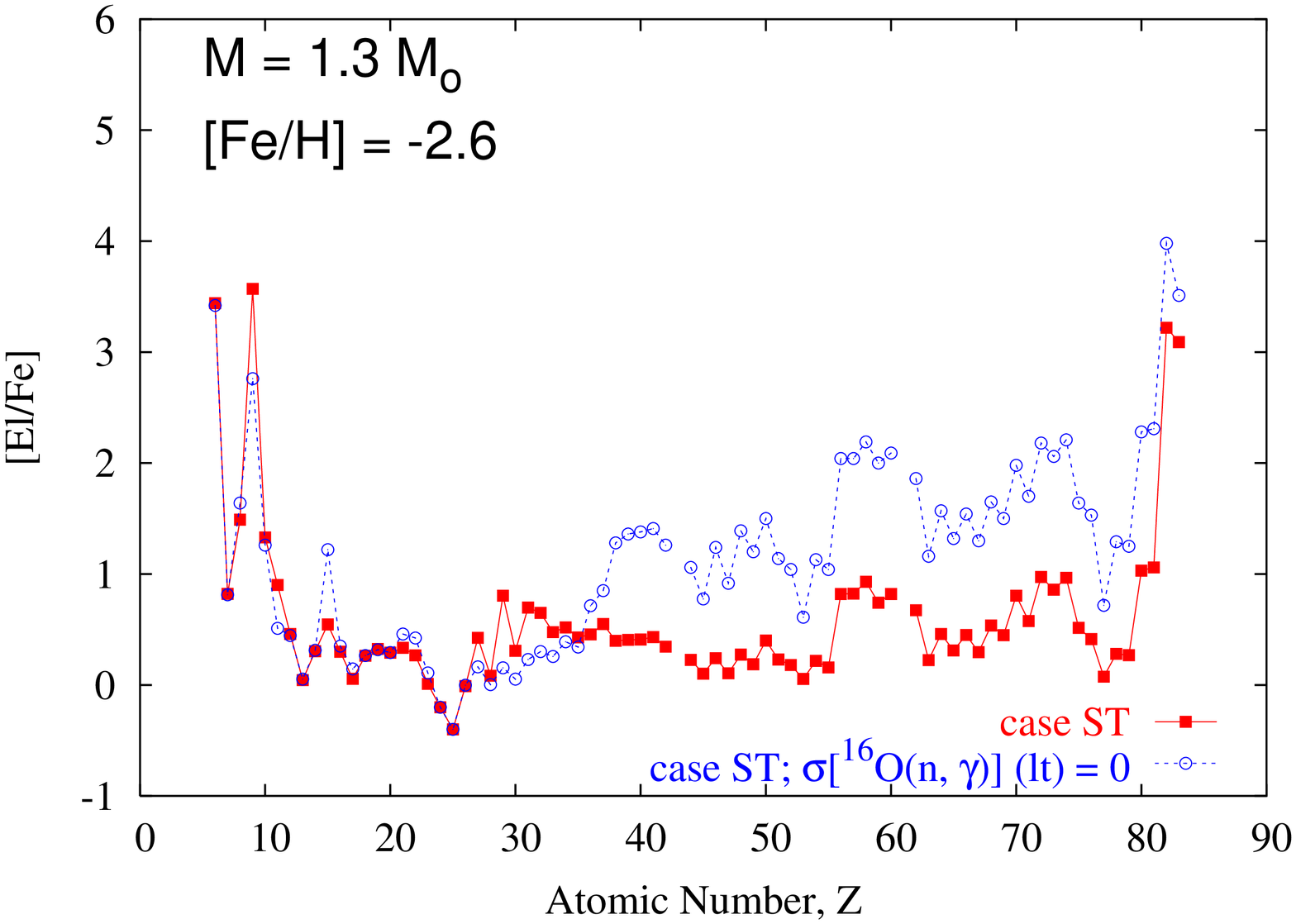}
\includegraphics[angle=-90,width=9cm]{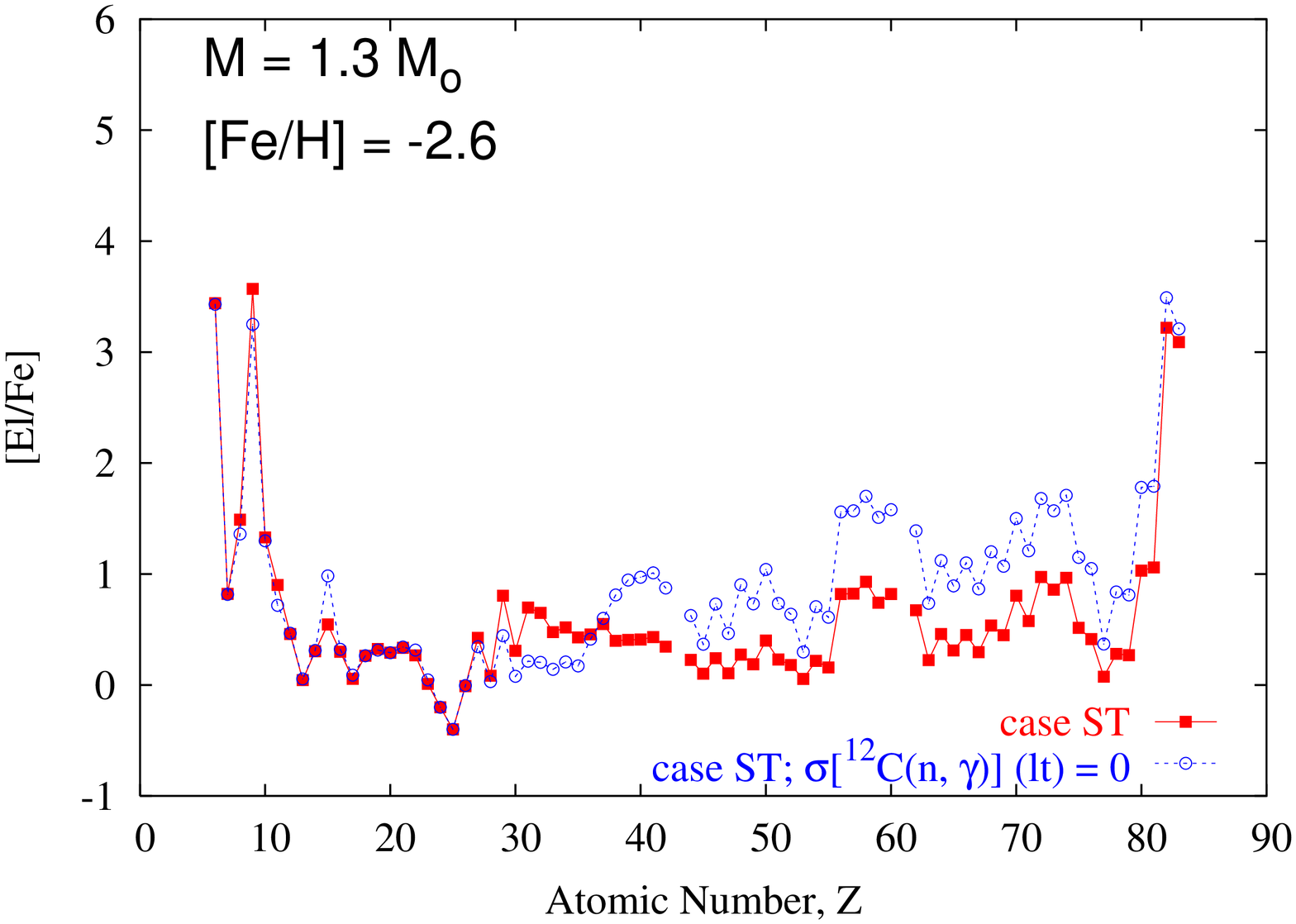}
\caption{{\it Top panel:} theoretical predictions for AGB models 
of $M$ = 1.3 $M_{\odot}$, a case ST and [Fe/H] = $-$2.6. The line with full squares stands for the ordinary case, while the line with empty circles corresponds to a case with the cross 
 section of $^{16}$O(n, $\gamma$) put to zero.
 {\it Bottom panel:} the same as top panel, but the line with empty circles corresponds to a case with the cross 
 section of $^{12}$C(n, $\gamma$) put to zero.}
\label{o16c12m1p3}
\end{figure}